\documentclass[12pt,twoside]{book}

\input epsf.sty

\usepackage{amsmath,amsfonts}
\usepackage{amssymb,amsfonts}

\usepackage{fancyheadings}
\usepackage{these}

\begin{document}
\cfoot{}

% Definitions for several articles (in TeX) 

%\def\one{\hbox{1\kern-.25em l}}

\def\Adotx{{\dot A}_x}
\def\Adot{{\dot A}}
\def\Asd{{\dot A}_\sigma}
\def\Gbar{\bar{G}}
\def\Lbar{\bar{L}}
\def\Mpl{M_{pl}}
\def\Mtp{{\tilde M}_{pl}}
\def\Mts{{\tilde M}_s}
\def\Nisf{{\cal N} =  4}
\def\Rele{R_{11}}
\def\Rn{R_9}
\def\Rtn{{\tilde R}_9}
\def\Rvet{\mathbb{R}}
\def\Tvet{\mathbb{T}}
\def\Tr{{\rm Tr}}
\def\Xbar{{\overline{X}}}
\def\Xdot{{\dot X}}
\def\Xtdot{\dot{\tilde X}}
\def\Xt{{\tilde X}}
\def\Zvet{\mathbb{Z}}
\def\cH{{\cal H}} 
\def\diagonal{{\rm diag}}
\def\gym{g_{YM}}
\def\hf{\frac{1}{2}}
\def\is{ & =}
\def\lpl{l_{pl}}
\def\lst{l_{st}}
\def\ltp{{\tilde l}_{pl}}
\def\mth{M-theory}
\def\one{\hbox{1\kern-.25em l}}
\def\pele{p_{11}}
\def\phat{{\hat p}}
\def\pperp{{{}_{\! \perp}}}
\def\qi{/\kern -0.65em \bar D}
\def\qA{/\kern -0.65em \tilde A}
\def\roughly#1{\raise.3ex\hbox{$#1$\kern-.75em\lower1ex\hbox{$\sim$}}}
\def\wbar{{\overline{w}}}
\def\zbar{{\overline{z}}}
\def\tr{{\rm tr}}

%%% Some clumsy trick for eqnarray 
\newlength{\baselinetruc}
\setlength{\baselinetruc}{\the\baselineskip}
\global\divide\baselinetruc by 5 
\global\multiply\baselinetruc by 8
%%% 

%%% Harvmac \eqn partially adapted to latex 

\def\eqn#1#2
{\begin{equation} #2 \end{equation}\xdef#1{(\theequation)\ }}

\def\eqnar#1#2
{\begin{eqnarray} #2 \end{eqnarray}\xdef#1{(\theequation)\ }}

%%%

%%% My own way of making references 
%%% simplified adaption of harmac to latex

\newcounter{citatieno}

\def\beginitem{\begin{list}{[\arabic{citatieno}]}{\usecounter{citatieno}}}
\def\enditem{\end{list}}

\newwrite\rfile

\global\newcount\referentie 
\global\referentie=0
\openout\rfile=referenties.tmp
\write\rfile{\noexpand\beginitem}

\def\lref#1#2
{\global\advance\referentie by 1
\write\rfile{{\noexpand\item #2\ }}
\xdef#1{[\number\referentie]}
}

\def\listrefs{
\newpage \ifodd\thepage \relax \else 
\phantom{bla} \newpage \fi 
\write\cfile{\noexpand\\[2mm]
\bf Bibliography &&\bf \thepage\noexpand\\[2mm]}
\lhead[\fancyplain{}{\sc Bibliography}]{\fancyplain{}{}}
\rhead[\fancyplain{}{}]{\fancyplain{}{\sc }}
\thispagestyle{empty}
{\phantom{tk} \hfill \Huge\rmfamily 
\bfseries\slshape Bibliography}\\[15mm]
%\noindent{\Large\bf Bibliography} \\ 
\immediate\write\rfile{\noexpand\enditem}
\immediate\closeout\rfile
\input referenties.tmp
}
%%%

%%% Definition of table of contents

\global\newcount\clineno \global\clineno=0 

\def\testclino{\global\advance\clineno by 1
\ifnum\the\clineno<33 \relax\else \global\clineno=0 
\write\cfile{\noexpand\end{tabular}\noexpand\newpage
\noexpand\noindent\noexpand\hskip-\the\tabcolsep
\noexpand\begin{tabular}{lcr}
}\fi}

\def\newtableofcontents{
\immediate\newread\leesfile
\immediate\openin\leesfile=contents.tmp \ifeof\leesfile\relax\else 
\lhead[\fancyplain{}{\sc Contents}]{\fancyplain{}{}}
{\phantom{tk} \hfill \Huge\rmfamily 
\bfseries\slshape Contents}
\\[15mm]
%\noindent\centerline{\Huge\bf Contents}\\[15mm]  
\input contents.tmp\fi
\closein\leesfile
\newwrite\cfile \openout\cfile=contents.tmp
\write\cfile{\noexpand\thispagestyle{empty} 
\noexpand\hskip-\the\tabcolsep
\noexpand\begin{tabular}{lcr}}
%\write\cfile{} 
%\closeout\cfile 
}

%%% Definition of figures 
\global\newcount\fignumber
\global\fignumber=0
\global\newcount\hfignumber
\global\hfignumber=0
\global\newcount\nfignumber
\global\nfignumber=0

\def\hfigure#1#2{
\global\advance\fignumber by 1 
\noindent\\ \centerline{\epsfbox{#1}} \\[2mm]
\centerline{\vbox{\baselineskip12pt
\advance\hsize by -1truein\noindent{\small 
{\bf Figure~\thechapter.\the\fignumber\ } #2}
}}
}
%%%
\def\hhfigure#1#2{
\global\advance\hfignumber by 1 
\noindent\\ \centerline{\epsfbox{#1}} \\[2mm]
\centerline{\vbox{\baselineskip12pt
\advance\hsize by -1truein\noindent{\small 
{\bf Figure~\the\hfignumber\ } #2}
}}
}
\def\nfigure#1#2{
\global\advance\nfignumber by 1 
\noindent\\ \centerline{\epsfbox{#1}} \\[2mm]
\centerline{\vbox{\baselineskip12pt
\advance\hsize by -1truein\noindent{\small 
{\bf Figuur~\the\nfignumber\ } #2}
}}
}

%%% Definition of tables 

\global\newcount\tablenumber
\global\tablenumber=0

\def\htable#1#2{
\global\advance\tablenumber by 1 
\noindent\\ \centerline{\epsfbox{#1}} \\[3mm]
\centerline{\vbox{\baselineskip12pt
\advance\hsize by -1truein\noindent{\small 
{\bf Table~\thechapter.\the\tablenumber\ } #2}
}}
}
%%% 

%%% Definition of chapter, section & subsection

\def\newchapter#1#2#3{\chapter{#2} \xdef#1{\thechapter\ } 
\global\fignumber=0 \setcounter{subsection}{0} 
\testclino 
\write\cfile{\noexpand\phantom{\bf 3~~}\noexpand\phantom{\bf 
High energy scattering in matrix string theory\hskip65pt} &&\noexpand\\
\bf\thechapter~~\bf #3 &  & 
\hskip42.5pt \bf \thepage\noexpand\\[1.5mm]}
 }
%%%
\def\newsection#1#2{\setcounter{subsection}{0} 
\addtocounter{section}{1}\markright{\small\sc\thesection~#2} 
\testclino
\write\cfile{\indent\thesection~~#2 && \thepage\noexpand\\[1mm]}
\section*{\bf\sc\thesection~~#2} \xdef#1{\thesection\ }
}
%%%
\def\newsubsection#1#2{\addtocounter{subsection}{1}
%\subsection*{\bf\sc\thesubsection~#2} \xdef#1{\thesubsection\ }
%\advance\newsubsectionno by 1
\testclino
%\write\cfile{&\hskip28pt}
\subsection*{\bf\sc\thesubsection~#2} \xdef#1{\thesubsection\ }
\write\cfile{\indent~~~~~~#1~~#2&&\thepage\noexpand\\[1mm]}
}
\def\newappendix#1{\section*{\bf\sc #1}}

%%% Read references. Note all references will be listed by using \listrefs

% References 
%
% Journals
%
\def\cmp#1#2#3{Comm.\ Math.\ Phys.\ {{\bf #1}} {(#2)} {#3}}
\def\pl#1#2#3{Phys.\ Lett.\ {{\bf #1}} {(#2)} {#3}}
\def\np#1#2#3{Nucl.\ Phys.\ {{\bf #1}} {(#2)} {#3}}
\def\prd#1#2#3{Phys.\ Rev.\ {{\bf #1}} {(#2)} {#3}}
\def\prl#1#2#3{Phys.\ Rev.\ Lett.\ {{\bf #1}} {(#2)} {#3}}
\def\ijmp#1#2#3{Int.\ J.\ Mod.\ Phys.\ {{\bf #1}} 
{(#2)} {#3}}
\def\mpl#1#2#3{Mod.\ Phys.\ Lett.\ {{\bf #1}} {(#2)} {#3}}
\def\jdg#1#2#3{J.\ Differ.\ Geom.\ {{\bf #1}} {(#2)} {#3}}
\def\pnas#1#2#3{Proc.\ Nat.\ Acad.\ Sci.\ USA.\ {{\bf #1}} 
{(#2)} {#3}}
\def\top#1#2#3{Topology {{\bf #1}} {(#2)} {#3}}
\def\zp#1#2#3{Z.\ Phys.\ {{\bf #1}} {(#2)} {#3}}
\def\prp#1#2#3{Phys.\ Rep.\ {{\bf #1}} {(#2)} {#3}}
\def\ap#1#2#3{Ann.\ Phys.\ {{\bf #1}} {(#2)} {#3}}
\def\ptrsls#1#2#3{Philos.\ Trans.\  Roy.\ Soc.\ London
{{\bf #1}} {(#2)} {#3}}
\def\prsls#1#2#3{Proc.\ Roy.\ Soc.\ London Ser.\
{{\bf #1}} {(#2)} {#3}}
\def\am#1#2#3{Ann.\ Math.\ {{\bf #1}} {(#2)} {#3}}
\def\mm#1#2#3{Manuscripta \ Math.\ {{\bf #1}} {(#2)} {#3}}
\def\ma#1#2#3{Math.\ Ann.\ {{\bf #1}} {(#2)} {#3}}
\def\cqg#1#2#3{Class. Quantum Grav..\ {{\bf #1}} {(#2)} {#3}}
\def\im#1#2#3{Invent.\ Math.\ {{\bf #1}} {(#2)} {#3}}
\def\plms#1#2#3{Proc.\ London Math.\ Soc.\ {{\bf #1}} 
{(#2)} {#3}}
\def\dmj#1#2#3{Duke Math.\  J.\ {{\bf #1}} {(#2)} {#3}}
\def\bams#1#2#3{Bull.\ Am.\ Math.\ Soc.\ {{\bf #1}} {(#2)} 
{#3}}
\def\jgp#1#2#3{J.\ Geom.\ Phys.\ {{\bf #1}} {(#2)} {#3}}
\def\ajou#1&#2(#3){\ \sl#1\bf#2\rm(19#3)}
\def\jou#1&#2(#3){,\ \sl#1\bf#2\rm(19#3)}
\def\hep#1{{hep-th/}#1}
\def\atmp#1#2#3{Adv.\ Theor.\ Math.\ Phys.\ {\bf #1} {(#2)} {#3}}
\def\jhep#1#2#3{J.\ High\ Energy\ Phys.\ {\bf #1} {(#2)} {#3}}
%
%
%
% References 
%
% More or less in alphabetic order
%
%
\lref\aganagic{
M. Aganagic, C. Popescu and J.H. Schwarz, 
{\it Gauge-Invariant and Gauge-Fixed D-Brane Actions}, 
\np{B495}{1997}{99}, \hep{9612080}. 
}
\lref\AiSe{
P.C. Aichelburg and R.U. Sexl, 
{\it On the gravitational field of a massless particle}, 
Gen.\ Rel.\ Grav.\ {\bf 2} (1971) 303.
}
\lref\Ven{
D. Amati, M. Ciafaloni, and G. Veneziano, 
{\it Superstring collisions at Planckian energies}, 
\pl{B197}{1987}{81}.
}
\lref\aruti{
G.E. Arutyunov and S.A. Frolov, 
{\it Virasoro Amplitude from the $S^NR^{24}$ Or\-bi\-fold Sigma Model}
, Theor.\ Math.\ Phys.\ {\bf 114} (1998) 43,  
\hep{9708129} 
; {\it Four Graviton Scattering Amplitude from $S^N {\bf R}^8$ 
Supersymmetric Orbifold Sigma Model}, \np{B524}{1998}{159}, 
\hep{9712061}. 
}
\lref\arutii{
G.E. Arutyunov S.A. Frolov and A. Polishchuk, 
{\it On Lorentz Invariance and Supersymmetry of Four Particle 
Scattering Amplitudes in $S^N R^8$ Orbifold Sigma Model}, 
\hep{9812119}.  
}
\lref\bachas{
C. Bachas, {\it D-brane dynamics}, \pl{B374}{1996}{37}, 
\hep{9511043}. 
}
\lref\bachasimp{
C. Bachas, {\it (Half) a lecture on D-branes}, 
in {\sl Gauge theories, Applied supersymmetry and Quantum Gravity II
}, eds.~A.~Sevrin, K.S.~Stelle, K.~Thielemans and A.~van~Proeyen, 
Imperial College Press (1997), \hep{9701019}.
}
\lref\BGL{
V. Balasubramanian, R. Gopakumar and F. Larsen, 
{\it Gauge theory, geometry and the large N limit},  
\np{B526}{1998}{415}, \hep{9712077}.
}
\lref\bfss{T. Banks, W. Fischler, S.H. Shenker and L. Susskind, 
{\it M theory as a Matrix Model: a Conjecture}, 
\prd{D55}{1997}{5112}, \hep{9610043}.}
\lref\tomnati{T. Banks and N. Seiberg, 
{\it Strings from Matrices},
\np{B497}{1997}{41}, \hep{9702187}.
} 
\lref\mmatrix{T. Banks, N. Seiberg, and S. Shenker,
{\it Branes from Matrices}, 
\np{B490}{1997}{91}, \hep{9612157}.
}
\lref\BaSu{T.Banks and L.Susskind, 
{\it Brane - anti-brane forces}, \hep{9511194}.
}
\lref\BeBe{K. Becker and M. Becker, 
{\it A two-loop test of M(atrix) theory}, 
\np{B506}{1997}{48}, \hep{9705091}. 
}
\lref\BBPT{K. Becker, M. Becker, J. Polchinski, and A. Tseytlin, 
{\it Higher order graviton scattering in M(atrix) theory},  
\prd{D56}{1997}{3174}, \hep{9706072}. 
}
\lref\BeCo{
D. Berenstein and R. Corrado, {\it M(atrix)-theory in various
dimensions}, \pl{B406}{1997}{37}, \hep{9702108}. 
}
\lref\bergshoeff{
E. Bergshoeff, C.M. Hull and T. Ortin, 
{\it Duality in the Type--II Superstring Effective Action}, 
\np{B451}{1995}{547}, \hep{9504081}. 
}
\lref\BiSu{
D. Bigatti and L. Susskind, {\it Review of matrix theory}, 
\hep{9712072}.
} 
\lref\bonelli{G. Bonelli, L. Bonora and F. Nesti, 
{\it String Interactions from Matrix String Theory}, 
\np{B538}{1999}{100}, \hep{9807232}. 
}
\lref\braam{P. J. Braam and P. van Baal, 
{\it Nahm's Transformation for Instantons}, \cmp{122}{1989}{267}; 
P. van Baal, {\it Instanton Moduli for $T^3\times R$},
Nucl. Phys. Proc. Suppl. 49 (1996) 238, \hep{9512223}.
}
\lref\connes{
A. Connes, M.R. Douglas and A. Schwarz, 
{\it Noncommutative geometry and Matrix theory: Compactification 
on tori}, \jhep{9802}{1998}{003}, \hep{9711162}.
}
\lref\cremmer{
E. Cremmer, B. Julia and J. Scherk, 
{\it Supergravity theory in eleven-dimensions}, 
\pl{76B}{1978}{409}. 
}
\lref\dailp{
J. Dai, R. Leigh and J. Polchinski, 
{\it New connections between string theories}, \mpl{A4}{1989}{2073}.   
}
\lref\danielsson{U.H. Danielsson, G. Ferretti and B. Sundborg, 
{\it D-particle Dynamics and Bound States}, 
\ijmp{A11}{1996}{5463}, \hep{9603081}.
}
\lref\dewiti{
B. de Wit, J. Hoppe and H. Nicolai, 
{\it On the quantum mechanics of supermembranes},
\np{B305}{1988}{545}. 
}
\lref\dewitii{
B. de Wit, M. L\"uscher and H. Nicolai, 
{\it The supermembrane is unstable}, 
\np{B320}{1989}{135}.
}
\lref\fivebraneo{R. Dijkgraaf, E. Verlinde and H. Verlinde,
{\it BPS Spectrum of the Five-Brane and Black Hole Entropy},
\np{B486}{1997}{77}, \hep{9603126}. 
}
\lref\fivebrane{R. Dijkgraaf, E. Verlinde and H. Verlinde, 
{\it BPS Quantization of the Five-Brane}, 
\np{B486}{1997}{89}, \hep{9604055}.}
\lref\dmvv{R. Dijkgraaf, G. Moore, E. Verlinde and H. Verlinde,
{\it Elliptic Genera of Symmetric Products and Second Quantized Strings}, 
\cmp{185}{1997}{197}, \hep{9608096}.
}
\lref\matrixstring{R. Dijkgraaf, E. Verlinde, and H. Verlinde, 
{\it Matrix String Theory}, \np{B500}{1997}{43}, \hep{9703030}.
}
\lref\mstreview{R. Dijkgraaf, E. Verlinde and H. Verlinde, 
{\it Notes on Matrix and Micro Strings}, 
\np{\rm B (Proc. Suppl.) 62}{1998}{348}, \hep{9709107}.} 
\lref\dinehs
{M. Dine, P. Huet and N. Seiberg, 
{\it Large and small radius in string theory},  
\np{B322}{1989}{301}.  
}
\lref\dixon{L. Dixon, J. Harvey, C. Vafa and E. Witten, 
{\it Strings on Orbifolds}, 
\np{B261}{1985}{678}; \np{B274}{1986}{285}.
}
\lref\douglashull{M. Douglas and C. Hull, 
{\it D-branes and the noncommutative torus}, 
\jhep{9802}{1998}{002}, \hep{9711165}. 
}
\lref\dkps{M. Douglas, D. Kabat, P. Pouliot and S. Shenker, 
{\it D-branes and Short Distance in String Theory}, 
\np{B485}{1997}{85}, \hep{9608024}. 
}
\lref\mikegreg{
M. Douglas and G. Moore, {\it D-branes, Quivers and ALE instantons},
\hep{9603167}. 
}
%
%\lref\riemann{H. Farkas and I. Kra, {\it Riemann Surfaces}, 
%Springer-Verlag, New-York, 1992.}
%
\lref\ganor{O. Ganor, S. Ramgoolam and W. Taylor,
{\it Branes, Fluxes and Duality in M(atrix) Theory},
\np{B492}{1997}{191}, \hep{9611202}
}
\lref\sbgthesis{
S.B. Giddings, {\it Conformal techniques in String theory and 
String field theory}, \prp{170}{1988}{167}.
}
\lref\SBGTasi{
S.B. Giddings, {\it Fundamental strings},
in {\sl Particles, Strings, and Supernovae},
eds.~A.~Jevicki and C.-I.~Tan, World Scientific (1989).}
\lref\GiWo{S.B. Giddings and S. Wolpert, 
{\it A Triangulation of Moduli Space from Light-cone String Theory},
 \cmp{109}{1987}{177}.} 
\lref\ghv{S.B. Giddings, F. Hacquebord and H. Verlinde, 
{\it High Energy Scattering and D-pair Creation in Matrix String Theory}, 
\np{B537}{1999}{260}, \hep{9804121}.}
\lref\ginsparg{
P. Ginsparg, {\it Applied Conformal Field Theory}, in {\it Fields, Strings
and Critical Phenomena}, ed. E. Br\'ezin and J. Zinn-Justin, 
North-Holland, 1988.
}
\lref\girardello{
L. Girardello, A. Giveon, M. Porrati and A. Zaffaroni,
{\it S-Duality in N=4 Yang-Mills Theories with General Gauge Groups}, 
\np{B448}{1995}{127}, \hep{9502057}. 
}
\lref\gpr{
A. Giveon, M. Porrati and E. Rabinovici, 
{\it Target space duality in string theory}, 
\prp{244}{1994}{77}, \hep{9401139}. 
}
\lref\Greo{M.B. Green, {\it Effects of D-instantons},  
\np{B498}{1997}{195}, \hep{9701093}. 
}
\lref\Gret{M.B. Green, 
{\it Connections between M-theory and superstrings}, 
\np{\rm B (Proc. Suppl.) 68}{1998}{242}, \hep{9712195}.
}
\lref\GGV{
M.B. Green, M. Gutperle, and P. Vanhove,
{\it One loop in eleven dimensions},
\pl{B409}{1997}{177}, \hep{9706175}.
}
\lref\gsw{M.B. Green, J.H. Schwarz and E. Witten, {\it Superstring Theory}
Vol.1 and 2, Cambridge University Press, Cambridge, 1987.
} 
\lref\GrVh{M.B. Green and P. Vanhove, 
{\it D-instantons, Strings and M-theory}, 
\pl{B408}{1997}{122}, \hep{9704145}. 
}
\lref\grossmende{D.~Gross and P.~Mende, 
{\it The high-energy behavior of string scattering amplitudes}, 
\pl{197B}{1987}{129} ; 
{\it String theory beyond the Planck scale}, \np{B303}{1988}{407}.
}
\lref\duality{F. Hacquebord and H. Verlinde, 
{\it Duality symmetry of N=4 Yang-Mills theory on $T^3$}, 
\np{B508}{1997}{609}, \hep{9707179}.
}
\lref\harvey{J. Harvey, G. Moore and A. Strominger, 
{\it Reducing $S$-duality to $T$-duality}, 
\prd{D52}{1995}{7161}, \hep{9501022}. 
}
\lref\hirzebruch{F. Hirzebruch and T. H\"ofer, 
{\it On the Euler number of an orbifold}, 
\ma{286}{1990}{255}.
}
\lref\christiaan{
C. Hofman and E. Verlinde, (1997) unpublished.
}
\lref\hofmani{
C. Hofman and E. Verlinde, 
{\it U-Duality of Born-Infeld on the Noncommutative Two-Torus},
\jhep{12}{1998}{010}, \hep{9810116}.
}
\lref\hofmanii{
C. Hofman and E. Verlinde,  
{\it Gauge Bundles and Born-Infeld on the Noncommutative Torus}, 
\hep{9810219}. 
}
\lref\hvz{C. Hofman, E. Verlinde and G. Zwart, 
{\it U-duality invariance of the four-dimensional Born-Infeld 
theory}, 
\jhep{10}{1998}{020}, \hep{9808128}.
}
\lref\dhoker{E. D'Hoker and S.B. Giddings, 
{\it Unitarity of the closed bosonic Polyakov string}, 
\np{B291}{1987}{90}.
}
\lref\hooft{G. 't Hooft, 
{\it A property of Electric and Magnetic Flux in Non-Abelian Gauge
Theories},
\np{B153}{1979}{141}. 
}
\lref\hulltown{C.M. Hull and P.K. Townsend, 
{\it Unity of Superstring Dualities}, 
\np{B438}{1995}{109}, \hep{9410167}. 
}
\lref\kabat{D. Kabat and P. Pouliot, 
{\it A comment on zero-brane quantum mechanics}, 
\prl{77}{1996}{1004}, \hep{9603127}. 
}
\lref\KVK{E. Keski-Vakkuri and P. Kraus, 
{\it M-momentum transfer between gravitons, membranes, and fivebranes 
as perturbative gauge theory processes}, 
\np{B530}{1998}{137}, \hep{9804067}.
}
\lref\aschwarz{A. Konechny and A. Schwarz, 
{\it 1/4-BPS states on noncommutative tori}, \noexpand\break
\hep{9907008}. 
}
% 
%
%\lref\kostov{I. Kostov and P. Vanhove, {\it Matrix String Partition
%Functions}, 
%\pl{B444}{1998}{196}, \hep{9809130}. 
%}
%
\lref\leigh{R. Leigh, 
{\it Dirac-Born-Infeld Action from Dirichlet Sigma Models},
\mpl{A4}{1989}{2767}.  
}
\lref\mandelstami{
S. Mandelstam, {\it Lorentz Properties of the Three-String Vertex}, 
\np{B83}{1974}{413}; {\it Interacting-String Picture of the Fermionic
String}, Prog. Theor. Phys. Suppl. {\bf 86} (1986) 163;
{\it Dual-resonance models}, \prp{13}{1974}{259}.
}
\lref\motl{L. Motl, 
{\it Proposals on nonperturbative superstring interactions},
\hep{9701025}.
}
\lref\montonen{C. Montonen and D. Olive, 
{\it Magnetic monopoles as gauge particles},  
\pl{72B}{1977}{117};  
P. Goddard, J. Nuyts and D. Olive, {\it Gauge theories and magnetic charge}, 
\np{B125}{1977}{1}.
}
\lref\wnahm{
W. Nahm, {\it Monopoles in quantum field theory}, Proceedings of the 
monopole meeting, ed. Craigie et al, World Scientific, Singapore,
1982. 
}
\lref\obers{N. Obers and B. Pioline, 
{\it U duality and M theory}, \hep{9809039}. 
}
\lref\osborn{
H. Osborn, {\it Topological charges for $N=4$ supersymmetric
gauge theories and monopoles of spin 1}, 
\pl{83B}{1979}{321}; 
A. Sen, {\it Dyon-monopole bound states, self-dual harmonic forms
\dots}, 
% on
%the multi-monopole moduli space, and $SL(2,Z)$ invariance in String
%theory} 
\pl{B329}{1994}{217},  \hep{9402032}. 
}
\lref\Polc{
J. Polchinski, {\it Dirichlet-branes and Ramond-Ramond charges},  
\prl{75}{1995}{4724}, \hep{9510017}.
}
\lref\poltasi{
J. Polchinski, {\it Tasi lectures on D-branes}, \hep{9611050}.  
}
\lref\polstring
{J. Polchinski, {\it String Theory} Vols. 1 and 2, Cambridge University
Press, Cambridge, 1998.
}
\lref\polpou{J. Polchinski and P. Pouliot, 
{\it Membrane scattering with M-momentum transfer}, 
\prd{D56}{1997}{6601}, \hep{9704029}.
}
\lref\rozali{M. Rozali, 
{\it Matrix theory and U-duality in seven dimensions}, 
\pl{B400}{1997}{260}. \hep{9702136}.
}
\lref\schwarz{
J. Schwarz, {\it An SL(2,Z) multiplet of type IIB superstrings}, 
\pl{B360}{1995}{13}; Erratum-ibid. {\bf B364} (1995) 252, 
\hep{9508143}.  
}
\lref\seibergsen{N. Seiberg, 
{\it Why is the Matrix Model correct?}, 
\prl{79}{1997}{3577}, \hep{9710009}; 
A. Sen, 
{\it D0 branes on $T^n$ and Matrix Theory}, 
\atmp{2}{1998}{51}, \hep{9709220}.
}  
%
%\lref\senudual{A. Sen, {\it A note on marginally stable bound states
%in type II String theory}, \prd{D54}{1996}{2964}, \hep{9510229};  
%{\it U-duality and intersecting D-branes}, \prd{D53}{1996}{2874}, 
%\hep{9511026}; {\it T-duality of p-branes}, 
%\mpl{A11}{1996}{827}, \hep{9512203}.  
%}
%
\lref\sethisuss{S. Sethi and L. Susskind, 
{\it Rotational invariance in the M(atrix) formulation of type IIB
theory}, \pl{B400}{1997}{265}, \hep{9702101}. 
}
\lref\Shen{
S. H. Shenker, {\it Another Length Scale in String Theory?}, 
\hep{9509132}.
}
\lref\susskind{L. Susskind, 
{\it T Duality in M(atrix) theory and S duality in field theory}, 
\noexpand\break \hep{9611164}.
}
\lref\dlcq{L. Susskind, 
{\it Another Conjecture about M(atrix) theory},
\hep{9704080}.
} 
\lref\taylor{W. Taylor, 
{\it D-brane Field Theory on Compact Spaces},
\pl{B394}{1997}{283}, \hep{9611042}.
} 
\lref\Tayll{
W. Taylor, {\it Lectures on D-branes, gauge theory and
M(atrices)}, \hep{9801182}.
}
\lref\townsend{
P. K. Townsend, {\it The eleven-dimensional supermembrane revisited}, 
\pl{B350}{1995}{184}, \hep{9501068}. 
}
\lref\tseytlin{A. Tseytlin, 
{\it On non-abelian generalisation of Born-Infeld action in
string theory}, \np{B501}{1997}{41}, \hep{9701125}; 
D. Brecher and M. J. Perry, 
{\it Bound states of D-Branes and the non-abelian Born-Infeld
action}, \np{B527}{1998}{121}, \hep{9801127}.
}
%
%\lref\vafaudual{
%C. Vafa, {\it Gas of D-Branes and Hagedorn Density of BPS States}, 
%\np{B463}{1996}{415}, \hep{9511088}. 
%}
%
\lref\vafawitten
{C. Vafa and E. Witten, {\it A strong coupling test of $S$-duality}, 
\np{B431}{1994}{3}, \hep{9408074}. 
}
\lref\wittenstd{E. Witten, 
{\it String theory dynamics in various dimensions}, 
\np{B443}{1995}{85}, \hep{9503124}
}
\lref\wittenbound{E. Witten, 
{\it Bound States of Strings and $p$-Branes},
\np{B460}{1996}{335}, \hep{9510135}.
}
\lref\wittenolive{
E. Witten and D. Olive, 
{\it  Supersymmetry algebras that include topological charges}, 
\pl{78B}{1978}{97}.
}
\lref\wynteri{T. Wynter, 
{\it Gauge Fields and Interactions in Matrix String Theory}, 
\pl{B415}{1997}{349}, \hep{9709029}.
}
\lref\wynterii{T. Wynter,   
{\it Anomalies and Large N Limits in Matrix String Theory}, 
\pl{B439}{1998}{37}, \hep{9806173}.
}  
\lref\wynteriii{T. Wynter, 
{\it High energy scattering amplitudes in matrix string theory}, 
\noexpand\break \hep{9905087}. 
}
\lref\gysbert{G. Zwart, 
{\it U-Duality in Supersymmetric Born-Infeld Theory},  
\jhep{06}{1999}{011}, \hep{9905068}. 
}

%%% Begin chapter Introduction

%%%%%%%%%%%%%%%%%%%%%%%%%%%%%%%%%%%%%%%%%%%%%%%%%%%%%%%%%%%%%%%%%%%%
%%%%%%%%%%%%%%%%%%%%%%%%%%%%%%%%%%%%%%%%%%%%%%%%%%%%%%%%%%%%%%%%%%%%
%%%                                                              %%%
%%%                                                              %%%
%%%             INTRODUCTION                                     %%%
%%%                                                              %%%
%%%                                                              %%%
%%%                                                              %%%
%%%%%%%%%%%%%%%%%%%%%%%%%%%%%%%%%%%%%%%%%%%%%%%%%%%%%%%%%%%%%%%%%%%%
%%%%%%%%%%%%%%%%%%%%%%%%%%%%%%%%%%%%%%%%%%%%%%%%%%%%%%%%%%%%%%%%%%%%

\thispagestyle{empty}\noindent
\begin{center}
{\large\sc Symmetries and Interactions}
\\[2mm]
{\large\sc in}
\\[2mm]
{\large\sc Matrix String Theory}
\\[17mm]
{\sc Feike Hacquebord} \\[5mm]
{\it Institute for theoretical physics}\\
{\it University of Amsterdam}\\
{\it 1018 XE Amsterdam}\\
{\it  The Netherlands}\\[3cm] 
{\sc Abstract} 
\end{center} 
This PhD-thesis reviews matrix string theory and recent developments 
therein. Emphasis is put on symmetries, interactions and scattering 
processes in the matrix model. We start with an introduction to 
matrix string theory and a review of the orbifold model that flows out
of matrix string theory in the strong YM coupling limit. Then we turn our
attention to the appearance of $U$-duality symmetry in gauge models, 
after a (very) short summary of string duality, D-branes and M-theory. 
The last chapter reviews matrix string interactions and scattering 
processes in the high energy limit. Also, pair production of 
D-particles is studied in detail. D-pair production is
expected to give important corrections to high energy scattering 
processes in string theory. 
\newpage \thispagestyle{empty}
\phantom{bla} \newpage 
\thispagestyle{empty}\noindent
\centerline{
\Huge\sc Symmetries and Interactions}
\\[5mm]
\centerline{
\Huge\sc in}
\\[5mm]
\centerline{
\Huge\sc Matrix String Theory}
\\[5cm]
\centerline{\large\sc ACADEMISCH PROEFSCHRIFT}
\\[1cm]
\centerline{\vbox{\baselineskip14pt
\advance\hsize by -2.8truein\noindent{\normalsize
ter verkrijging van de graad van doctor aan de 
Universiteit van Amsterdam op gezag van de Rector 
Magnificus prof.~dr~J.J.M.~Franse ten overstaan 
van een door het college voor \nobreak promoties ingestelde 
commissie in het \nobreak openbaar te verdedigen in de Aula der 
Universiteit op}
\centerline{woensdag 22 september 1999 te 11.00 uur}
}}
\\[3cm]
\centerline{door}
\\[2cm]
\centerline{\large\sc Feike Hayo Hacquebord}
\\[2cm]
\centerline{geboren te Epe}
\newpage \thispagestyle{empty} 
\noindent
\hskip-\the\tabcolsep
\begin{tabular}{ll}
Promotor: & \hskip12pt Prof.~dr H.L.~Verlinde \\[1cm]
Promotiecommissie: & \hskip12ptProf.~dr P.~van Baal \\
& \hskip12pt Prof.~dr ir F.A.~Bais\\
& \hskip12pt Prof.~dr R.H.~Dijkgraaf \\
& \hskip12pt Prof.~dr J.~Smit \\
& \hskip12pt Prof.~dr E.P.~Verlinde \\
& \hskip12pt Prof.~dr B.Q.P.J~de Wit\\
\end{tabular}
\\[2cm]
Faculteit der Wiskunde, Informatica, Natuur- en Sterrenkunde
\\[5mm]
Instituut voor Theoretische Fysica
\\[3mm]
ISBN 90-9013055-1\\
NUGI 812 \\
\vfill\noindent
\epsfysize=13.5mm
\epsfbox{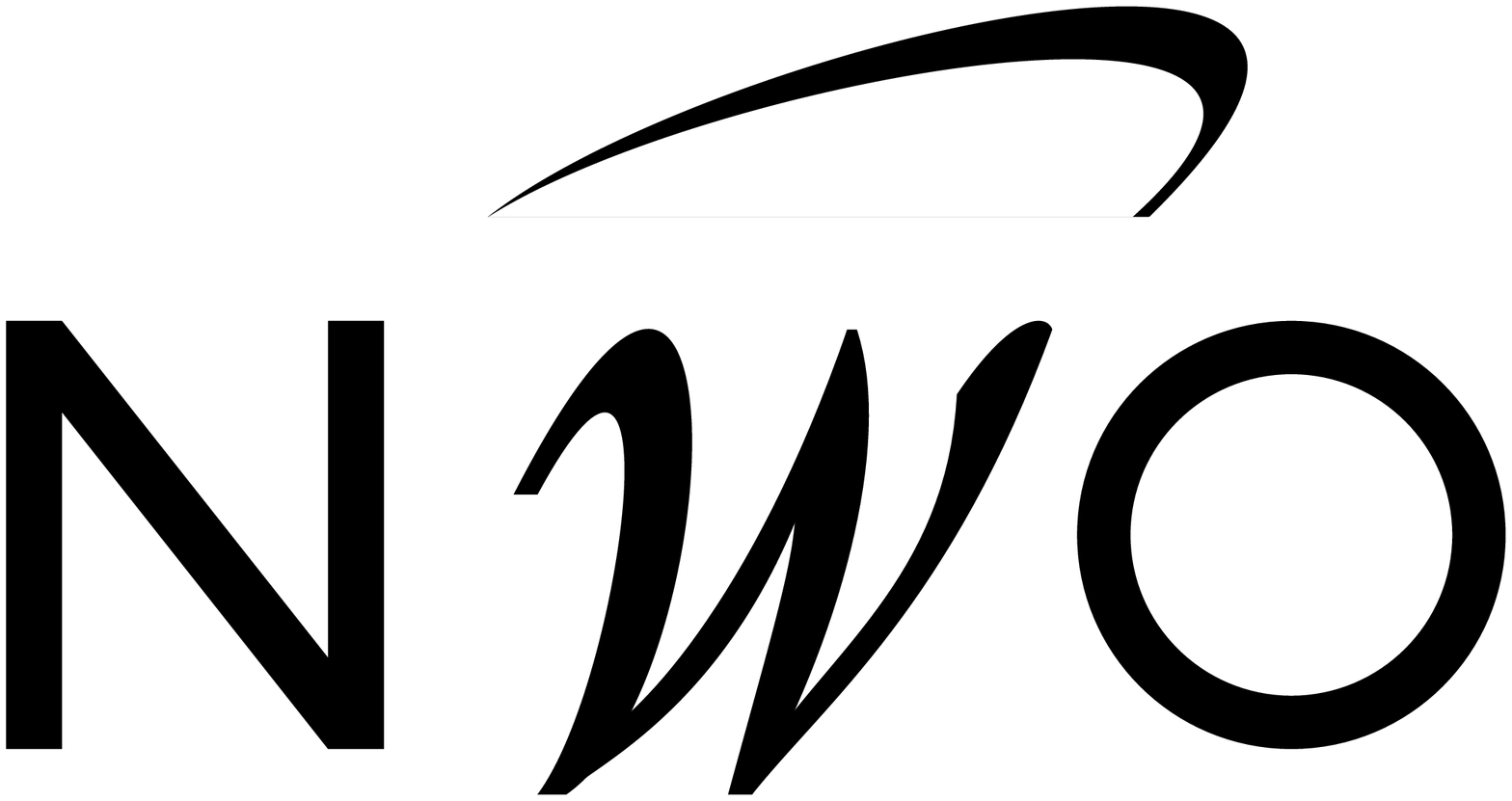}
\\
The research described in this thesis is financially supported by a
Pionier fund of NWO.
\newpage

\noindent
\newtableofcontents

\newpage \ifodd\thepage \relax \else \phantom{bla} \newpage \fi 

\lhead[\fancyplain{}{\sc Introduction}]{\fancyplain{}{}}

\thispagestyle{empty}

\noindent 
\\[50pt]
\testclino
\write\cfile{\bf Introduction &&\bf \thepage\noexpand\\[1.5mm]}
{\phantom{tk} \hfill 
\Huge\rmfamily\bfseries\slshape Introduction}
\\[2.5cm]
\noindent
The subject of this thesis is a new effective formulation of string 
theory in terms of a $1+1$ dimensional Yang-Mills theory. 
This approach will 
prove to be useful for the investigation of the short-distance
behavior of strings. Before we explain in some more detail what the
coming chapters are about, we will first spend some words on the 
motivation to do string theory. 
\\ \indent
The motivation to study string theory is twofold: one is the
desire of physicists to construct a theory that unifies all 
elementary forces in nature. A second important motivation is 
to learn more about gravity. Newtonian mechanics and general
relativity give precise predictions for gravitational interactions 
at large distances. For example, we can in principle launch a
spacecraft, % into the universe, 
explore a planet in our solar system  
and return safely to the earth again. Without our knowledge of 
gravity this would be a rather hazardous adventure.  
\\ \indent
For very tiny distance scales (like $10^{-33}$cm, the Planck 
length) however, we expect that Newtonian mechanics 
or general relativity no longer gives an accurate description of gravity. 
The physical conditions at the Planck scale are so extreme that 
they cannot be realized in laboratories. This makes it hard to 
do actual experiments (but they are not ruled out). 
At present consistency is the only tool available . 
It is of interest to try to formulate a theory
of gravity in the kinematic regime of the Planck scale, because it is 
relevant for the study of black holes (their existence is confirmed 
by convincing observational evidence) 
and the study of the early universe. 
\global\advance\hfignumber by 1 
\xdef\smoothgrav{\the\hfignumber} \global\advance\hfignumber by -1 
\\ \indent
At the Planck scale quantum effects for gravitational interactions
are important. When one tries to formulate a quantum field 
theory of gravity, for example by a perturbation expansion of the 
Einstein Hilbert action with a coupling term to a scalar field, 
increasingly severe short-distance divergencies appear, 
which cannot be removed by renormalization procedures. 
Therefore one has to find a way out; the singularities should 
be somehow smeared out by a drastic adaption of the theory. 
\\ \indent
At present there is only one consistent way known how to do this. 
This way is string theory, a theory where the fundamental objects are 
one dimensional, instead of zero dimensional as for point particles. 
Interactions in string theory have a geometrical interpretation in
terms of smooth Riemann surfaces, as indicated in figure \smoothgrav. 
This already gives a heuristic explanation why strings are able to 
smear out the short-distance divergences. 
\\ \indent
A key idea of string theory is that the different vibration modes 
of a string correspond to different particle states. 
The massless states are the most relevant ones, as they should form 
the particles known today in the standard
model (which are massless or very light compared with the Planck mass). 
The effective theories of these states can in principle be derived
from string theory, but in practice that can be hard. 
\\
\epsfysize=5cm
\hhfigure{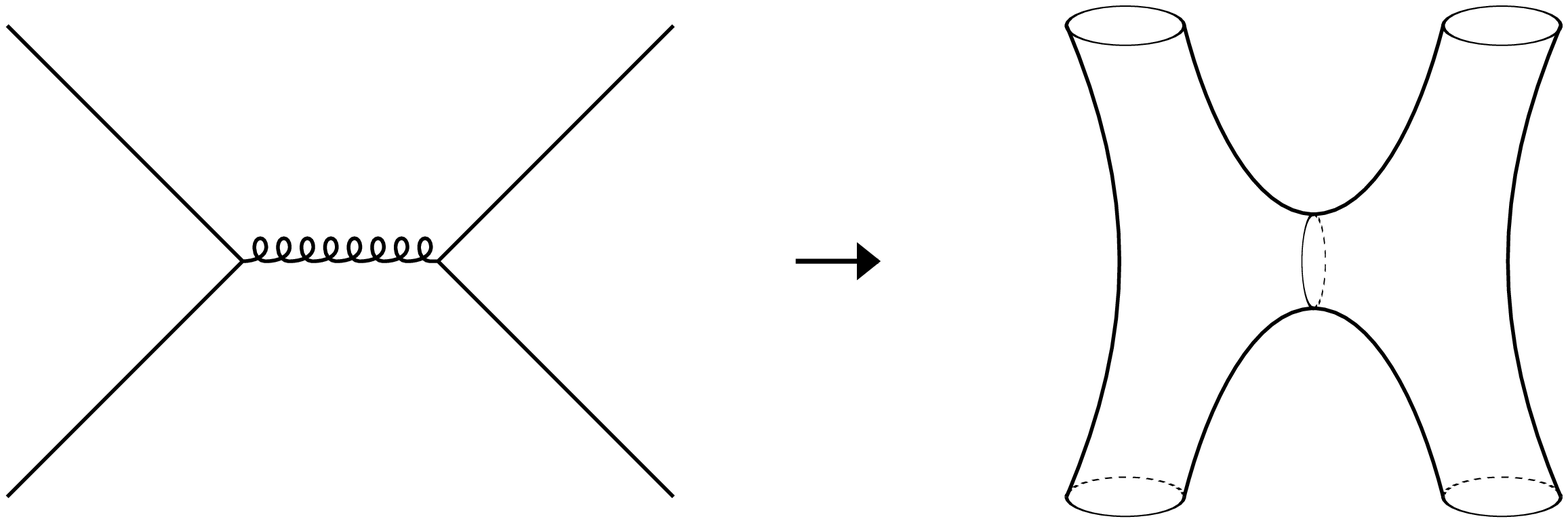}
{String theory smoothens out the short-distance divergences 
in field theories of gravity (like Einstein-Hilbert gravity
coupled to a scalar field). 
The figure shows a tree level diagram in field theory which describes 
two scalar particles exchanging a graviton, and its analogue in string 
theory. Both diagrams are finite, but higher loop contributions are 
finite only in string theory.} 
\\ 
It is not a priori clear why string theory should be the right way 
to proceed. One could for instance try membranes, objects with two
spatial dimensions, which might be able to smear out the divergences 
as well. The study of membranes revealed some problems 
however, while string theory proved to have particular attractive 
features. 
Among these features are the following: string theory contains a 
massless spin-2 state (the graviton) whose low energy effective 
description is general relativity; string theory has a 
consistent perturbation expansion; it is rich enough to be a candidate
for a unification of all known forces in nature. In particular string
theory has intimate relations with gauge theories.  
There are no free parameters in string theory, instead there are many 
(classical) ground states parameterized by the expectation values the 
scalar fields take. 
\\ \indent 
Consistency of string theory requires supersymmetry, a symmetry
between the bosonic and fermionic degrees of freedom. 
%Bosons are 
%states (fields) with integer spin, whereas fermions have half-integer 
%spin. 
Supersymmetry is expected to be a symmetry of nature for 
sufficient high energies only. The order of these energies is (almost) 
within reach of the elementary particle colliders of today, so in
principle the superpartners of known particles states 
could be detected there. Detection of the superpartners would 
be an experimental proof of supersymmetry and strong evidence 
for string theory to be the right way to proceed.
\\ \indent
Superstrings live in ten dimensions. This may be viewed as a drawback,
but compactifications to lower dimensions give rise to a rich
structure of Kaluza Klein fields. In ten dimensions we can define 5 
superstring theories: type IIA and IIB, which are theories of closed 
strings; type I describing unoriented open strings (plus closed
strings) and two types of the heterotic string. We will be mainly 
interested in type IIA and IIB so we will not dwell on the precise 
definition of other string theories. The important point here 
is that the theories are all related by duality transformations. 
It was conjectured by Witten that the five string theories are in 
fact manifestations of one eleven-dimensional theory, that goes with 
the name {\it M-theory}. The dualities and the existence
of M-theory are not proven yet, but substantial evidence 
has been collected in recent years. 
\\ \indent
The string duality symmetries were known from earlier studies 
as {\it classical} symmetries of supergravity theories. 
These models failed to be consistent quantum theories of gravity; 
the addition of supersymmetry did not resolve the short-distance 
divergences. Later supergravity theories 
appeared again as low energy effective descriptions
of the graviton state (and its superpartner the gravitino) 
of string theory and it is 
expected that their classical symmetries are genuine quantum 
mechanical duality symmetries in string theory. 

An example of a string duality transformation is the mapping that 
relates strings at large distances to strings at small distances. 
This is somewhat similar to the more familiar electromagnetic duality in 
(supersymmetric) gauge theories, that relates the weak coupling 
to the strong coupling regime. Both duality transformations map a 
region of the theory where a perturbation expansion suffices, 
to a region where this expansion breaks down. 

To understand (and to prove) these dualities we have to learn about 
the non-per\-tur\-ba\-tive degrees of freedom. These degrees of freedom 
in string theory have been mysterious for a long time, till in the
fall of 1995 Polchinski realized that the so-called D-branes 
carry them. D-branes were known before as hyper-surfaces 
in space-time on which open strings can end, but in fact they are 
dynamical non-perturbative objects in string theory. 

At low energy the dynamics of D-branes are described by 
supersymmetric Yang-Mills theories. These theories have 
matrix valued scalar fields, that commute when the D-branes are
widely separated. One can then diagonalize the matrices simultaneously
and interpret the eigenvalues on the diagonals as the D-brane
positions in the traditional sense. When the D-branes come 
close something remarkable happens: in general the matrices no longer 
commute and the interpretation of the D-brane 
positions gets obscured. Space-time becomes fuzzy: its coordinates no 
longer commute. This non-commutativity is somewhat similar to what
happens with the classical phase space in quantum mechanics and is 
therefore highly suggestive. 

The supersymmetric Yang-Mills model in $1+1$ dimensions compactified
on a circle is an effective theory of one dimensional D-branes (called
D-strings), but it can also be viewed as a theory of fundamental
strings. This gauge theory, called matrix string theory, is defined on
a cylinder that is covered by the world-sheet of a string one or more
times.

\global\advance\hfignumber by 1 
\xdef\layman{\the\hfignumber} 
\global\advance\hfignumber by -1 

Again one can think of the eigenvalues of the matrix valued scalars 
as the coordinates of the strings. These 
coordinates fields are not necessarily periodic along the Yang-Mills circle,  
but instead they may satisfy particular 
non-trivial boundary conditions. The simplest non-trivial example 
of a string of length 2 is illustrated in figure \layman. 
In the lower left corner
the two eigenvalues of a $2 \times 2$ matrix are mapped to each by
going around the circle once. So though it looks like we have two 
short strings we in fact have one long string. In the lower 
right corner the configuration of eigenvalues has changed. 
Here the two eigenvalues of the matrix are periodic and we have two 
short strings. The one string configuration transforms to the other
by changing the non-trivial boundary conditions at a certain 
time and place on the world-sheet. This is illustrated in the 
diagram in the middle of figure \layman. We added in the figure 
the usual geometric representation of the interactions 
(the joining or splitting of strings). 
\\ \indent
Compared with the known perturbative string theories, matrix string 
theory has the advantage that non-perturbative degrees of freedom
are contained in the model as well. Namely, the gauge model has besides
matrix valued scalar fields, a two dimensional gauge field. 
This gauge field makes it possible to adorn the string states with an
extra quantum number: the D-particle number (zero dimensional branes). 
\\
\epsfysize=7.3cm
\hhfigure{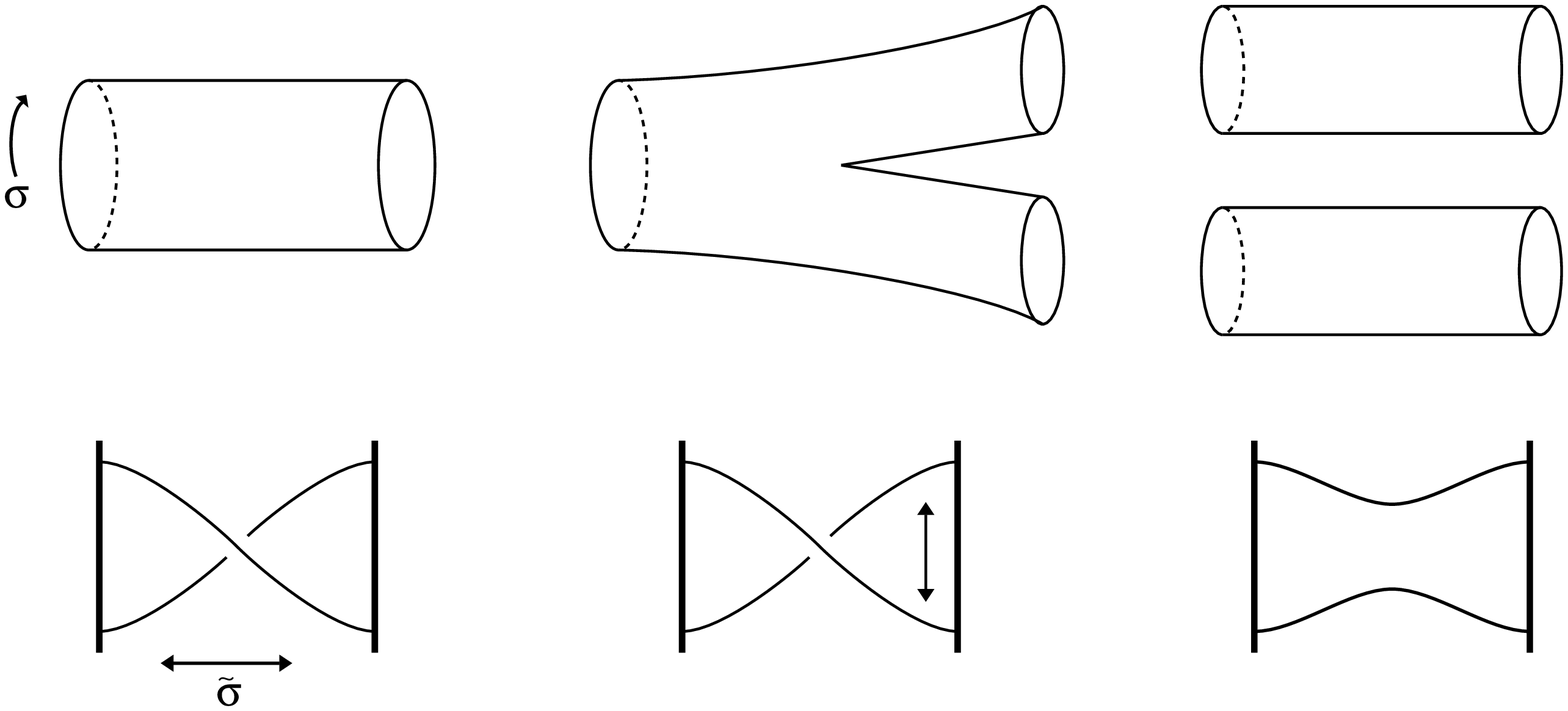}
{The smooth Riemann surfaces illustrate how a single closed string can split
in two strings in perturbative string theory.
The lower pictures show the eigenvalues of matrix valued 
scalar fields in matrix string theory as a function of $\tilde{\sigma}$, 
where $\tilde{\sigma}$ runs over {\it half} the interval of 
the spatial world-sheet coordinate $\sigma$. These eigenvalues 
describe the coordinates of strings. They may satisfy non-trivial 
boundary conditions on the $\tilde{\sigma}$ circle, so that the eigenvalues 
can describe different collections of strings. 
In the lower left corner we have one long string; in the
lower right corner we have two short strings. These two configurations
transform to each other when the boundary conditions of the
eigenvalues are changed, as indicated by the arrow in the middlest
diagram. The conventions are chosen such that the added string
lengths are conserved during interactions.}
\\
D-particles are special, because they can be interpreted as 
Kaluza Klein particles in  
eleven dimensional M-theory compactified on a circle. 
It has been conjectured that the low energy effective 
model of D-particles (a $U(N)$ matrix quantum mechanics) is 
equivalent to M-theory in a particular gauge. This theory called 
matrix theory is closely related to matrix string theory. 
\\ \indent
The fact that matrix string theory contains the degrees of freedom 
of D-particles in a natural way, makes it 
possible to calculate non-perturbative corrections to processes 
in string theory. These corrections are important for the 
investigation of the short-distance behavior of strings, because 
in this regime the perturbation expansion of string theory appears 
to break down. 
\\ \indent 
% Organization of thesis
In chapter three we will show how string interactions can be realized
in matrix string theory. Interactions arise as instanton type
solutions of the Yang-Mills theory equations of motion. With these 
instantons known results from perturbative string theory can be 
reproduced. We will also start a calculation of non-perturbative 
corrections. 
\\ \indent 
In chapter two we will investigate in how far the symmetries of string
theory are present in (supersymmetric) gauge theories. In particular 
we will show that there are quantum states in the theory that have
degeneracies consistent with a large string duality symmetry. 
\\ \indent 
Chapter one reviews geometrical aspects of perturbative string 
theory in the light-cone gauge and in the discrete light-cone gauge
quantization (DLCQ). 
We discuss a reformulation of DLCQ string theory in 
terms of an orbifold model. We end the chapter by introducing 
matrix string theory. 
\\ \indent 
Although we have tried to add as much explanation of relevant basic 
concepts in string theory as possible, 
this thesis is not meant to be self-contained. 
As background material
we refer to the textbooks on string theory \gsw\polstring\ 
and quick introductions to light-cone string theory \sbgthesis\SBGTasi, 
to D-branes \poltasi\bachasimp\ and to matrix (string) theory 
\BiSu\mstreview. For some basic concepts of conformal field theory 
we will need, we refer to  \ginsparg.

%\tableofcontents

\newchapter\introintro{Towards Matrix String Theory}{Towards 
Matrix String Theory}
\lhead[\fancyplain{}{\leftmark}]{\fancyplain{}{}}

\thispagestyle{empty} 

\newsection\noname{Perturbative string theory}

\newsubsection\noname{The Polyakov approach to string theory}

As strings move through space-time, they sweep out a two-dimensional surface.  
On this surface, called the world-sheet, the coordinates of the strings
are defined. The history of a collection of strings is thus described
by a map from the two-dimensional world-sheet into the
$d$-dimensional space-time. This space-time is endowed with a metric 
that in principle should be derived from the string configuration. 
We will take the strings however in a fixed background metric, usually
just flat. 
\\ \indent 
In the Polyakov approach to string theory, the string perturbation 
expansion appears as a sum over two-dimensional Riemann surfaces. 
These Riemann surfaces are of a certain genus $g$
(the counting parameter of the perturbation expansion) and have a 
number of boundary curves that correspond to the external states. 
They are the Feynman diagram representations of scattering amplitudes 
at $g$-loop order.  
An actual calculation of the contribution to the scattering amplitude 
in principle involves 
a path-integral over all maps of the surface into the space-time 
manifold, as well as an independent integration over all 
two-dimensional metrics $g_{ab}$ on the genus $g$ surface. 
\\
%%%
\epsfysize=4cm
\hfigure{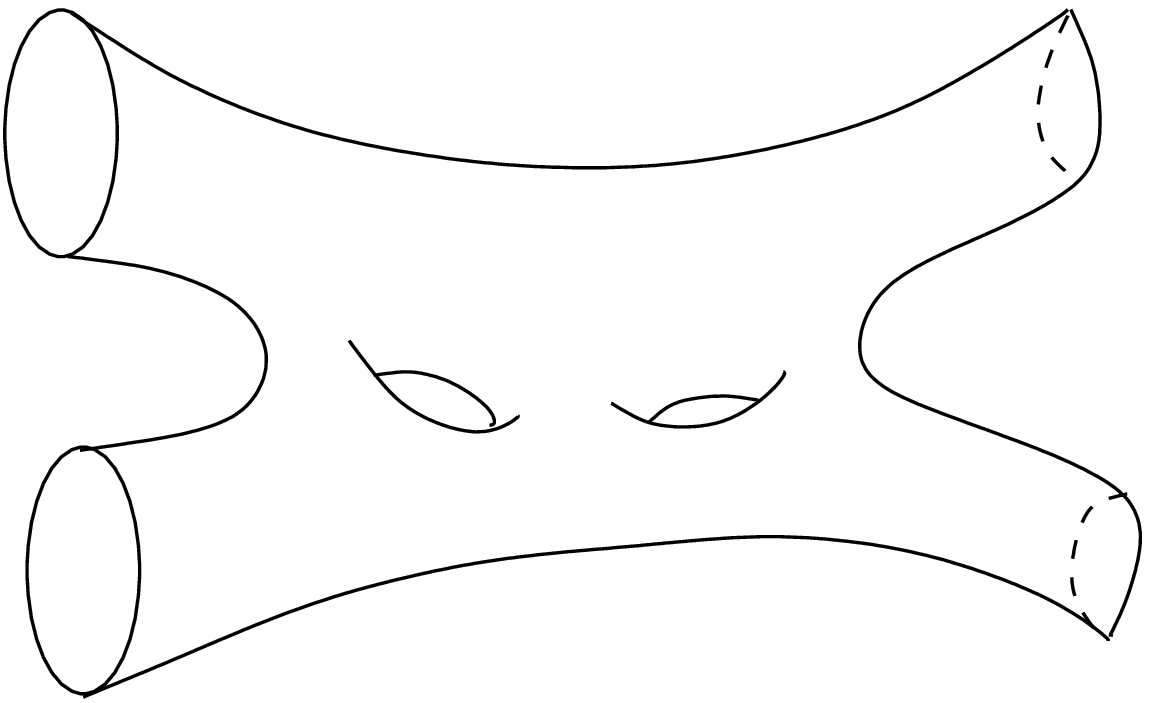}
{A typical world-sheet of genus two. The four tubes correspond
with the external strings and extend to infinity.} 
%%%
\xdef\genustwoi{\thechapter.\the\fignumber\ } 
\\
More concretely in the Polyakov path integral approach, 
a general $g$-loop amplitude is given by the expression 
\eqn\feynpath{
A_g(\Psi_1,\cdots, \Psi_n) = {1 \over {\cal N}} 
\int {\cal D}g_{ab} \int_{\Psi_1,\cdots \Psi_n} 
{\cal D}X^{\mu}  \exp{(-S)}.
}
Here ${\cal N}$ is a normalization factor and 
the $\Psi$'s indicate the external state wave functionals, 
that are defined on the boundaries of the world-sheet at infinity. 
For bosonic strings the action $S$ is given by 
\eqn\polyakovaction{
S = {1 \over 2\pi} 
\int d\sigma d\tau \sqrt{g}g^{ab}G_{\mu\nu}
\partial_a X^\mu \partial_b X^\nu. 
}
Here $\sigma$ and $\tau$ are world-sheet coordinates 
taking values on a cylinder 
%, $g_{ab}$ is the world-sheet metric 
and $G_{\mu\nu}$ is the metric of space-time.

The action \polyakovaction has local symmetries, namely 
reparametrizations (diffeomorphisms)
and Weyl rescalings. Weyl rescalings act on the world-sheet 
metric as  
\eqn\noname{
g_{ab} \rightarrow e^\Omega g_{ab} 
}
and it easy to verify that the action \polyakovaction is invariant 
under this transformation. 

One can use this Weyl invariance to make, at least locally, 
the world-sheet metric to be flat. 
Then there is still a gauge degree of freedom left over, 
namely arbitrary holomorphic coordinate transformations acting 
on the complex coordinate $w=\sigma +i\tau$ combined with an 
appropriate Weyl rescaling, so that the metric remains flat. 
These combined transformations are precisely conformal
transformations, that will prove to 
be an important tool in string theory. 
\\
%%%
\epsfysize=3cm
\hfigure{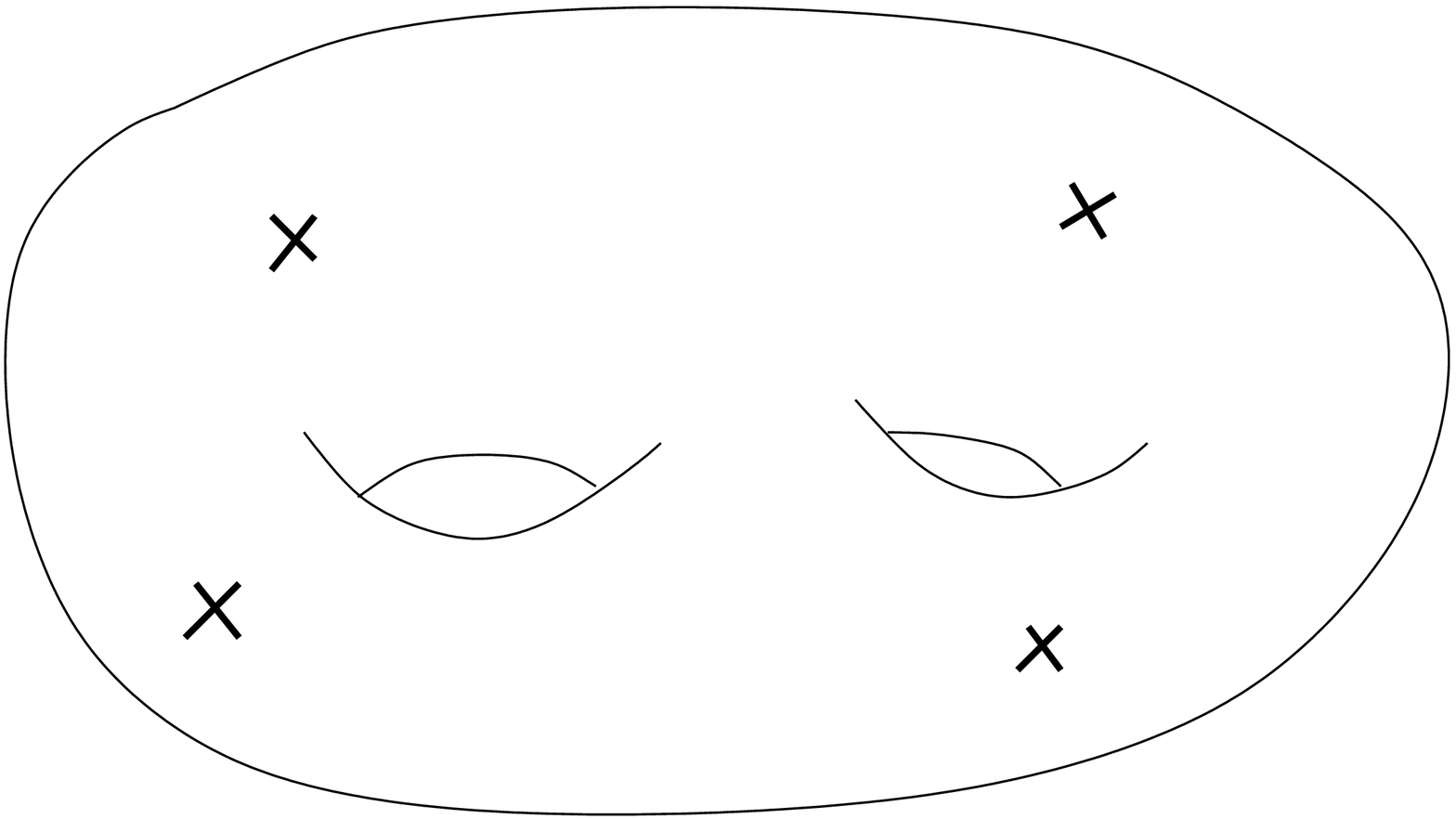}
{The world-sheet in figure \genustwoi is conformally equivalent to 
a genus two Riemann surface with four punctures.} 
\xdef\genustwoii{\thechapter.\the\fignumber\ } 
%%%
\\
The most obvious parameterization of a world-sheet 
has boundaries at infinity (we assumed this up to now), 
like in figure \genustwoi\!\!. By a particular conformal transformation 
this surface can be mapped to a compact Riemann surface with 
punctures, that correspond with the external string states.  
%
%This can be visualized by locally applying 
%the exponential mapping that transforms a 
\\
%%%
\epsfysize=2cm
\hfigure{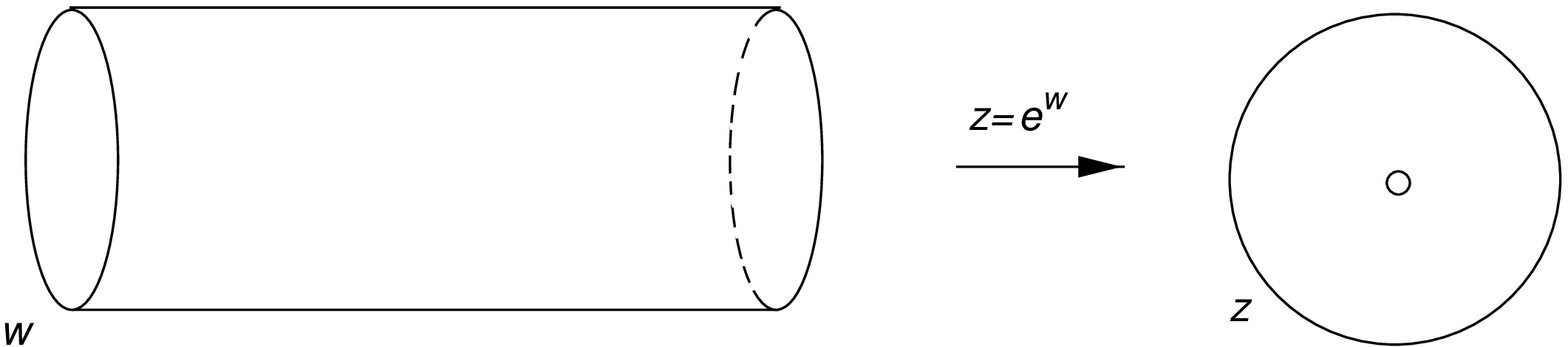}
{The conformal transformation $w \rightarrow z= e^w$, that maps
cylinders to annuli in the complex plane.} 
\xdef\longtube{\thechapter.\the\fignumber\ } 
%%%
\\
This can be visualized by locally applying 
the exponential mapping that transforms a 
tube to an annulus in the complex plane. 
A world-sheet as in figure \genustwoi  
is thus conformally equivalent to the 
Riemann surface in figure \genustwoii\!\!. 

The conformal mapping that transforms the Riemann
surface \genustwoi to the one in figure \genustwoii should be combined 
with the replacement of the boundary states $\Psi_i(X)$ by local 
operators $V_i$ on the world-sheet. These operators called vertex 
operators create the external states on the world-sheet by introducing 
extra momentum and/or other quantum numbers. 
The most simple vertex function is the one that represents the emission 
of a tachyon, with no other degrees of freedom then the momentum.
It is given by the expression  
\eqn\tachyonvertex{
V = e^{ip\cdot X},
}
where $p^2 = 8$ is the on shell condition for tachyons.  
In terms of the vertex operators
the amplitude \feynpath becomes 
\eqn\noname{
A_g(1,\cdots,n) = {1 \over {\cal N}} 
\int {\cal D}g_{ab} \int {\cal D}X^{\mu}  \exp{(-S)} 
V_1 \cdots V_n.
}
As mentioned before making use of the local symmetries of 
string theory greatly reduces the moduli space of world-sheet 
metrics over which one has to integrate. 
This simplification of the path integral is obtained after 
constructing a good slice through the space of metrics
on the Riemann surface, that modulo Weyl transformations and 
diffeomorphisms covers all of moduli space only once. 
%Generally speaking this will lead us to the expression 
% 
%
%\eqn\ampli{
%A_g(1,\cdots,n) = g_s^{2g+2}\int_{{\cal M}^g_n} [dm] ({\rm det's}) 
%\left<V_1,\cdots,V_n\right>,
%}
%where $[dm]$ is the measure on the finite dimensional slice 
%${\cal M}^g_n $ through the moduli space. In the integrand there
%are determinants associated to the Weyl rescalings and diffeomorphisms
%orthogonal to the moduli that parametrize ${\cal M}^g_n$. 
%%
For external tachyons whose vertex operators are given by 
\tachyonvertex the expectation value of the vertices is a 
simple Gaussian integral over the fields $X$, that can readily
be calculated to be 
\eqn\expectation{
A(1,\cdots,n) = g_s^{2g+2} \int [dm] \prod_i \int d^2 z_i \sqrt{g(z_i)} 
({\rm det's}) \exp[-{1 \over 2} \sum p_i \cdot p_j G_M(z_i,z_j)],
} 
where $G_m(z_i,z_j)$ is the scalar Green function 
on the genus $g$ world-sheet (depending on the moduli $m$, which 
have to be integrated over). In the integrand of \expectation 
we abbreviated the fluctuation determinants that appear after
integrating out the coordinate fields and reducing the path integral
to a finite dimensional integral. 
In the limit that {\it all} $p_i \cdot p_j$ become large, 
the leading behavior of the integral can be derived by 
saddle point techniques \grossmende. Later on in chapter 3 these 
saddle points will prove to be crucial when we establish an alternative 
approach to string theory, called matrix string theory 
\motl\tomnati\matrixstring, 
that is also able to include non-perturbative corrections.

\newsubsection\lightconebasics{Light-cone string theory}

In this section we will discuss some geometric facts of yet
another approach to perturbative string theory:  light-cone
string theory. As background 
material we refer to \gsw\SBGTasi\ and \sbgthesis.  

Quantization of strings in the light-cone gauge formalism 
has two particular features. One is the absence of ghosts, 
only physical degrees of freedom are present, the other is 
the existence of a globally well-defined world-sheet time \GiWo. 
Covariance is broken in light-cone string theory, but the formalism 
has been proven to be equivalent to the covariant Polyakov path 
integral formulation of string theory \dhoker. 

In the light-cone gauge, light-cone time and world-sheet time 
are identified via 
\eqn\lcgauge{
X^+(z,\bar{z}) = p^+\tau, 
}
where $p^+$ is one of the light-cone momenta. 
This parameterization together with taking the world-sheet metric to
be flat %(this can be achieved by an appropriate rescaling of 
%$(\sigma,\tau)$) 
leads to the mass-shell condition
\eqn\noname{
2p^+p^- = \int d\sigma  ( (\partial_\tau X^I)^2 + 
(\partial_\sigma X^I)^2)  := H, 
}
with $H$ the world-sheet Hamiltonian and
$p_-$ and $p^+$ the integrated light-cone momenta. 
The index $I$ runs over all directions transversal to the
light-cone directions. 
\\
%%%%%%
\epsfysize=5cm
\hfigure{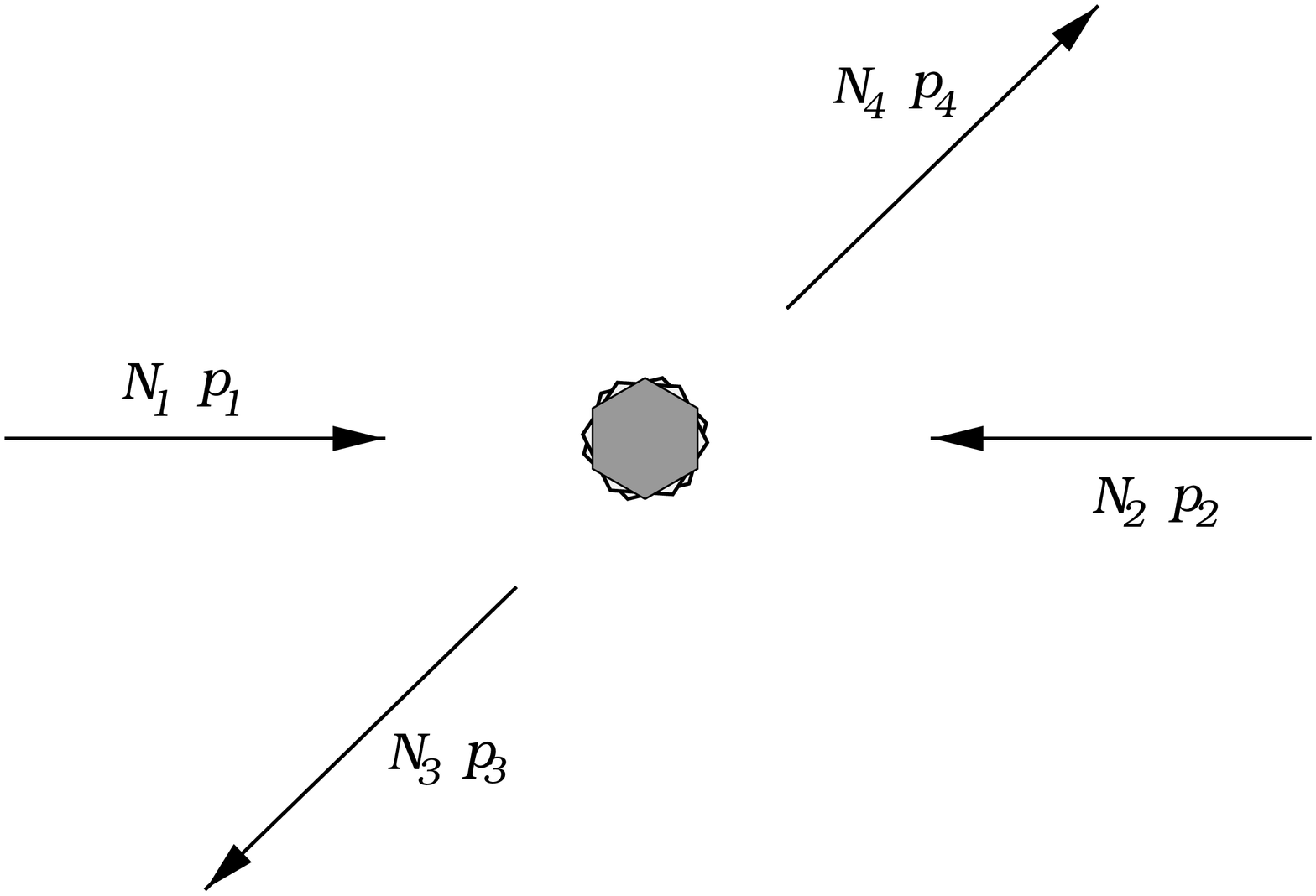}
{This figure indicates the kinematics of a two particle 
scattering process in string theory. The $p_i$ are the 
eight dimensional transverse momenta that lie in one plane; 
the $N$'s stand for the $p^+$ momenta.}
\xdef\kinematics{\thechapter.\the\fignumber\ } 
%%%%%%
\\
A two particle scattering process like in figure \kinematics can
be conveniently summarized in light-cone string theory
by drawing a so called light-cone diagram or Mandelstam diagram. 
A light-cone diagram is a geometrical 
way of representing a scattering process in string theory. 
The external states are represented by tubes that extend to 
infinity. Together with a collection of other cylinders that 
correspond to internal strings they are glued together at 
interaction points. Thus a two-dimensional surface is formed
that contains all essential information about the interaction 
process. 

\global\advance\fignumber by 1 

It is customary to rescale the spatial coordinate $\sigma$ 
on the world-sheet, so that its range becomes 
$0 \leq \sigma \leq 2\pi p^+$. 
Because of this rescaling the radii $\alpha_i$ 
of the cylinders are proportional to 
the light-cone momenta $p_i^+$ of the corresponding strings. 
The total momentum $p^+$ is conserved during a scattering
process, and therefore the sum of the radii of the cylinders 
is also conserved. An example of a light-cone diagram that 
describes the tree level contribution to the scattering process 
indicated in figure 
\kinematics\!\!, is shown in figure \thechapter.\the\fignumber\  
where we have two incoming strings that join to one string and
split again into two strings.  
In the diagram the interaction times and possible twist angles are 
indicated. These twist angles are the angles under which the internal
tubes are allowed to rotate before undergoing another interaction. 
As these twistings cannot be undone by a conformal transformation or 
reparametrization, they belong to the moduli of the 
light-cone string diagrams. \footnote{A twisting over an angle of 
$2\pi$, called a Dehn twist, is a global symmetry of the light-cone 
string action. The twist angles are therefore defined modulo $2\pi$.}

\global\advance\fignumber by -1 

%
%%%
\noindent \\[4mm]
\epsfysize=5cm
\hfigure{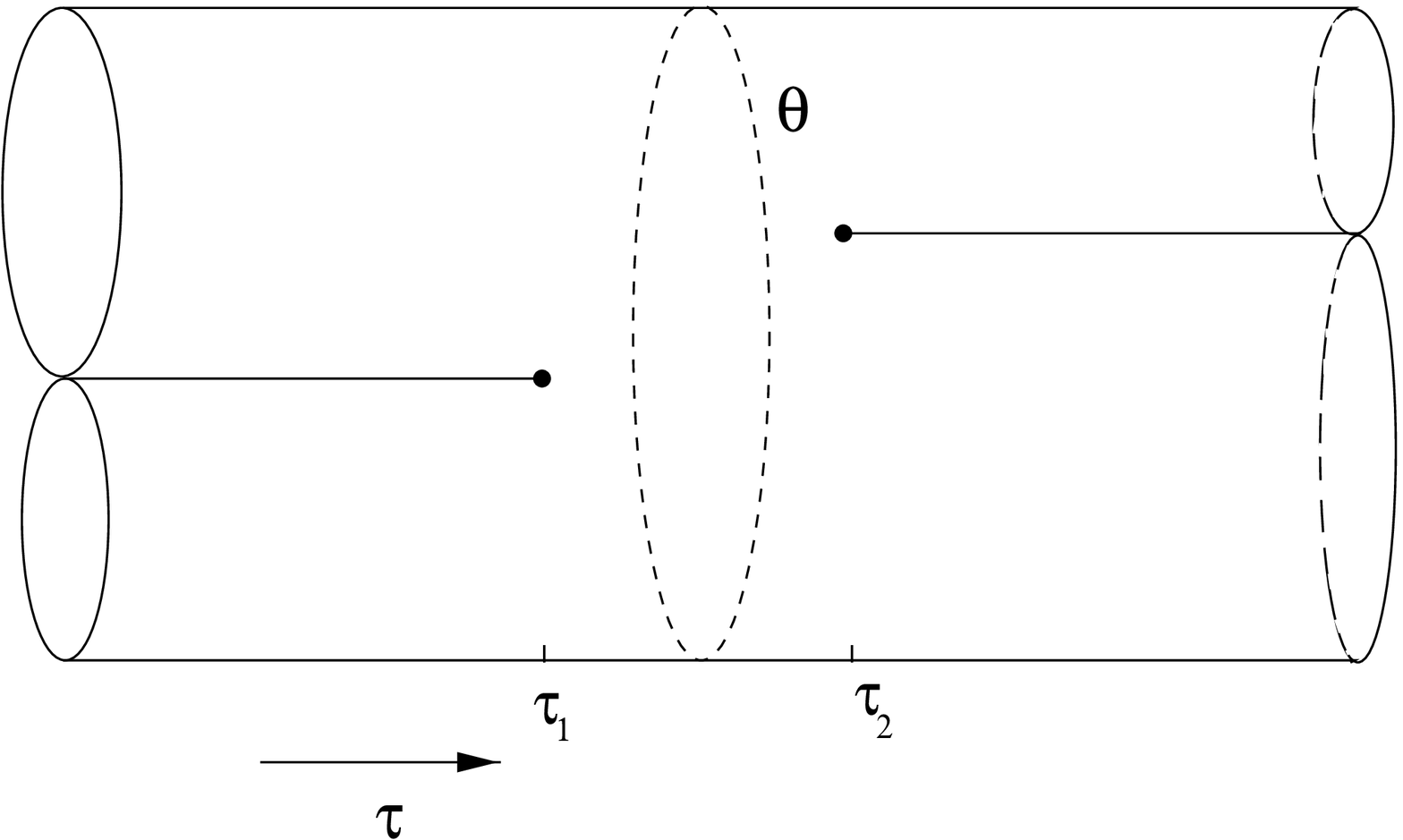}{A tree level light-cone diagram. 
Two incoming strings join to one string and then split again into 
two outgoing strings. The interaction times $\tau$ are indicated, 
as well as one twist angle $\theta$, over which the
internal tube is allowed to rotate before undergoing another 
interaction.}
\xdef\lcdiagram{\thechapter.\the\fignumber\ } 
%%%
\\[3mm]
For fixed external momenta $p^+$, 
a light-cone diagram is uniquely characterized by its moduli: 
the interaction times, the twist angles  
and the internal $p^+$ momentum fractions.
In calculating the contribution to the scattering amplitude of a 
diagram like in figure \lcdiagram one has to do a path integral over
these moduli. This path integral is equal to the analogous 
expression in the Polyakov approach \feynpath\!\!. 
The general form of an $h$-loop contribution
to the amplitude of $n$ scattering strings is  
\eqn\noname{
A_n = {1 \over {\cal N}} \int [d\tau][d\alpha][d\theta] 
\int_{\Psi_1 \cdots \Psi_n} {\cal D} X^I e^{-S_{lc}},
}
where the functional integral is taken over the moduli of the light-cone
diagram. The integral is weighted with the exponential of the 
light-cone action that is just the action of free strings 
\eqn\lcgaugeaction{
S_{lc} = \int d\sigma d\tau 
( -(\partial_\tau X^I)^2 + (\partial_\sigma X^I)^2 ). 
}
%
%Here $I$ runs over the transversal coordinates. 
%
As discussed in the previous section  
light-cone diagrams like figure \lcdiagram are conformally equivalent to  
Riemann surfaces of genus equal to the number of internal strings, and
a number of punctures on it, that correspond to the external strings. 

An important feature of these Riemann surfaces is that the $X^+$ 
coordinate defines a one-form $\omega$ on it, that 
contains all relevant geometrical information.
We can turn this around: each Riemann surface has a 
unique one-form  that can be used to define a local coordinate 
on the string world-sheet $w = \tau + i\sigma$ via 
\eqn\localcoord{
\omega = dw = dX^+(z).  
}
This gives a precise relation between the parameterizations
of the world-sheet $z,w$ and the holomorphic component 
$X^+(z)$ of $X^+(z,\bar{z})$. 

The abelian differential 
$\omega$ has simple poles at the punctures, 
with real residues that add up to zero. It moreover has purely 
imaginary periods on any homology cycle. 
Because the periods are purely imaginary the world-sheet 
time defined via \localcoord is well-defined, and can therefore 
be extended to all over the world-sheet of any genus \GiWo. 
The other coordinate $\sigma$ in \localcoord is 
the multivalued space-like world-sheet coordinate. 

The simple poles of the abelian differential \localcoord mark 
the locations of the vertex operators of the external states on
the Riemann surface. This can be seen by noting that when the 
abelian differential \localcoord has a simple pole at a certain 
point $w_i$, with residue $p^+_i$ 
\eqn\simplepole{
dw \sim {p^+_idz \over (z-z_i)},
}  
we have for the local coordinate $w$ in a neighborhood of the pole 
\eqn\externalstring{
w-w_i \sim p_i^+\log(z-z_i).
}
Hence a punctured neighborhood of $z_i$ in the complex plane is 
mapped by the logarithm to infinity, the inverse mapping of 
the conformal mapping illustrated in figure \longtube\!\!. 

There are also specific points on the world-sheet 
at which strings join or split. 
These interactions take place at zeros of $\omega$, that is critical
points $z=z_0$ of the light-cone coordinate $X^+$. 
In the neighborhood of a simple zero of the abelian differential 
$\omega$ we have 
\eqn\doubleu{
dw \sim (z-z_0)dz \longrightarrow  (w-w_0) \sim (z-z_0)^2. 
}
From this equation we see that if we follow a contour around $z_0$
in the complex $z$-plane once, an angle $4\pi$ is swept out in $w$. 
This angle is the same one that is swept out when circling around an 
interaction point in the light-cone picture. We conclude that zeroes
of $\omega$ correspond to points on the world-sheet where string 
interactions take place. Higher order zeroes correspond to higher 
order interactions. 
\\ 
%%%
\epsfysize=3cm
\hfigure{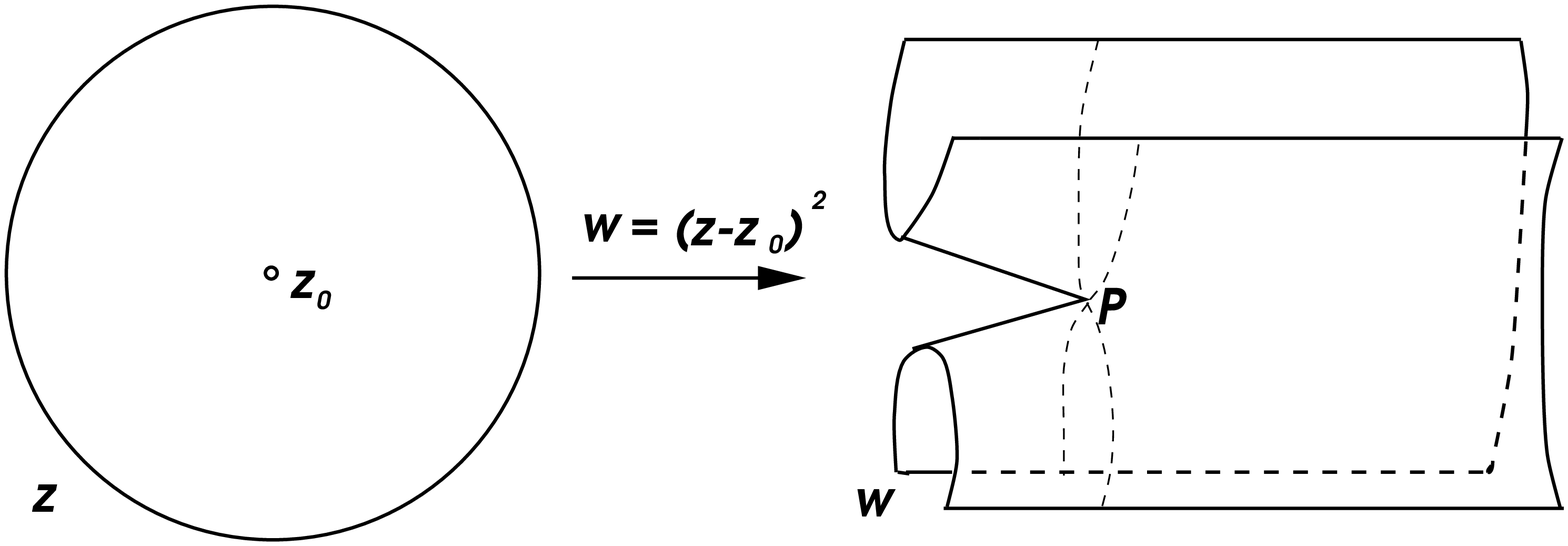}
{A critical point of $\omega$ corresponds with a string
interaction point.}
%%%
\\
The abelian differential also contains information about the internal 
$p^+$ momenta and the twist angles via its integral along the $a$- 
and $b$-cycles of the Riemann surface. We have 
\eqn\noname
{
\oint_{\alpha_i} \omega = ip_i^+,
\qquad 
\oint_{\beta_i} \omega = ip^+_j {\cal M}_{ja}^i \theta_a, 
}
where ${\cal M}_{ja}^i$ is a real-valued matrix that can be chosen to 
have integer coefficients.  

As an example for the location of the interaction points, 
we consider an $n$ particle  scattering process at tree level.  
For a given set of locations $z^i$ of the corresponding
vertex operators, the classical location of the world-sheet 
is described by 
\eqn\xplusn{
X^+(z,\bar{z}) = {1 \over 2} \sum_i \epsilon_i p^+_i 
\log|z-z_i|^2,
}
\eqn\xtrans{
X^I(z,\bar{z}) = {1 \over 2} \sum_i \epsilon_i p_i^I \log |z-z_i|^2,
}
where $\epsilon=1$ for incoming and $-1$ for outgoing particles. 
The interactions take place at critical points $z=z_0$ of the 
light-cone coordinate $X^+$, cf \doubleu
\eqn\noname{
dX^+\mid_{z=z_0}=0.
}
Inserting the explicit form \xplusn\ for $X^+$ gives
\eqn\zsolve{
\sum_{i=1}^n {\epsilon_i p^+_i \over z_0 - z_i} = 0 \ .
}
In case of $n$-point scattering, this condition can be reduced to an
equation of degree $n-2$ in $z_0$, after a conformal transformation
that maps one of the locations of the vertex operators to infinity. 
The $n-2$ zeroes correspond with elementary splittings or joinings 
of strings that occur at the interaction points
of the corresponding tree level light-cone diagram. 

\newsubsection\noname{High energy scattering in string theory}

In chapter three 
we will be particularly interested in high energy four string 
scattering. The motivation to study string theory in this kinematic 
regime is to explore strings at short distances and to learn more 
about the structure of string theory. 
The high energy behavior of scattering strings has been studied in the
past \Ven\grossmende\ by making use of perturbation techniques. 
An important result of this study, is that the dominant world-sheets at
every order of the genus expansion, are all determined by the same 
saddle-point. As a consequence the dominant world-sheets  
have the same form at any genus (up to scaling factors), 
and their contribution 
to the scattering amplitude can be calculated in principle. 

However the analysis revealed two apparent difficulties as well. 
One is that the size of scattering strings tends to grow 
with increasing energy. Strings are therefore not suitable as probes
to investigate the short distance behavior of string theory.
The other difficulty is the fact that the higher order 
amplitudes grow as an exponential of the energy. 
This behavior is completely different from most field theories. 
It leads to the disturbing conclusion that higher order corrections 
are important. This means, in less mild words, that 
the perturbation expansion breaks down for high-energy string 
scattering processes. 

These two problems might be cured when we take non-perturbative 
effects into account. Firstly non-perturbative objects called
D-branes can be used as probes for much smaller distance scales then 
the string scale $\alpha'$ \Shen\danielsson\dkps. Secondly non-perturbative
effects will give important corrections to scattering amplitudes, 
as already was suspected in \grossmende. 
 
The need of a better knowledge of non-perturbative effects in string theory
will be one of the main motivations to introduce the matrix string theory 
model. Before we do this, we will first explain
some features of string theory in the so called discrete light-cone
gauge (DLCQ) formalism. We will moreover review a reformulation of it
in terms of an orbifold model.

\newsubsection\towardsmst{String theory in the DLCQ formalism}

In the DLCQ gauge one compactifies one of the light-like directions
on a circle of radius $R$, so one identifies 
\eqn\compactxminus{
X^{-} \sim X^- + 2\pi R.
}
The other light-like coordinate 
$X^+$ gets the interpretation of time. 
Compactification of $X^-$ modifies the theory in two
ways: the Hilbert space gets truncated to sectors with total 
$p^+$ momentum that is integer valued 
(in $1/R$ units) 
\eqn\noname{
p^+ = N. 
}
In each sector the $p^+$ momenta of states take values in a 
finite positive set only (namely the integers $0 \dots N$). 
Another modification of the theory is that we allow for winding
modes around the $X^-$ direction. These winding modes again 
decouple from the theory in the large $N$ limit, which should 
lead us back to light-cone gauge string theory. 
\\
%
%%%
\epsfysize=5cm
\hfigure{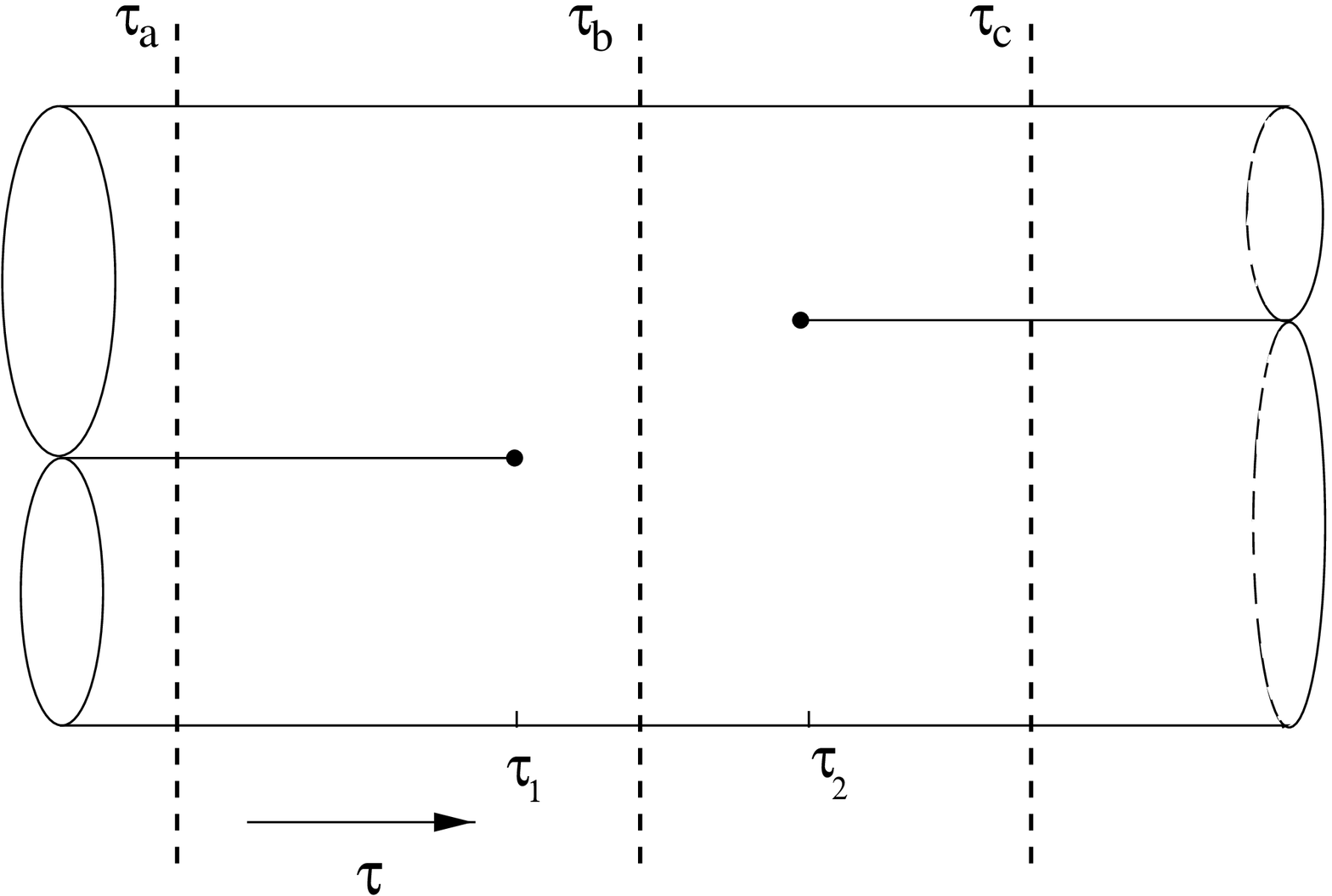}
{In this light-cone diagram we indicated three time slices. In the 
first time slice we have the two incoming strings, in the second the 
intermediate string and in the last slice we have the outgoing string
configuration.}
\xdef\matrixstringeps{\thechapter.\the\fignumber\ } 
%%%
\\
The fact that in DLCQ string theory $p^+$ momenta are integer 
valued, is an important first step towards defining matrix string
theory. It will enable us to define matrices out of the string 
coordinates.  
\\ \indent
Consider the light-cone diagram, illustrated in figure \matrixstringeps 
with total $p^+$ momentum equal to $N$. 
Out of the eight transversal coordinate fields of the strings we 
construct eight diagonal matrices in the following way. 

%%%
The spatial world-sheet 
coordinate $\sigma$ runs over an interval $[0,N]$. 
We cut this interval into $N$ equal pieces of length one 
and define new fields $X^I_i(\sigma,\tau)$ 
\eqn\noname{
X^I_i(\sigma,\tau) = X^I(\sigma+(i-1),\tau),
}
where $i$ runs from $1, \cdots,N$ and $I=1,\cdots,8$ are 
the transversal directions and $\sigma$ now runs over the interval
$[0,1]$. These fields are then interpreted as the eigenvalues of an  
$N\times N$ matrix 
\eqn\matrixcoo{
X^I(\sigma,\tau) = 
\left(
\begin{matrix}
X^I_1(\sigma,\tau) \cr
& X^I_2(\sigma,\tau) \cr
&&\ddots\cr
&&&\ddots\cr
&&&&X_N^I(\sigma,\tau)
\end{matrix}
\right).
}
The light-cone action on the cylinder \lcgaugeaction 
becomes in terms of these matrix eigenvalues 
\eqn\cftaction{
S= {1 \over 2\pi} \int d\sigma d\tau \left(
\partial X^I_i \bar{\partial} X^I_i +
i\theta^{\alpha}_i\partial\theta^{\alpha}_i
+i\theta^{\dot{\alpha}}_i
\bar{\partial}\theta^{\dot{\alpha}}_i \right), 
}
where we included $16N$ fermionic fields 
$\theta^{\alpha}_i, \theta^{\dot{\alpha}}_i$
which together form $N$ 16-component Majorana-Weyl spinors. 
The action \cftaction is left invariant by two space-time left-moving 
and two right-moving supersymmetries. The corresponding left-moving 
supercharge is 
\eqn\gscharge
{Q^\alpha = \sqrt{N} \oint d\sigma \sum_{i=1}^N 
\theta^\alpha_i, \qquad \qquad Q^{\dot{\alpha}} = 
{1\over \sqrt{N}} \oint d\sigma G^{\dot{\alpha}}, 
}
where 
\eqn\Gdef{
G^{\dot{\alpha}}(z) = \sum_{i=1}^N \gamma^I_{\alpha\dot{\alpha}} 
\theta^{\alpha}_i \partial X^I_i. 
}
The right-moving charges have analogous definitions.

The eigenvalues in the matrix \matrixcoo are not single-valued, 
but multivalued with respect to the coordinate $\sigma$. 
For different time slices the eigenvalues $X_i$ 
in \matrixcoo satisfy different quasi-periodic boundary conditions. 
For example we have for the first time slice $\tau_a$ 
two blocks of eigenvalues as indicated in the next equation
\eqn\blockprodigy{
X^I(\sigma,\tau_a) = 
\left(\ 
\begin{matrix}
\fbox{$
\begin{matrix}
X^I_1 \cr
& \cdots \cr
&& X^I_{n_1}  
\end{matrix}
$} & \emptyset
\cr
\emptyset &\fbox{$
\begin{matrix}
X^I_{n_1+1} \cr
& \cdots \cr
&& \cdots \cr
&&& X^I_{N}  
\end{matrix}
$}
\end{matrix}\ 
\right).
}
The collection of eigenvalues in the blocks each 
satisfy cyclic boundary conditions. The 
eigenvalues of the first block satisfy 
\eqnar\cconditions{
X^I_i(\sigma + 1)  &=& X^I_{i+1}(\sigma) \qquad i=1\cdots n_1-1
\nonumber \\ \\[-\the\baselinetruc] , \cr
X^I_{n_1}(\sigma+1) &=& X^I_1(\sigma). \nonumber
}
We can write this condition as an $n_1 \times n_1$ matrix 
equation
\eqn\perio{
X^I(\sigma+1) = V X^I(\sigma) V^{-1},
}
with $V$ the cyclic permutation matrix on the $n_1$ eigenvalues,
\eqn\cycle{  
V= \left(
\begin{matrix}
&~1~&&& \emptyset \ \ \cr &&~1~&&\cr & %\emptyset 
&&\ddots &\cr &\emptyset &&&1 \ \cr ~1~&&&&
\end{matrix}
\right).
}
At time slice $\tau_c$ in figure \matrixstringeps 
the blocks have changed; we then have an
$n_3 \times n_3$ matrix and an $n_4 \times n_4$ matrix, 
representing the outgoing states with $p^+$ momenta $n_3$ 
respectively $n_4$. 
%
%
%%%
\epsfysize=4cm
\hfigure{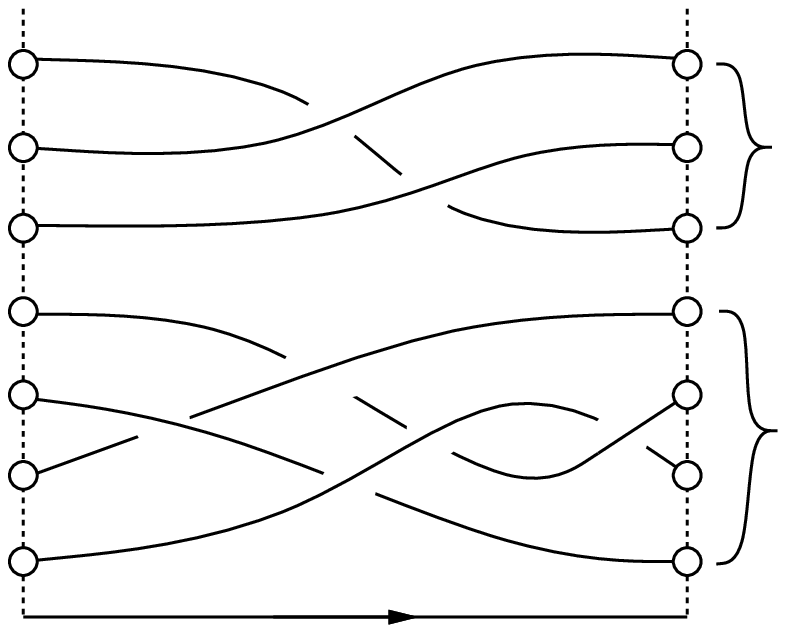}
{A configuration of long strings is determined by the particular 
boundary conditions on the fields $X^I_i$. %To each long string 
%a permutation element of the symmetric group $S_N$ is associated. 
Here we have two 
strings of length 3 and 4. Figure taken from \matrixstring
.}
\xdef\examplesym{\thechapter.\the\fignumber\ } 
%%%
%
\\
For a general string configuration the coordinate matrix \matrixcoo 
will satisfy a periodicity condition of the form \perio with $V$ a 
block diagonal matrix consisting of (say) $s$ blocks of order $n_i$, 
such that
% \eqn{\sum_{i=1}^s N^{i} = N.} 
each block can be taken of the form \cycle\!\!, and thus, 
as described above, defines a string of length $n_{i}$.

Each sector contributes to the total world-sheet energy 
and momentum via  
\eqn\decomp{
H = \sum_i {1\over n_i}
(L_0^{(i)} + \Lbar_0^{(i)}) \qquad\quad P = \sum_i {1\over n_i}
(L_0^{(i)} - \Lbar_0^{(i)}),
}
where we normalized the $L^{(i)}_0$ operators of each 
separate string, such that the oscillator levels are canonically 
counted. They moreover satisfy the level matching condition in 
the DLCQ formalism 
\eqn\levelmatch{
L^{(i)}_0 - \bar{L}^{(i)}_0 = n_im_i,
}
so that each contribution to the total world-sheet momentum is 
integer valued. Here $m_i$ are 
the winding numbers along the light-like circle $X^-$. 
In the large $N$ limit the usual level matching condition 
$L_0 -\bar{L}_0 = 0$ gets restored \matrixstring, because in 
this limit the winding states decouple. 
This arises, because when $N \rightarrow \infty$, 
only long strings survive with $p^i_+ = n_i / N$ finite. These 
strings necessarily have zero winding number, because else they become
infinitely massive compared to the energy of strings with vanishing 
winding number.

\newsection\stringorbifold{DLCQ string theory from an orbifold model}

The numbers $n_i$ that stand for the $p^+$ momenta of
the separate strings in the preceding section, form a partition
of total rank $N$. Thus they define a conjugacy class of the 
symmetric group $S_N$, that is the permutation group of $N$ 
elements. One can write each group element $g$ of $S_N$ as
a product of irreducible cyclic permutations. By conjugating 
$g$ with an appropriate group element of $S_N$ 
its (disjunct) cycle decomposition has the particular simple form 
\eqn\noname{
g = (1 \cdots n_1) (n_{1}\!\!+\!1 \cdots n_1\!\!+\!n_2) 
\cdots\cdots (N\!\!-\!n_{k}\!\!+\!1\cdots N).
}
Other elements in the conjugacy class of $g$ have an equivalent 
decomposition, that is their cycle decompositions all have the 
same number $N_n$ of cycles of length $n$, with the obvious 
restriction 
\eqn\partition{
\sum_n nN_n=N,
}
so that the conjugacy classes of $S_N$ are in one-to-one 
relation to partitions of $N$. 

The now established fact that in the DLCQ formalism, one can
associate to each collection of free strings, a conjugacy class of the 
symmetric group $S_N$, naturally suggests the following conjecture: 
free moving strings in the DLCQ sector $p^+=N$ 
can be reformulated in terms
of a supersymmetric two dimensional conformal field theory
with $8N$ free bosons $X^i_I$, (here $i=1 \dots N$ and $I=1 \cdots 8$)
defined on the symmetric product orbifold 
\eqn\nonameorbi{
S^N \Rvet^8 = (\Rvet^8)^N/S_N, 
}
that together with their fermionic partners are described by the 
action of free strings \hbox{\cftaction\!\!,} where it is understood that all field 
configurations $(X,\theta)$ related by $S_N$ transformations 
are identified $X \sim gX$, $\theta \sim g\theta $ for permutations
$g$. 

In the next subsections we will review evidence for this 
conjecture. First we will show that the model is able to recover the 
complete Fock space of second quantized type IIA string theory in the 
large $N$ limit. Subsequently we will perturb the action of the model
\nonameorbi by an appropriate interaction term that describes
elementary splitting and joining of strings. 
After this we will review work that was done in \aruti\ and
\arutii, where the authors were able to reproduce all tree level four 
particle scattering amplitudes in string theory. 

We will start however by making some general remarks about orbifold 
theories and their partition functions. 
The results we derive will be useful both for understanding matrix 
string theory, and for the next chapter, where we will calculate 
the degeneracy formula of BPS states in $\Nisf$ Yang-Mills theory 
on a three torus. 

\newsubsection\hilbertorbifold{Hilbert space of an orbifold theory} 

Before we derive the Hilbert space of the orbifold model \nonameorbi\!\!,
we will first discuss the Hilbert space in the more general case of
a conformal field theory defined on an orbifold target space $M/G$, 
that consists of a smooth manifold $M$ divided out by a discrete 
group $G$. Again we take the two dimensional world-sheet of our model 
to be a cylinder parameterized by coordinates $(\sigma,\tau)$, where 
$0< \sigma < 1 $. As we divide out by the group $G$ 
all field configurations $(X,\theta)$ are identified when they can be
mapped to each other by a group element, $X \sim gX$, 
$\theta \sim g\theta$ for $g \in G$. Because of this identification 
the coordinate fields of the string are no longer necessarily 
periodic in the spatial coordinate $\sigma$. Instead 
it is allowed that they satisfy so called twisted boundary conditions
\eqn\twistedb{
X(\sigma + 1) = g X(\sigma) . 
}
When we act with a group element $h$ on the coordinate field $X$ 
satisfying \twistedb we get another coordinate field $Y=hX$ with
boundary condition $Y(\sigma+1)= hgh^{-1} Y(\sigma)$. 
The field $Y$ is to be identified with $X$, so  
the twisted boundary conditions are well defined for conjugacy classes 
only. The group element $g$ in \twistedb must therefore be thought of 
as a representative of its conjugacy class $[g]$. 

In order to construct the $G$-invariant Hilbert space of the orbifold model, 
we first consider the subspace $H_g$ which consists of states with 
sigma winding $g$ (that is the coordinate fields satisfy the 
boundary condition \twistedb\!\!). 
The subspace $H_g$ will be mapped to $H_{hgh^{-1}}$, when acting 
with a element $h$ on the states, so we have to include the subsector 
$H_{hgh^{-1}}$ as well. In case that $h$ is an
element of the centralizer of $g$, $H_g$ is mapped to itself. This 
implies that all states in $H_g$ have to be invariant under the action
of the centralizer $C_g$ of $g$. 

We thus conclude that the total Hilbert space of 
the orbifold model is determined by the conjugacy classes of the 
group $G$, with each twisted sector invariant under the centralizer.  
In other words the Hilbert space of an orbifold conformal field theory
has the decomposition 
\eqn\hilbertorb{
\cH(M/G)=\bigoplus_{[g]}\cH^{C_g}_g(M),
}
where $g$ is an arbitrary representative of the conjugacy class 
$[g]$ and where $\cH^{C_g}_g$ is the $g$-twisted subsector 
of the total Hilbert space that is invariant under conjugation
by elements in the centralizer $C_g$.

At the level of partition functions we can see the just described
structure of the Hilbert space as follows. We start with the path-integral
representation of the orbifold partition function on a torus with two 
homology cycles along spatial and time direction 
\eqn\partitioni{
Z = {1 \over |G|} \sum_{g,h \in G \atop [g,h] =1} Z(g,h) .
}
Here $Z(g,h)$ represents the partition function evaluated with 
twisted boundary conditions by the group elements $g$ and $h$ 
along the two different cycles. For consistency the elements $g$ and
$h$ have to commute. The next step is then to recognize that the
partition functions $Z(g,h)$ depend only on the conjugacy classes of $G$ 
: $Z(xgx^{-1},xhx^{-1}) = Z(g,h)$ for all $x \in G$.  
Hence we can write \partitioni as follows
\eqn\partitionii{
Z= \sum_{[g]} {1 \over |C_g|} \sum_{h \in C_g} Z(g,h),
}
where we used the identity $| C_g || [g] | = |G| $.  
As the operator
\eqn\projection{
P_{[g]} = {1 \over |C_g|} \sum_{h \in C_g} h 
}
projects onto $C_g$ invariant states we conclude that the 
partition function can indeed be written 
\eqn\partitioniii{ 
Z= \sum_{[g]} \tr_{H_{[g]}} P_{[g]} q^{L_{0}} \bar{q}^{\bar{L}_0}.
}
The form of this partition function precisely corresponds with 
the structure of the Hilbert space of the orbifold model we have 
just discussed.

\newsubsection\noname{The long string picture} 

We have seen in the previous section that the Hilbert space of 
an orbifold model is determined by the conjugacy classes of the 
symmetry group that is divided out. For the symmetric product 
orbifold model \nonameorbi this group is the permutation group
$S_N$. 

As explained before a conjugacy class of the symmetric group
$S_N$ is labeled by a partition $\sum nN_n =N$ of $N$. 
The corresponding centralizer subgroup takes the form 
$C_g = \prod_n \Zvet^{N_n}_{n} \rtimes S_{N_n}$
where $S_{N_n}$ permutes the $N_n$ cyclic permutations of length $n$, 
and where each subfactor $\Zvet_n$
acts within one particular cyclic permutation. 
Due to the factorization of the conjugacy classes in irreducible 
elements of $S_N$, one can express the twisted sectors of the Hilbert
spaces $\cH^{C_g}_g$ as a product over the sectors of 
graded $N_n$-fold symmetric tensor products of smaller
Hilbert spaces $\cH^{\Zvet_n}_{(n)}$
\eqn\smallhil{
{\cH}^{C_g}_g = \bigotimes_{n>0} \, S^{N_n} \cH_{(n)}^{\Zvet_n}.
}
Combining the ingredients we get the decomposition for the total
Hilbert space 
\eqn\decompose{
\cH(S^N X)=\bigoplus_{N_n \atop \sum_n nN_n=N} 
\bigotimes_{n>0} \, S^{N_n} \cH_{(n)}^{\Zvet_n}. 
}
The space $\cH^{\Zvet_n}_{(n)}$ is a particular 
$\Zvet_n$-invariant subsector of the total Hilbert space. It can
be interpreted as the space of states of a single string living on 
$X \times S^1$, winded $n$ times around the circle. 
%
%%%%%%
\epsfysize=6cm
\hfigure{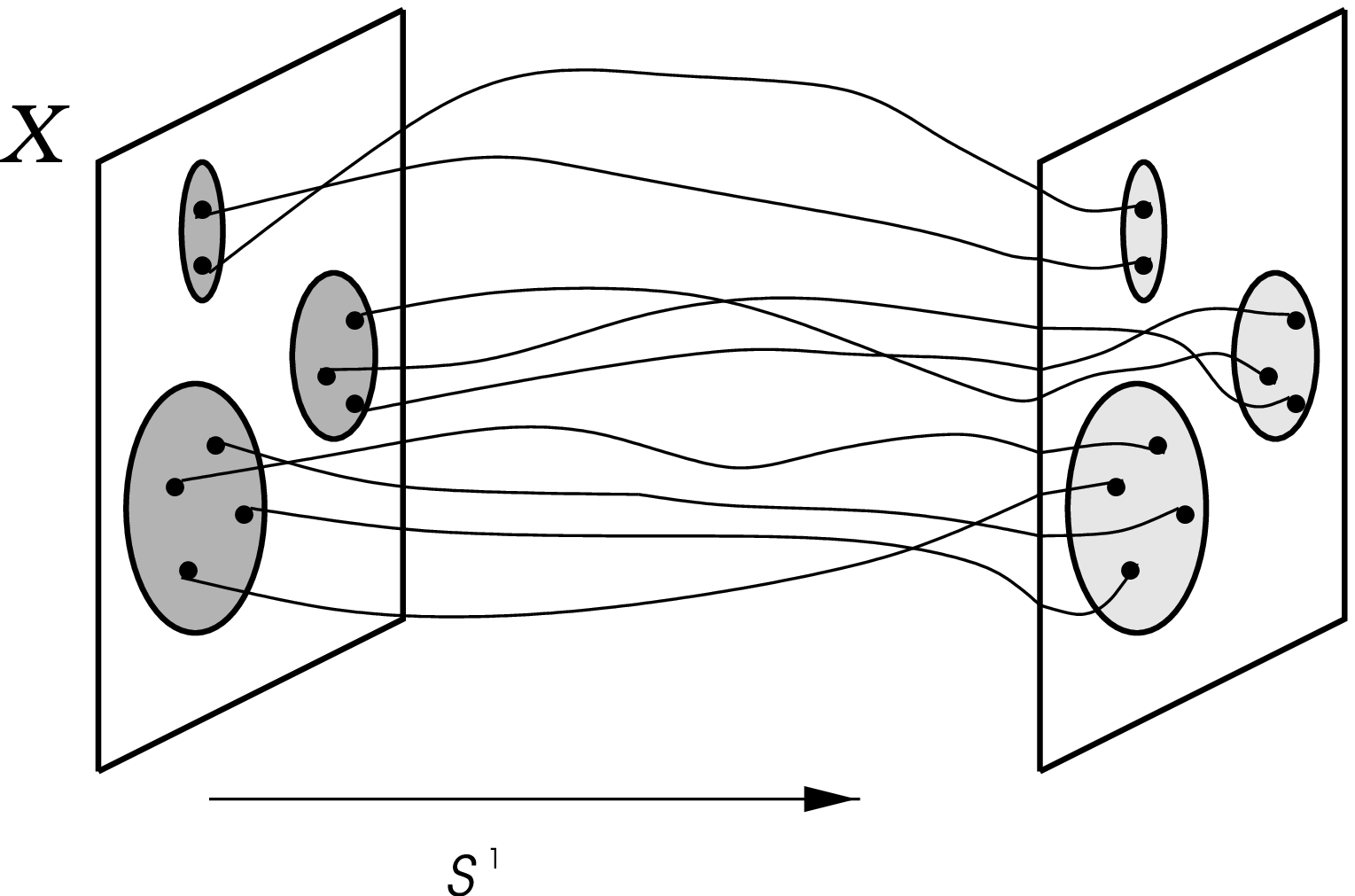}
{An illustration of the long string picture introduced in 
\dmvv\matrixstring. Three strings are shown, of length 2,3 and 4.
}
%%%%%%
%
\xdef\longstringpic{\thechapter.\the\fignumber\ } 
\\
This single string has a description in $n$ coordinate fields 
$X_I \in X$, with the cyclic boundary conditions \cconditions. 
Thus we are back to the string coordinate 
configurations of section \towardsmst\!\!. 
One can glue the coordinates $X_I$ of a 
single string into one single field $X(\sigma)$ via 
\eqn\gluing{
X(\sigma+k) = X_{k+1}(\sigma), 
}
where now the argument of the field $X$ runs over a circle 
of radius $N$.

This string coordinate has fractional oscillators modes, but
invariance under the group $\Zvet_n$ implies that the fractional 
left moving minus right moving oscillator numbers add up to an integer. 
This can be
easily seen by noting that $\Zvet_n$ acts by cyclic permutations 
on the fields $X_I$ and therefore acts by translations 
$\sigma \rightarrow \sigma + 1$ on the field $X$ of the single 
long string. In other words the $\Zvet_n$ invariance 
implies that the contribution from each single string sector 
to the total world-sheet momentum $P=L_0-\bar{L}_0$ is integer-valued. 
So we are back to the decomposition of the total world-sheet energy
and momentum in \decomp\!\!, combined with the level matching condition
\levelmatch\!\!.

%\newsection\lcstorbi{LCG string theory  from an Orbifold Model}%
%
%Now we will review the orbifold model \nonameorbi that,  
%as we already announced, is able to reproduce all four particle 
%scattering amplitudes in string theory. 
%

\newsubsection\twistedvac{Twisted vacua and excitations} 

In this subsection we will discuss the twisted vacua of the model
\nonameorbi\!\!. It is convenient 
to change coordinates from the cylinder $w=\tau+i\sigma$ to the complex
plane $w \rightarrow z= e^w $ (cf. figure \longtube\!\!). 
%As usual, the equations of motion of the model \nonameorbi 
%allow writing the fields 
%$X^i_I$ as a sum of left-moving (holomorphic) and 
%right-moving (anti-holomorphic) components
%%
%\eqn\leftright{
%X^I_i(z,\bar{z}) = X^I_i(z)+\bar{X}^I_i(\bar{z}).
%}
%
%We can therefore concentrate on the left-moving part in the following.
%
%
%In matrix string theory strings arise as diagonal matrices 
%that satisfy the equations of motion, and whose eigenvalues 
%are multivalued on the YM cylinder. Different boundary conditions
%correspond to different configurations of strings, and they can be 
%labelled by conjugacy classes of the symmetric group $S_N$, as 
%discussed in the first chapter. 

A string of maximal length 
(i.e. a long string, whose length equals the total $p^+$ momentum) 
has coordinates that satisfy the quasi-periodic boundary conditions 
\perio\!\!.
In the CFT we are considering these boundary conditions can be used to 
define a bosonic twist field $\sigma_n (z, \bar{z})$ \ginsparg\ 
via the monodromy relation 
\eqn\monodromy{
X^I(e^{2\pi i}z, e^{-2\pi i}\bar{z}) \sigma_{n}(0) = 
\omega_n X^I(z,\bar{z})\sigma_{(n)}(0),
}
where now $X$ should be viewed as an operator and 
where $\omega_n$ is the generator of the $\Zvet_n$ group that acts 
cyclically on the long string sector of length $n$. The twist operator
$\sigma_n$ defined by \monodromy can be used to create the 
vacuum in the twisted sector by acting with it on the untwisted 
vacuum  
\eqn\vacuum{
\left|(n)\right> = \sigma_{(n)}(0,0)\left|0\right>.
}
In this twisted sector the left-moving component of the 
field $X(z,\bar{z})$ has the Laurent type power series expansion
\eqn\laurent{
\partial X^I_{j} = -{i \over n}\sum_m \alpha^I_m 
e^{-{2\pi i\over n}jm} z^{-{m\over n}-1}, 
}
where the $\alpha^I_m$ are the usual creation and annihilation operators
that satisfy the commutation relation
\eqn\commutations{
[\alpha_m^I , \alpha^J_n] = m\delta^{IJ}\delta_{m+n,0}.
}
The conformal weight of the twist field $\sigma_{(n)}$ can
be found by reading off the singular term in its OPE with the 
bosonic stress-energy tensor. 
This stress-energy tensor is defined as
\eqn\stresstensor{
T(z) = -{1\over 2} \sum_I^8\sum_i^n %\left( 
:\partial X^I_i(z) \partial X^I_i(z): .%\right) 
}
By a straightforward calculation one finds the conformal 
weight of $\sigma_{(n)}$. The result is 
\eqn\cftweight{
h^b_n = {1\over 3}(n- {1 \over n}). 
}
The most singular term in the operator product expansion of the 
bosonic twist field $\sigma_{(n)}$ and the field $\partial X_i^I$ 
is 
\eqn\btwist
{\partial X^I_j(z) \cdot \sigma_{(n)}(w,\bar{w}) 
\sim (z-w)^{-(1-{1\over n})}
e^{{2\pi i\over n}j} \tau^I_{(n)}(w,\bar{w}),
}
where 
\eqn\noname{
\tau^I_{(n)}(0,0) \left|0 \right> 
= -{i\over n}\alpha^I_{-1} \left|(n)\right>,
}
is the first excited state in the twisted sector 
$\left|(n)\right>$ . By dimensional counting we find that the 
conformal dimension of $\tau^I_{(n)}$ is 
${1\over 3}(n+{2\over n})$. The operator $\tau^I$ will be used 
later for the construction of the interaction vertex. 

One obtains excitations of the twisted vacuum states by 
applying the usual vertex operators. For instance a scalar particle with 
momentum $k^I$ in a sector $|n>$ is obtained by acting on this vacuum 
state with the vertex operator 
\eqn\scalar{
: e^{ik^I_i X^I_i(0,0)} : |(n)>, 
}
Here $k^I = \sum_{i=1}^n k^I_i$ is the total momentum of the 
collection of long strings in the transversal direction $I$. 

In a long string sector of length $n$ the expansions of 
the left and right-moving components of the fermions
are 
\eqn\fermionexpi{
\theta^{\alpha}_j(z) = {1 \over \sqrt{n}} 
\sum_m \theta^{\alpha}_m 
e^{-{2\pi i \over n}jm} z^{-{m\over n}-{1\over 2}}. 
}
\eqn\fermionexpii{
\theta^{\dot{\alpha}}_j(\bar{z}) = {1 \over \sqrt{n}}
\sum_m \theta^{\dot{\alpha}}_m e^{{2\pi i \over n}jm} 
\bar{z}^{-{m\over n}-{1\over 2}}.
}
The creation and annihilation operators satisfy the commutation 
relations 
\eqn\fermcom{
\left\{ \theta^{\alpha}_m , \theta^{\beta}_n \right\} = 
\delta^{\alpha\beta} \delta_{m+n,0}.
}
These commutation relations imply that the zero modes form
a Clifford algebra. This means that the vacuum state must represent 
this algebra. Because of triality in the transversal spatial index $I$
and the spin indices $\alpha$ and $\dot{\alpha}$ the vacuum state can 
be chosen as a 16 component vector with components $\left| I \right>$
and $\left|\dot{\alpha}\right>$ 
normalized in the standard way, that moreover satisfy
the relations \gsw\
\eqn\fvacua{
\theta^{\alpha}_0 \left|I \right> = {1 \over \sqrt{2}} 
\gamma^I_{\alpha\dot{\alpha}}\left| \dot{\alpha}\right>, 
\quad \theta^{\alpha}_0 \left|\dot{\alpha}\right> = 
{1\over \sqrt{2}} \gamma^I_{\alpha\dot{\alpha}} \left|I \right>.
}
The vacua $\left|I\right>$ and $\left|\dot{\alpha}\right>$ are 
created by primary fermionic twist fields (spin fields) 
which we denote by $\Sigma^I_{(n)}$ respectively 
$\Sigma^{\dot{\alpha}}_{(n)}$. 

The most singular terms of the OPE of the fermionic twist fields and
the fermionic fields $\theta^{\alpha}$ are 
\eqn\ftwisti
{\theta^{\alpha}_i(z) \cdot \Sigma^I_{(n)} (w) \sim {1\over \sqrt{2n}} 
(z-w)^{-1/2} \gamma^I_{\alpha\dot{\alpha}}
\Sigma^{\dot{\alpha}}_{(n)}(w), 
}
\eqn\ftwistii
{\theta^{\alpha}_i(z) \cdot \Sigma^{\dot{\alpha}}_{(n)}(w) \sim
{1\over \sqrt{2n}} 
(z-w)^{-1/2} \gamma^I_{\alpha\dot{\alpha}}
\Sigma^I_{(n)} (w).
}
The stress energy tensor for the fermionic fields is defined 
as 
\eqn\stressf{
T^F(z) = -{1\over 2} \sum_{\alpha=1}^8\sum_{i=1}^n 
: \theta^{\alpha}_i(z)\partial \theta^{\alpha}_i(z) :.
}
From the OPE of the twist operator $\Sigma_{(n)}$ with the 
fermionic stress energy tensor one can read off its 
conformal weight.

It equals $\Delta^f_{n} = {n \over 6} + {1 \over 3n}$.

Now we have defined the bosonic and fermionic twist operators 
we can construct vertex operators, that create the vacuum ground states
of arbitrary twisted sectors. Each sector is represented by a
particular conjugacy class of the symmetric group $S_N$.  
For a given conjugacy class $[g]$ the vertex operator 
can be written as a product of vertex operators in correspondence 
to the decomposition of a representative $g$ in irreducible 
cyclic permutations of length $n$
\eqn\vertexproduct
{
V_{[g]} = \prod_{n} V_{(n)} .
}
Each vertex operator $V_{(n)}$ in equation \vertexproduct
can be represented by products of bosonic and fermionic twist 
fields \btwist\!\!, \ftwisti\!\!-\ftwistii\!\!, that together create 
the vacua of the twisted sectors. Explicitly the vertex operator 
\vertexproduct reads 
\eqn\vertexo
{V_{[g]} = {1 \over N!} \sum_{h\in S_N} V_{h^{-1}gh} (z,\bar{z}) .
}
The expression \vertexo is invariant under conjugation of
elements of the symmetric group and therefore well defined.
We assume that the vertex operators can be written as the tensor
product of a left-moving part and a right-moving part, that can each
be decomposed in a fermionic twist operator and a bosonic operator. 
%We define the operator product of two vertex operators 
%$V_{g_1}$ and $V_{g_2}$ such that the decomposition in left/right-moving
%and bosonic/fermionic parts is preserved, e.g.
%%
%\eqn\noname{
%(V^b_{g_1}\otimes V^f_{g_1}) \cdot (V^b_{g_2}\otimes V^f_{g_2}) = 
%c^{h_1h_2}_{g_1g_2}\ V^b_{h_1h_2}\otimes V^f_{h_1h_2} 
%}
%%
%where the $c$'s are structure constants. 

%that are written as tensor
%products of fermionic and bosonic vertex operators, might have 
%non-diagonal terms, namely 
%%
%\eqn\unwanted{
%(V^b_{g_1}\otimes V^f_{g_1}) \cdot (V^b_{g_2}\otimes V^f_{g_2}) = 
%c^{h_1h_2}_{g_1g_2}\ V^b_{h_1h_2}\otimes V^f_{h_1h_2} 
%+
%\tilde{c}^{h_1h_2}_{g_1g_2}\ V^b_{h_2h_1}\otimes V^f_{h_2h_1} 
%+ {\rm extra \ terms}
%%V_{g_1}V_{g_2} = c^{h_1h_2}_{g_1g_2} V_{h_1h_2}+ {\rm extra\ terms} 
%}
%%
%The extra terms in \unwanted can arise, 
%due to the non-abelian nature
%of the orbifold model we are considering. 
%{\it The problem might disappear when taking the sum over all conjugacy
%classes, we usually do, have to check this}. 
%However as shown in \aruti\ the factorization can be assumed, provided
%we include certain normalization factors at the final stages of the
%calculation of the scattering amplitudes. 
%As we will not present detailed calculations of scattering
%processes here, we will proceed and refer the reader to \aruti\ 
%for the subtleties of tensoring the vertex operators in fermionic/bosonic
%and left-moving/right-moving components. 
%
%%%%%%%%%%%%%%%%%%%%%%%%%%%%%%%%% 
Combining the fermionic and bosonic states, 
all the 256 massless states of IIA supergravity are obtained. For example
in the long string sector a graviton with momentum $k^I$ and 
polarization $\zeta$ is created by the vertex operator
\eqn\graviton{
V_{(n)}[k^I,\zeta](z,\bar{z}) = \zeta_{IJ} 
\sigma_{(n)}[k^I](z,\bar{z})
\Sigma^I_{(n)}(z) \bar{\Sigma}_{(n)}^J(\bar{z}).
}
Here $\sigma_{(n)}[k^I]$ is shorthand for the product 
of the twist operator $\sigma_n$ and the vertex operator 
\scalar that introduces momentum $k^I$. 

In the next section we will consider 
string interactions that can be described by the orbifold model
after adding a deformation term.

\newsection\noname{Reproducing tree level string scattering}

\newsubsection\mstinteraction{Interaction vertex} 

The question arises, how interactions between
strings can be formulated in the orbifold model \nonameorbi\!\!. 
To include splitting and joining processes of strings we have to
deform the model by adding an interaction term. 
It is clear from figure \matrixstringeps that an interaction should 
somehow connect the block diagonal field configurations \blockprodigy
of different ranks. %, that represent the asymptotic states. 
In other words an interaction changes the quasi-periodic boundary 
conditions the matrix fields satisfy.
Two eigenvalues of the field $X_I$ will be interchanged 
at some intermediate stage, so that a group element $g_1$ 
of $S_N$ that represents the initial string configuration is 
changed into another element $g_2$ (see figure 
\global\advance\fignumber by 1 \thechapter.\the\fignumber).
\global\advance\fignumber by -1 
Then one naturally 
associates the group element $g= g_1^{-1} g_2$ to the interaction 
vertex, though of course this group element is not unique. 

As an example we take the scattering process of figure 
\matrixstringeps\!\!.
The incoming state can be represented by an $S_N$ group element that 
consists of two permutations $(12\cdots n_1)(n_1+1 \cdots N)$, 
whereas the intermediate string state 
has maximal length and therefore has 
an associated permutation element $(12\cdots N)$. 
These two group elements are related by a simple permutation
\eqn\cyclecycle{
(12\cdots n_1)(n_1\!\!+\!1 \cdots N) (n_1N) = 
(12\cdots N),
}
or by any other transposition that exchanges an element of the
integers $1 \cdots n_1$ with another element of 
$n_1\!\!+\!1 \cdots N$. 

Hence we see that the joining process in figure \matrixstringeps 
can be described by simple permutations. 
Likewise the splitting of one string into two strings can be described
with the help of transpositions. 
\\
%%%
\epsfysize=3.3cm
\hfigure{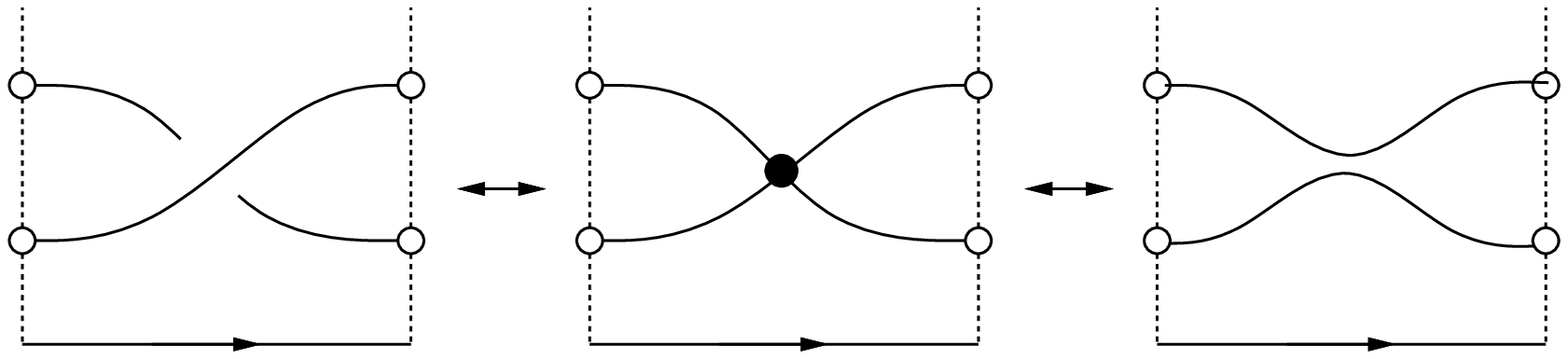}
{A splitting process of a long string corresponds to a 
change of the boundary conditions of the coordinate fields. 
To an elementary splitting of strings a 
simple permutation is associated, 
cf equation \cyclecycle\!\!. Figure taken from \matrixstring
.}
\xdef\matrixstringeps{\thechapter.\the\fignumber\ } 
%%%
%
\\
It is quite natural to use twist fields \ginsparg\ 
for the definition of 
the interaction vertex, as these twist fields change 
the boundary conditions of bosonic or fermionic fields. 
Joining and splitting of strings have a description in terms
of $\Zvet_2$ twist operators, that are special cases of the 
operators defined in 
\btwist\!\!, \ftwisti and \ftwistii\!\!. 
They are defined via the operator product expansions 
\eqn\btwistz
{\partial X_-^I(z) \cdot \sigma (w,\bar{w}) 
\sim (z-w)^{-{1\over 2}} \tau^I (w,\bar{w}),
}
\eqn\ftwistzi
{\theta^{\alpha}_-(z) \cdot \Sigma^I_{(n)} (w) \sim {1\over 2} 
(z-w)^{-{1\over 2}} \gamma^I_{\alpha\dot{\alpha}}
\Sigma^{\dot{\alpha}}(w), 
}
\eqn\ftwistzii
{\theta^{\alpha}_-(z) \cdot \Sigma^{\dot{\alpha}}(w) \sim
{1\over 2} (z-w)^{-{1\over 2}} \gamma^I_{\alpha\dot{\alpha}}
\Sigma^I(w).
}
Here the index $-$ refers to the element of $\Zvet_2$ with order 2. 

With the twist fields \btwistz\!-- \ftwistzii one can construct a
vertex that 
describes the joining and splitting of strings, and that is moreover 
supersymmetric and manifestly $SO(8)$-invariant \matrixstring, 
\eqn\dvvvertex
{
V_{int} = {\lambda N\over 2\pi} \sum_{i<j} \int d^2z 
\left(\tau^I(z) \Sigma^I(z) \otimes 
\bar{\tau}^J(\bar{z})\bar{\Sigma}^J(\bar{z})\right)_{ij}.
}
Here $\lambda$ is a coupling constant of mass dimension $-1$, 
as the integrand is a weight $({3\over 2},{3\over 2})$ 
conformal field. We take $\lambda$ to be proportional to the 
string coupling constant $g_s$, so that in the zero string 
coupling limit the interaction vertex vanishes.
The vertex operator is manifestly invariant under the supercharges
$Q^\alpha$ \gscharge\!\!, as the $Q^\alpha$'s only depend on the zero modes 
$\theta^\alpha_+$ . 
The other supercharge $Q^{\dot{\alpha}}$ in \gscharge 
acts in a non-trivial way on the
terms in \dvvvertex but leaves the vertex as a whole invariant 
\matrixstring: 
First we note that (no summation over $\dot{\alpha}$)
\eqn\descendent{
\left[G_{-{1\over 2}}^{\dot{\alpha}}, \sigma \Sigma^{\dot{\alpha}}\right]
={1\over 2}\tau^I \Sigma^I, 
}
where $G_{-{1\over 2}}^{\dot{\alpha}}$ 
is the Fourier mode with a simple pole 
in the expansion of the supersymmetry operator $G^{\dot{\alpha}}$ 
which has conformal weight $3/2$
\eqn\gexpansion{
G^{\dot{\alpha}} = 
\sum_{n \in \Zvet + {1\over 2}} G_{-n} z^{n-3/2}.
}
Taking the commutator with $G_{-{1\over 2}}^{\dot{\alpha}}$ 
on both sides
of \descendent\!\!, and using the Jacobi identity we get 
\eqn\totalder{
[G_{-{1\over 2}}^{\dot{\alpha}}, \tau^I\Sigma^I ] = 
[L_{-1}, \sigma\Sigma^{\dot{\alpha}}] = 
\partial_z (\sigma \Sigma^{\dot{\alpha}}). 
}
Because the right-hand side of \totalder is a total derivative we 
conclude that the vertex operator is invariant under 
the supercharge $Q^{\dot{\alpha}}$.  

The interaction vertex \dvvvertex is unique in the sense that 
it is the least irrelevant supersymmetric $SO(8)$-invariant 
operator we can form with the twist operators. 

Long time ago Mandelstam already proposed the vertex
\dvvvertex in the context of interactions formulated in the NSR 
approach to superstrings \mandelstami. In this formulation the 
twist fields $\Sigma^i$ play the role of the fermionic variables 
of the string. In the light-cone gauge the obvious geometrical 
three vertex function 
\eqn\noname{
\prod_{\sigma, I} \delta(X^I_{\rm f}(\sigma) - X^I_{\rm i}(\sigma)),
}
is not  $SO(9,1)$ invariant. 
Here $X_{\rm i}$ and $X_{\rm f}$ are the string states before and 
after the interaction. A Lorentz invariant vertex is obtained 
for the NSR formalism by adding an extra term factor at the joining
point \mandelstami
\eqn\extrafactor{
\Sigma^I \partial X^I. 
}
Remarkably when we consider the geometric joining and splitting 
that in the formalism of \matrixstring\ takes the form of the
twist operator $\sigma(0)$ and take into account the extra factor
\extrafactor we precisely arrive at the vertex operator
\dvvvertex
\eqn\noname{
\oint {dz \over z^{1/2}} \Sigma^I \partial X^I (z) \sigma(0) 
= \tau^I \Sigma^I(0).
}
The fact that in the NSR formulation the factor \extrafactor makes
the three-vertex $SO(9,1)$ invariant, gives a hint that 
matrix string theory also has ten dimensional Lorentz invariance, 
although of course we expect full Lorentz invariance only 
in the large $N$ limit.

\newsubsection\noname{Tree level string scattering} 

Now we have derived the vertex operator for elementary 
joining and splitting processes we can in principle calculate 
amplitudes of string scattering processes and compare the 
results with light-cone string theory. 

In \aruti\ 
the four graviton scattering amplitude was calculated at tree level 
and in \arutii\ this result was extended to all (tree level) four 
particle scattering amplitudes. To compare these results  with 
light-cone string theory, the large $N$ limit should be
taken, but it appears that for tree level amplitudes this limit is 
straightforward.   

As the calculations of the amplitudes are 
rather technical we will only sketch here the general 
ideas, and refer to \aruti\arutii\ for the details.  

For a four graviton scattering process the $S$-matrix is up to 
quadratic order in $\lambda$ \aruti 
\eqn\smatrixorb{
\left<f \right| S\left|i\right> = -{1\over 2}
\left({\lambda N \over 2\pi}\right)^2
\left<f \right| \int d^2z_1d^2z_2 |z_1||z_2| {\cal T} \sum_{i<j} 
V_{ij}(z_1,\bar{z}_1) 
\sum_{k<l}V_{kl}(z_2,\bar{z}_2)  \left| i\right>, 
}
where $V_{ij}$ is defined as the integrand of the integral in 
\dvvvertex\!\!, 
${\cal T}$ is the ordering operator used in radial quantization 
\ginsparg\ and 
$\left|i\right>$ and 
$\left|f\right>$ are the initial and final states, each consisting of two
gravitons with certain momenta and polarization,  
\eqn\instate{
\left|i\right> = V_{[g_0]}[\vec{k}_1,\zeta_1;\vec{k}_2,\zeta_2](0,0) 
\left|0\right>
}
and
\eqn\outstate{ 
\left<f\right| =\lim_{z\rightarrow\infty} 
\left<0\right|V_{[g_{\infty}]}[\vec{k}_3,\zeta_3;\vec{k}_4,\zeta_4]
(z,\bar{z}) z^{2h}\bar{z}^{2\bar{h}}.  
}
The conjugacy class $[g_0]$ 
can be represented as the product of two irreducible (disjunct) cycles
$g_0 = (n_0)(N-n_0)$, with according to the long string picture 
$n_0$ and $N-n_0$ the $p^+$-momenta of the two scattering gravitons. 
Likewise $[g_\infty]$ has a representative $(n_\infty)(N-n_\infty)$. 

The expression \smatrixorb gets simplified after the conformal 
transformation $z \rightarrow {z \over z_1}$ 
and the introduction of a new integration variable $u={z_2 \over z_1}$. 
The integral over $z_1$ contributes a delta-function of the $k_-$
momenta, so that the S-matrix reduces to a single integral 
\eqn\tussenresultaat{
\left<f\right|S\left|i\right> =
-i2\lambda^2N^3 \delta(\sum_i k^-_i) 
\sum_{i<j, k<l}
\int d^2u |u| 
\left<f\right|{\cal T}\ V_{ij}(1,1) V_{kl}(u,\bar{u}) 
\left| i \right>. 
}
The expression in \tussenresultaat involves correlation functions
of four vertex functions
\eqn\fourvertex{
\left<V_{g_\infty}(\infty)V_{h_2}(1,1)V_{h_1}(u,\bar{u})
V_{g_0}(0,0)\right>. 
}
The vertex functions $V_{g_\infty}$, $V_{g_0}$ correspond to 
the twisted in- and out-states, the other two describe elementary
splitting respectively joining of strings (and therefore the 
group elements $h_{1,2}$ are two-cycles). 

The vertex functions \fourvertex vanish unless the $S_N$ 
group elements that label the vertex functions, satisfy the 
appropriate physical conditions. These conditions are determined
by the particular process by which the incoming state gets transferred
to the outgoing state. For the four particle scattering process we 
consider, we have two possibilities: One of the two incoming strings
splits into two strings; out of the three intermediate strings two 
join together again, so that there are two outgoing strings. 
In the other process two incoming strings join to one string; 
this intermediate string propagates and splits again in two outgoing 
strings. 
\global\advance\fignumber by 1 
To these possibilities, that are illustrated in figure
\thechapter.\the\fignumber, 
particular group elements of $S_N$ are associated, for which the 
expectation value \fourvertex does not vanish. 
\global\advance\fignumber by -1 
\\
\epsfysize=4cm
\hfigure{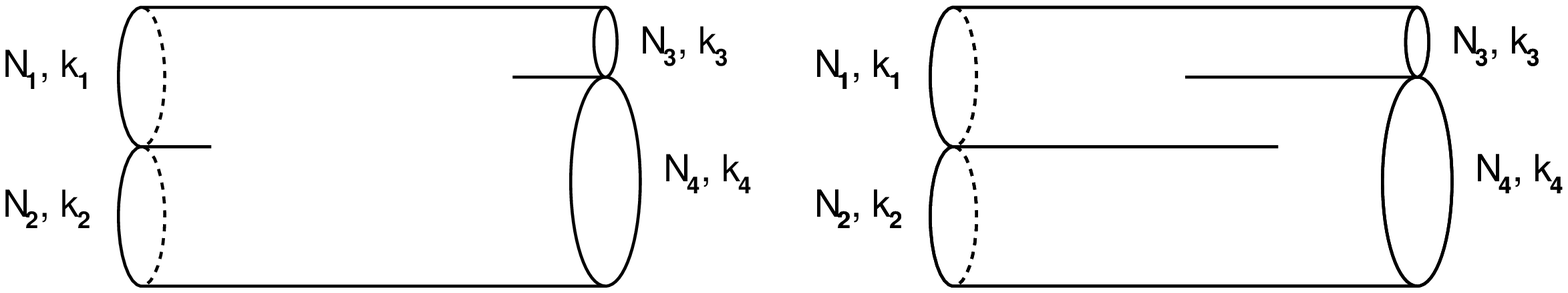}{Two typical diagrams that contribute to
the tree level amplitudes.}
\\
The rest of the calculation is rather tedious; for the full details we
refer to the original papers \aruti\arutii. There, firstly the correlation
functions are factorized in left-moving/right-moving and
bosonic/fermionic parts. Then these bosonic and fermionic correlation
functions are determined explicitly, together with the appropriate 
normalizations. 

The end result for four-graviton scattering at tree level is 
\eqn\resulttedious{
A = \lambda^2 2^{-8} \int d^2z |z|^{{1\over 2}k_1k_4-2} 
|1-z|^{{1\over 2}k_3k_4-2} K(z,\bar{z},\zeta), 
}
where $K$ is a kinematic factor depending on the polarizations
and momenta known from tree amplitudes in string theory. 
The result \resulttedious coincides with known results of 
string theory \gsw.

For the above result the large $N$ limit is needed. For tree level
this limit is relatively simple, because there are no subleading terms
in the scattering amplitude like $1/N$ corrections. 
This is the case as all 
strings in the diagrams of figure \thechapter.\the\fignumber\ 
satisfy the standard level matching condition $L_0 -\bar{L}_0= 0$. 
Note that this is no longer necessarily true for higher loop string 
diagrams, where the intermediate strings may {\it not} satisfy the 
usual level matching condition. We therefore expect that the large
$N$ limit will be less trivial for higher genus contributions to the
scattering amplitude. 

In \arutii\ the results of \aruti\ were further extended to all four 
particle scattering amplitudes of IIA string theory. They are all 
reproduced by the orbifold model. This gives clear evidence for the 
conjecture that the orbifold model is a reformulation of 
light-cone string theory. 

It is however not entirely clear
-with the knowledge reviewed in this chapter- 
how non-perturbative corrections may be added. 
Therefore we will now turn our attention to a more direct way 
to add the non-perturbative degrees of freedom. This approach 
that will be explained starting in the coming section, and 
in more detail in the next chapters, has the orbifold model
as a special limit. Using this relation it is possible to propose
a way to include the non-perturbative degrees of freedom to the 
orbifold model, but we will postpone this till the end of chapter three.

\global\advance\fignumber by -1

\newsubsection\noname{Towards matrix string theory} 
 
To get a quick understanding of matrix string theory 
(for a review see \mstreview) we will 
give here a somewhat heuristic introduction of the model. 
In the course of the next chapters we will become more precise. 
The ideas presented here are nevertheless important ingredients
of matrix string theory, and we will need them throughout this 
thesis.

In principle it is possible to say in one sentence what 
matrix string theory is all about: 
matrix string theory is a supersymmetric gauge theory that contains 
DLCQ string theory and
extra degrees of freedom that represent the non-perturbative objects
in string theory. The presence of these additional degrees of freedom
is the very motivation to study matrix strings: their presence gives us a 
way to calculate non-perturbative corrections to processes in 
perturbative string theory.

In the preceding sections we only did some reformulations of free
strings and their interactions. The orbifold model we used for this 
alternative description of perturbative strings can be shown to flow out
of a gauge theory in the infra-red. This gauge theory is 
${\cal N}=8$ two dimensional supersymmetric $U(N)$ Yang-Mills 
theory defined on a cylinder, that can be viewed as the dimensional 
reduction of ${\cal N}=1$ SYM from ten down to two dimensions.
The idea is that the matrices \matrixcoo\!\!, which were formed out 
of the string coordinates in section \towardsmst\!\!, 
are identified as diagonal field
configurations of Higgs fields in the gauge model, when the strings 
are widely separated. Then the differences
between the eigenvalues of the Higgs fields $X^I$ are large, so all
charged fields in the SYM theory become very massive (as compared to 
the Higgs scalars) and they effectively decouple from the dynamics.
Widely separated
strings break the gauge symmetry $U(N)$ to $(U(1))^N$, but when they
approach each other and/or interact, part of the broken gauge symmetry
is restored. A description of interactions therefore needs the complete
non-abelian theory. 

The action of the gauge model is given by 
\eqnar\mstaction{
S= \int \! d\tau\! \int \! \! {d\sigma} \, 
{\rm Tr} \Bigl\{- 
{g_s^2\over 4} F_{\alpha\beta}^2 - {1 \over 2} (D_\alpha X^I)^2 + 
{1\over 4g_s^2}[X^I,X^J]^2
\qquad\nonumber \\ \qquad %\quad 
+ i{\bar \psi}\not\kern-0.3em D\psi - {1\over g_s}
\bar\psi\Gamma^I\left[X_I, \psi\right]\Bigr\}\ . 
}
We identified the Yang-Mills coupling constant with the inverse of
the string coupling constant
\eqn\gaugestringc{
g_{YM}^2 = 1/g_s^2, 
}
so that the relation between the gauge model \mstaction and 
DLCQ string theory is an example of a strong-weak coupling 
duality. String interactions are perturbative processes in string theory, 
but non-perturbative in matrix string theory. 

In equation \gaugestringc we wrote the 
Yang-Mills coupling constant 
(which has dimensions of inverse length)  
in units where the radius of the Yang-Mills cylinder is equal to one. 
The zero string coupling limit is therefore equivalent to the 
strong coupling limit and/or infra-red limit. 
We postpone a more concrete derivation why the gauge model \mstaction
is related to DLCQ string theory, including the coupling constant 
identifications \gaugestringc till the next chapter.

The basic dynamical variables of the theory are $N \times N$ hermitian 
matrices and include 8 scalar fields $X^I$ and 8 fermion fields 
$\psi^a_L$ and $\psi^{\dot{a}}_R$.
In the free string limit $g_s \rightarrow 0$ 
(or strong Yang-Mills coupling limit) the gauge fields $A_\alpha$ 
decouple from the theory and the Higgs fields satisfy the 
commutation relations 
\eqn\commzero{
[X^I,X^J] = 0. 
}
The equations \commzero imply that the Higgs fields 
can be diagonalized simultaneously. 
They still can satisfy non-trivial boundary 
conditions, that are classified by the Weyl group of the gauge 
group. For $U(N)$ this Weyl group is the symmetric group $S_N$, 
so we are back to the orbifold model \nonameorbi we discussed 
before. 
We are thus lead to the conjecture that in the zero string 
coupling limit the gauge theory \mstaction 
flows to a supersymmetric conformal field theory 
that is defined on the orbifold \nonameorbi\!\!.

When we relax the limit $g_s \rightarrow 0$ and take the string 
coupling constant to be finite, there can be non-diagonal terms 
in the matrix string configurations. These terms are 
responsible for the extra degrees of freedom present in matrix 
string theory as compared to light-cone string theory. 
%They will 
%give us a way to describe the non-perturbative 
%D-branes, and their corrections to perturbative string processes. 

For finite $g_s$ interactions will be included automatically, but 
we still have to find the appropriate solutions of the equations 
of motion that yield a (classical) description of joining and splitting
of strings. Indeed in chapter 3 we will find an instanton-like 
solution that describes these elementary processes in matrix string 
theory. This instanton solution should be glued into a global solution 
that equals the asymptotic states far away from the interaction
region. For a concrete comparison with string theory, the
quantum fluctuations around the classical solution have to be taken 
into account. We will discuss this and other issues in chapter three.

\newappendix{Appendix A: Partition function of a 
symmetric product orbifold model}

\markright{\small\sc Appendix A}
\testclino
\write\cfile{\indent Appendix A && \thepage\noexpand\\[1mm]}

%Partition function of a symmetric 
%product orbifold model} 

\setcounter{equation}{0}

\xdef\thechapter{A} %%% I admit: a very very ugly trick. Terrible! 

In \dmvv\ an identity was proven that equates the elliptic genus 
partition function of a supersymmetric sigma model defined on an 
$N$-fold symmetric product $X^N/S_N$ to the partition function 
of a second quantized string theory on the space $X\times S^1$. 
Here $X$ is a K\"ahler manifold. 
The elliptic genus of a supersymmetric sigma model is 
defined as the trace over the R-R sector 
of the evolution operator $q^{L_0}$ times $(-1)^Fy^{F_L}$. 
Here $F=F_L+F_R$ is the sum of left and right moving fermion number, 
and $y$ is a complex parameter that counts
the left moving fermion number. By virtue of supersymmetry 
the right moving sector contributes only via the ground states. 
In the special case that 
$y=1$ the elliptic genus reduces to the Euler number.  
The orbifold Euler number can be read of from 
the generating functional
\hirzebruch\vafawitten\ 
\eqn\eulersym{
\sum_{N \geq 0} p^N \chi(S^NX) = 
\prod_{n>0}{1 \over (1-p^n)^{\chi(X)}}.
}
The full partition function of the supersymmetric sigma model 
%defined on a symmetric product space 
can also be related to a second
quantized string theory.  
More concretely we claim that the following identity holds, 
\eqn\partitionsym{
\sum_{N\geq 0}p^N\chi(S^N X;q,\bar{q})=
\prod_{n>0 \atop l,m\geq 0}
{1 \over (1-p^n (q\bar{q})^{l/n}q^m)}{}_{d(mn+l,l)}.
}
Here $d(m,l)$ are the degeneracies of the single partition 
function, that contribute at level $k$ and $l$ to the operators 
$L_0$ and $\bar{L}_0$, 
\eqn\partexp{
\chi(\cH;q,\bar{q})=\sum_{k,l} d(k,l) q^{k} \bar{q}^{l}.
}
The assumption we make is that the single Hilbert space $\cH$ is 
discrete, i.e. the indices $l,m$ in \partexp 
run over a discrete set, but are not necessarily 
integer-valued (this in contrast to the case of the elliptic genus
where all levels are integer-valued). 
Note that in \partexp there are both zero mode contributions and 
oscillator modes contributions to the levels of $q$ and $\bar{q}$, 
but we did not make this explicit in order to keep the formulas as 
simple as possible. 

Before we prove \partitionsym let us comment on the physical 
interpretation of the right-hand side of
\partitionsym\!\!. The right-hand side can be interpreted as the 
second quantized partition function of a string, whose Fock space
is made up by applying creation operators $\psi^i_{l,m,n}$, 
$i=1 \dots d(m,l)$. The parameter $p$ counts the light-cone momentum
$p^+$, $|q|$ and $q$ count two quantum numbers (not necessarily 
integer-valued). 

In the remaining part of this section we will proof
\partitionsym\!\!, thereby heavily relying on the original proof 
for the case of the elliptic genus partition function \dmvv. 

In the preceding subsection we have seen that the Hilbert
space of a closed string theory on a symmetric product space 
can be written as a direct sum of direct products of smaller 
Hilbert spaces, see equation \decompose\!\!. 
Repeatedly using the following rules for partition functions of
direct sums respectively tensor products of Hilbert spaces  
%
%\begin{eqnarray} 
\eqnar\parrules{
\chi(\cH \oplus \cH'; q,\bar{q}) &
= \chi(\cH ;q,\bar{q}) +\chi(\cH' ;q,\bar{q}),
\cr 
\chi(\cH \otimes \cH'; q,\bar{q}) & 
= \chi(\cH ;q,\bar{q}) \cdot\chi(\cH' ;q,\bar{q}),
}
%\end{eqnarray}
%
%\def\parrules{(\theequation)\ }
%\\[-\baselineskip]
we get for partition function of the orbifold model 
\eqn\partitie{
\chi(\cH(S^N X);q,\bar{q})= \sum_{N_n \atop \sum nN_n=N} \prod_{n>0} 
\chi(S^{N_n}\cH^{\Zvet_n}_{(n)};q,\bar{q}), 
}
where we used the notations of the previous subsection. 
To proceed we first proof that the partition function of a 
symmetrized tensor product of identical Hilbert spaces can be 
written in terms of the partition function of one single Hilbert 
space only. This will enable us to reduce further the righthand side 
of \partitie\!\!. 

As already indicated we assume that the single Hilbert space $\cH$ 
has a discrete spectrum. 
The partition function of the symmetrized tensor product of $\cH$ is
given by the generating function 
\eqn\someen{
\sum_{N \geq 0}p^N\chi(S^N\cH;q,\bar{q})=
\prod_{k,l}{1 \over (1-pq^{k}\bar{q}^{l})}{}_{d(k,l)}.
}   
This identity can be proven as follows \dmvv. 
We define the $V_{k,l}$ to be the vector space whose dimension 
equals the degeneracy number $d(k,l)$ of the single Hilbert 
space $\cH$. We have 
\eqn\som{
\sum_{N=0}^{\infty} p^N \chi(S^N \cH;q,\bar{q}) = 
\sum_{N=0}^{\infty} p^N 
\hskip-4mm
\sum_{N_{k,l} \atop \sum N_{k,l} =N}
\hskip-3mm
\prod_{k,l} (q^{k} \bar{q}^{l} )^{N_{k,l}} 
\dim(S^{N_{k,l}} V_{k,l}), 
}
where $N_{k,l}$ runs over all partitions of $N$. 
The summations in the expression of the right hand side in 
\som can easily be rewritten as 
\eqn\rewritesum{
\sum_{N=0}^{\infty} p^N \chi(S^N \cH;q,\bar{q}) = 
\prod_{k,l} \sum_{N} p^N (q^{k}\bar{q}^{l})^N 
\dim S^N (V_{k,l}).
}
Finally by using the identity
\eqn\dimsen{
\dim(S^N V_{k,l} ) = \left({d(k,l) + N-1 \atop N}\right), 
}
we arrive at the result \someen\!\!. 

Now we return back to the goal of this section, namely calculating the
partition function of the symmetric product orbifold model. The partition
function of the single Hilbert space $\cH^{\Zvet_n}_n$ is given by 
\eqn\partsingle{
\chi(q,\bar{q},\cH^{\Zvet_n}_n) = \sum_{k,l \geq 0 \atop k-l =nm} 
d(k,l) q^{k/n}\bar{q}^{l/n} =\sum_{l,m} d(l+nm,l)(q\bar{q})^{l/n} q^m .  
}
Combining this result with \rewritesum we arrive at the result 
\partitionsym
\eqnar\part{
\sum_{N\geq 0}p^N\chi(S^N X;q,\bar{q})=\sum_{N\geq 0} 
p^N \sum_{N_n \atop \sum nN_n=N}
\prod_{n>0}\chi(S^{N_n}\cH^{\Zvet_n}_{(n)};q,\bar{q}) \qquad \qquad
\cr
=\prod_{n>0}\sum_{N\geq 0}p^{kN}\chi(S^N\cH^{\Zvet_n}_{(n)};q,\bar{q})
=\prod_{n>0 \atop l,m\geq 0}
{1 \over (1-p^n (q\bar{q})^{l/n}q^m)}{}_{d(l+nm,l)}, 
}
where again 
it is understood that the labels $m,l$ run over a discrete set,
but are not necessarily integer valued. We will use the result 
\partitionsym in chapter 3 when we calculate the degeneracies
of BPS-states in supersymmetric Yang-Mills theory compactified 
on a three-torus. 

\xdef\thechapter{2}
%%%%%
\newchapter\dualitysym{
\boldmath $U$\!-duality in \unboldmath N=4 \break
 Yang-Mills theory on \boldmath $T^3$ \unboldmath} 
{\boldmath $U$\!-duality in \unboldmath N=4 
 Yang-Mills theory on \boldmath $T^3$ \unboldmath}
\lhead[\fancyplain{}{\sc $U$-duality symmetry in N=4 
Yang-Mills theory on $T^3$ }]{\fancyplain{}{}}
\thispagestyle{empty} 
%%%%%%%%%%%%%%%%%%%%%%%%%%%%%%%%%%%%%%%%%%%%%%%%%%%%%%%%%%%%%%%%%%%%%%

Toroidal compactified type II string theory is conjectured to be 
invariant under a discrete symmetry group, called $U$-duality 
\hulltown. %This duality mixes fundamental string states and 
%D-brane states, and it is therefore a non-perturbative 
%duality symmetry. 
%\\ \indent
Relations between gauge models and string theory suggest 
that this string duality should be reflected in gauge theories 
as well. 
In this chapter we will review why this is indeed the case. 
We will moreover show that gauge theories know about extended 
$U$-duality symmetries. That is, certain 
properties of the theories, like BPS mass spectra, or BPS state 
degeneracies, can be shown to be invariant under a larger symmetry group, 
than one might at first sight expect. 

In the gauge theory
interpretation this extended symmetry group is a combination of 
electro-magnetic duality, the mapping class group of tori, 
and Nahm-type dualities.

We will mainly concentrate on a supersymmetric Yang-Mills model 
defined on a three-torus.
We will study in detail the degeneracies of BPS states in this model, 
and show that they exhibit a $U$-duality symmetry. In the last section
we explain in detail the appearance of the duality symmetry, and its
absence in the BPS mass spectrum. 

%, as predicted by M-theory. 
%
First, however, we will introduce some basic concepts. We start with a
brief explanation of string duality. Then we give a  
review of D-branes, and their low energy description.
After this we introduce M-theory, 
a conjectured eleven-dimensional model that unifies all known string
theories. The notions of D-branes and M-theory
naturally lead us to the matrix theory proposal of \bfss, 
which says that M-theory in a special regime has a particularly simple 
description in terms of a matrix quantum mechanics. Matrix string theory
is closely related to this model, and following \seibergsen\ 
we will give arguments why it yields a description of non-perturbative
IIA strings. 

%After we have established these basic concepts, 
%we turn to the actual calculation of the 
%multiplicities of BPS states in the above mentioned model, and give 
%an explanation why they exhibit a larger $U$-duality symmetry. 

\newsection\dbranedual{D-branes and string dualities}

% For background material we refer to 
%\poltasi\bachasimp\Tayll\polstring. 

\newsubsection\nonamedual{String duality}

First we will briefly discuss string duality in closed string theory.
The bosonic massless modes of closed string theory come from 
two sectors, the RR sector and the NS-NS sector: 
\\
\eqnar\noname{
&{\rm NS-NS} \qquad & |\mu>_L \otimes \; |\nu>_R 
\ \longrightarrow\ G_{\mu\nu}, B_{\mu\nu}, \phi 
\nonumber \\ \\[-\the\baselinetruc] \cr \nonumber
&{\rm R-R} \qquad & |\alpha >_L \otimes \; |\beta>_R 
\ \longrightarrow\ {\rm antisymmetric \ forms} \ A^i 
}
Here $G$ is the ten-dimensional space-time metric, 
$B$ an antisymmetric two form, $\phi$ the dilaton field 
and the $A^i$ are antisymmetric RR forms. 
The RR gauge fields that satisfy definite chirality conditions, 
form together with the NS-NS fields the ground states of either 
type IIA or IIB string theory.  
This is illustrated in the next table 
\global\advance\tablenumber by 1  
\\
\eqn\noname{ \nonumber 
\begin{array}{|l|l|l|}
\hline 
& {\rm NS-NS}& {\rm R-R} 
\\[2mm] \hline 
{\rm IIA} & G_{\mu\nu}, \phi, B_{\mu\nu} & A^1, A^3
\\[2mm] \hline 
{\rm IIB} & G_{\mu\nu}, \phi, B_{\mu\nu} & A^0, A^2, A^4 
\\[2mm] \hline
\end{array}
}
\\[0mm]
\centerline{\vbox{\baselineskip12pt
\advance\hsize by -3truein\noindent{\small 
{\bf Table~\thechapter.\the\tablenumber\ } The bosonic massless 
fields of type IIA and IIB string theory}
}}
\\[4mm]
The low energy dynamics of these massless modes have a field theory
description in terms of supergravity theories. 

A remarkable property of these field theories are the so-called 
duality symmetries. These classical symmetries are expected to be
quantum mechanical symmetries in string theory. They come in two
types: $T$-duality and $S$-duality. A single $T$-duality transformation
(for a review of T(arget space)-duality see \gpr) changes the
chirality of the theory from type IIA to type IIB \dailp\dinehs\ 
and in the  
opposite direction. In its simplest non-trivial form it  
inverts the radius $R$ of a compact direction
to $\alpha'/R$, and exchanges the momentum and winding modes of 
strings winded along the compact direction, thereby leaving the total 
mass spectrum invariant. The RR forms get an extra index or lose
one, depending on whether the index was already there or not. 
It thus changes the rank of the RR tensor fields by $\pm 1$, 
so that we get a map from type IIA string theory to IIB and 
vice versa. Together with the mapping that acts by integral shifts 
in the fields, the 
$R \rightarrow \alpha'/R$ duality forms an $SL(2,\Zvet)$ symmetry group. 
Generalized to $d$ compactified dimensions the $T$-duality group
becomes $SO(d,d)$. For a detailed account of the action of $T$ duality 
on the fields we refer to \bergshoeff\ and \gpr.  
\\
\eqn\noname{ \nonumber
\begin{array}{|r|l|c|c|}
\hline
D & d & {\rm Sugra} & {\rm string~theory}
\cr
\hline
10 & 1 & & 
\cr
9 & 2 & SL(2,\Rvet) & SL(2,\Zvet)
\cr
8 & 3 & SL(2,\Rvet) \times SL(3,\Rvet) &SL(2,\Zvet) \times SL(3,\Zvet)
\cr
7 & 4 & SL(5,\Rvet) & SL(5,\Zvet) 
\cr
6 & 5 & SO(5,5,\Rvet) &  SO(5,5,\Zvet)
\cr
5 & 6 & E_6(\Rvet) & E_6(\Zvet)
\cr
\hline
\end{array}
}
\\[3mm]
\global\advance\tablenumber by 1 
\centerline{\vbox{\baselineskip12pt
\advance\hsize by -2truein\noindent{\small 
{\bf Table~\thechapter.\the\tablenumber\ } The $U$-duality groups
of type II string theory in different dimensions}
}}
\xdef\stugroups{\thechapter.\the\tablenumber\ } 
\\
Ten dimensional type IIB string theory is conjectured to have 
$S$-duality as a genuine symmetry \hulltown. 
It is a non-perturbative $SL(2,\Zvet)$
symmetry; it relates the weak coupling regime to the strong coupling 
regime, and it is therefore reminiscent of electro-magnetic duality in 
gauge theories. The dilaton forms together with the RR-field $A^0$ a
complex scalar that transforms under fractional $SL(2,\Zvet)$ 
transformations. The two-forms $B$ and $A^2$ transform as a doublet.

The generators of both types of duality groups do not commute, 
together they form a larger group that goes under the name of 
$U$-duality \hulltown. The $U$-duality groups become larger in lower
dimensions and are discrete: As indicated in table \stugroups\!\!, 
in eight dimensions the symmetry group is
$SL(2,\Zvet)\times SL(3,\Zvet)$, in seven $SL(5,\Zvet)$ and 
in six dimensions $SO(5,5,\Zvet)$.

\newsubsection\nonamedbrane{D-branes} 

An immediate corollary of the conjecture that IIB string theory 
has $S$-duality invariance is the existence of a dual string that 
transforms together with the elementary string in a doublet. 
This dual string has been mysterious for a long time, 
until Polchinski realized that the D-string could play 
this role \Polc. A D-string 
is a particular example of $p+1$ dimensional D-branes, which were 
previously known as hyper-surfaces with the property that open 
strings can end on them \dailp. Polchinski also clarified the role of the 
RR-fields by noting that they should couple to the D-branes \Polc. 

One explains the possibility that strings can end on a hyper-surface 
by considering the boundary conditions that
the coordinate fields of open strings have to satisfy at their 
endpoints. There are two types of possible boundary conditions 
for the open string,  
\eqnar\noname{
\partial_\perp X^\mu =0 & {\rm (Neumann)}, 
\nonumber \\ \\[-\the\baselinetruc] \cr \nonumber
\delta X^\mu = 0 & {\rm (Dirichlet)}.  
}
Dirichlet boundary conditions break Poincar\'e invariance, and hence
represent topological defects in space-time. 
These topological defects are called D-branes, which is shorthand for 
Dirichlet branes. A static $D$-brane with $p$ spatial dimensions is 
described by the boundary conditions 
\eqn\noname{
\partial_\perp X^{0,1,\dots, p}=0,  \qquad X^{p+1, \dots, 9} =0.
}
The massless modes of the open string tied to the world-volume of
the brane have a description in terms of a low energy field theory 
on the D-brane. Thus we get a field $A^\mu(\xi^\alpha)$, where 
$\xi^\alpha$ is a parameterization of the world-volume, that 
decomposes into a parallel gauge field $A^\alpha(\xi^\alpha)$ 
on the brane and transversal components 
$X^I(\xi^\alpha)$ that form scalar fields. 

The effective action of a (fluctuating) D-brane was originally derived 
by requiring that its equations of motion reproduce the 
conditions that are implied by conformal invariance of open strings 
in the D-brane background \dailp\leigh. Another, less technical,  
way to derive the action is by noting that the D-brane 
low energy effective action is essentially fixed 
by Lorentz invariance and T-duality \bachasimp.
To see this we start with the example of a D-particle moving 
through flat spacetime \bachasimp. 
We take its world-line through ten-dimensional space-time 
parameterized in the following way 
\eqn\worldlinep{
X^0(\tau) = \tau \qquad X^I = X^I(\tau).
}
The point particle Lagrangian is  
\eqn\vsimpleaction{
S = \tau_0 \int d\tau \! \ \sqrt{1-(\partial_0{X}_I)^2}.
% +\mu_0 \int d\tau \! \ (C_0+\dot{f}_IC_1^I) 
} 
The action \vsimpleaction is an effective action in the sense 
that higher order derivatives with respect to time, e.g. 
acceleration terms, are neglected. We have put the RR fields to zero, 
which would otherwise couple to the D-particle through Wess-Zumino terms.   

By a $T$-duality transformation in the 1 direction the boundary
condition of the open string coordinate field $X^1$ turns from 
Neumann to Dirichlet, the D-brane scalar field $X^1$ becomes a 
gauge field $A^1$ and hence the D-particle becomes a D-string. 
The velocity of the D-particle in the 1 direction plays the role 
of a field strength on the D-string
\eqn\noname{
F_{01} = \partial_0 A_1 = \dot{X}_1/2\pi\alpha'. 
}
%
%The RR background 1-form gets transforms to a RR scalar and components 
%of a RR two-form 
%%
%\eqn\noname{
%C_1 \rightarrow C  \qquad C_\mu \rightarrow C_{\mu 1} \quad 
%( \mu \neq 1 )
%}
%%
So we get for the effective world-volume action of the D-string  
\eqn\surprise{
S = \tau_1 \int d^2\xi \! \ 
\sqrt{1- (\partial_0 X_I)^2 - (2\pi\alpha'F_{01})^2 },
%+ \mu_1 \int d^2\xi \! \ (C_{01} + 2\pi\alpha' F_{01} C +
%\partial_0 X^{I} c_{I1}  )
}
where $I$ now runs from $2 \dots 9$. 
The same argument can be repeated for other directions, so
that we get higher dimensional D-branes. 
This leads us to the form of the 
bosonic low energy effective action of a $p+1$ dimensional D-brane
\eqn\borninfeldaction{
S= T_p \int d^{p+1}\xi \! \ e^{-\phi} 
\sqrt{\det (G_{\alpha\beta} + B_{\alpha\beta}+ 
2\pi\alpha' F_{\alpha\beta})}. 
%+ \mu_p \int d^{p+1}\xi \! \ A^{p+1}
}
Here $T_p$ is the brane tension. 
We included the pull-back of the antisymmetric 
NS-NS $B_{\mu\nu}$-field. 
Together with the pull-back of the RR field strength it forms 
the $U(1)$ gauge invariant world-volume field strength 
$2\pi \alpha' {\cal F}_{\alpha\beta} = B_{\alpha\beta} + 
2\pi \alpha' F_{\alpha\beta}$ that lives on the D-brane world-volume.  
The metric $\hat{G}$ is the pullback of the space-time metric to 
the D-brane volume
\eqn\noname{
\hat{G}_{\alpha\beta} = 
G^{\mu\nu} \partial_\alpha X_\mu \partial_\beta X_\nu. 
}
In addition to the action \borninfeldaction there are couplings of 
the D-brane to RR fields in the form of Wess Zumino terms. 

The Born-Infeld action \borninfeldaction as an effective description
of D-branes can also be verified by a perturbative string calculation 
\poltasi. This open string calculation moreover gives the actual 
value of the brane tension $T_p$; the result is \poltasi 
\eqn\noname{
\tau_p = {T_p \over g_s} = {1 \over g_s \sqrt{\alpha'}} 
{1\over (2\pi\sqrt{\alpha'})^p}, 
}
where the string coupling $g_s$ is related to expectation value of
the dilaton $g_s = e^{<\phi>}$. As expected the brane tension is 
proportional to the inverse of the string coupling constant, so
that the D-branes are indeed non-perturbative objects. 

%\newsubsection\noname{Yang-Mills description of D-branes}

With some further restricting assumptions the form of the BI action 
\borninfeldaction simplifies considerably: in the 
zero-slope limit $\alpha' \rightarrow 0$ combined with additional 
restrictions, the BI action reduces to abelian Yang-Mills theory. 

To show this we take for simplicity the background metric and the 
D-brane to be flat. The pullback of the metric to the brane is then 
\eqn\noname{
G_{\alpha\beta} = \eta_{\alpha\beta} + 
\partial_\alpha X^I \partial_\beta X^I + \dots .
}
Furthermore, we set the antisymmetric tensor $B$ to zero, and assume that
the terms $2\pi\alpha'F_{\alpha\beta}$ and $\partial_\alpha X^I$ are 
small. Then we can expand the Born-Infeld action as follows 
\eqn\nonameaction{
S = T_pV_p + {1 \over 4g_{\rm YM}^2} \int d^{p+1}\xi \! \ 
\left(F_{\alpha\beta}^2 +
{2 \over (2\pi\alpha')^2} \partial_\alpha X^I \partial_\beta X^I 
\right)+ \dots, 
}
where we made the identification 
\eqn\yangmstring{
g^2_{\rm YM} = {1 \over 4\pi^2(\alpha')^2\tau_p}.
}
The action \nonameaction is the bosonic part of 10-dimensional 
${\cal N}=1$ SYM, dimensionally reduced to $p+1$ dimensions. 
Though this theory is a {\it truncated} model of D-branes, it 
has some peculiar properties in favor when compared with 
Born-Infeld theory. One can easily add fermionic degrees of freedom,  
and especially one can replace in a straightforward way the abelian
gauge group by non-abelian $U(N)$ groups. 
This extension to non-abelian groups is useful, because when 
we have more than one D-brane, the dynamics get naturally 
a description in terms of $U(N)$ gauge theory. 

Thus D-branes shed new light on the intimate relation between string
theory and (non-abelian) gauge theories. 
We will explain this in more detail in the next section.

\newsubsection\nonamebound{Bound states of $N$ D-branes}

In the simplest
case of two parallel D-branes there are two hyper-surfaces on which
open strings can end. These possibilities can be included in string 
theory by adorning the open strings with extra degrees of freedom at 
their endpoints, called Chan Paton factors, that simply indicate 
which brane the endpoint is restricted to. With the inclusion of these 
Chan Paton factors, string wave functions have decompositions like 
\eqn\noname{
\left|k;a\right> = \left|k;ij\right> \lambda^a_{ij},
}
where the $\lambda^a_{ij}$ are, in this case, $2\times 2$ matrices
that label the Chan Paton factors. The matrices $\lambda^a_{ij}$ 
are allowed to be simultaneously conjugated by $U(2)$ 
transformations, as in string amplitudes only traces of  
products of Chan Paton factors appear.\footnote{We consider 
oriented open strings. For non-oriented strings the gauge group
changes to $SO(N)$ or $USp(N)$.} 

On each of the two D-branes we have a $U(1)$ gauge field that 
couples to open strings. 
These fields can be thought of as two abelian directions in a
larger $U(2)$ group. This $U(2)$ gauge group is equal to 
$U(1)\times SU(2)/\Zvet_2$. The $U(1)$ subgroup describes the center 
of mass motion of the branes, while the non-abelian $SU(2)$ part
determines the relative motion. The Weyl group $\Zvet_2$ of $SU(2)$ 
acts by permuting the branes, corresponding to the fact that they are 
indistinguishable.
\\
%
%%
%%%
\epsfysize=5.7cm
\hfigure{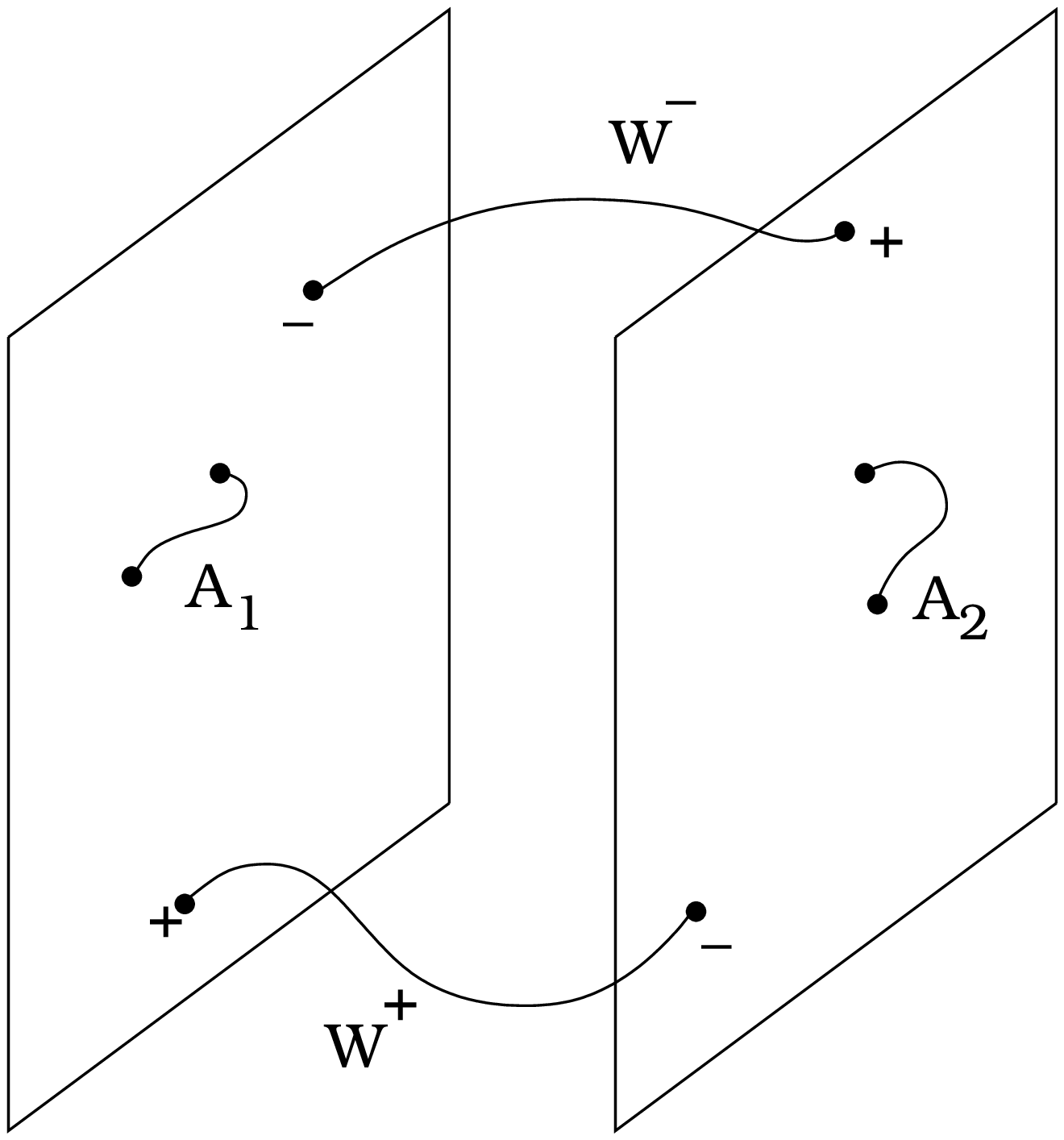}
{The low energy dynamics of two separated branes has a  description 
in terms of a $U(1)\times U(1)$ broken sector of $U(2)$ gauge
theory. The open strings that begin on one brane and end on another, 
are coupled to massive charged $W$-bosons. The open strings beginning
and starting on the same brane couple to the abelian $U(1)$ gauge fields
on the brane. When the branes are on top of each other the complete gauge 
symmetry gets restored \wittenstd.}
\xdef\gaugebreakfig{\thechapter.\the\fignumber\ } 
%%%
\\
When the branes are separated the $SU(2)$ gauge group is broken to 
$U(1)$
\eqn\noname{
U(1) \times SU(2) \longrightarrow U(1) \times U(1), 
}
much like in spontaneously broken gauge theories. For example the 
role of the charged $W^\pm$ bosons is played 
by the ground states of strings beginning on one brane and ending on 
another brane, and the masses of the $W^\pm$ particles are proportional
to the distance between the D-branes, and thus vanish when they
are on top of each other (in this case the complete gauge group $U(2)$
gets restored). 
Thus spontaneous symmetry breaking in gauge theory can be visualized 
in string theory by pulling D-branes apart. 
This is an example of a technique which is called geometric engineering, 
which can be used to give phenomena in gauge theories a geometric 
interpretation.

In case of $N$ parallel D-branes the total gauge symmetry group is
$U(N)$, which is broken to a subgroup, 
depending on the relative locations of the separate D-branes. 
The effective description of $N$ D-branes has an 
obvious generalization in terms of non-abelian Yang-Mills theory. 
Also the fermionic degrees of freedom can be included in a 
straightforward way. In this way we arrive at ten-dimensional 
${\cal N}=1$ supersymmetric Yang-Mills theory, 
dimensionally reduced to the dimension of the D-branes, 
which has precisely the right field content.  

As noted above this Yang-Mills description of D-branes is only 
valid in the zero-slope limit $\alpha' \rightarrow 0$. When we 
add up all $\alpha'$ corrections we would expect to arrive at a 
generalization of Born-Infeld theory, like in the abelian case. 
The generalization of Born-Infeld theory to non-abelian
gauge groups with or without inclusion of fermionic degrees of freedom 
is however not yet fully established.
\footnote{The appropriate supersymmetric version of the Born-Infeld action, 
that describes both the fermionic degrees of freedom and the 
bosonic degrees of freedom has been constructed for the abelian 
case \aganagic.}  
There have been proposals for a non-abelian generalization 
\tseytlin. By considering the equations of motion implied by the
action, the right one can be singled out almost uniquely. There
appears, however, to be an ambiguity in the choice of representation 
that is to be used for the trace over the gauge group. 
For a discussion of these matters see \tseytlin.

\newsubsection\boundstates{Bound states within D-branes} 

In supersymmetric gauge theories electric charged particles can
combine with magnetic charged solitons to dyonic bound states. 
These states usually break part of the supersymmetry and  
saturate a BPS mass bound. Their masses are smaller than the added 
masses of single electric and magnetic objects; in other words 
the binding energy is non-zero and the dyons are therefore truly bound.

In a quite analogous way D-branes can form 
bound states with p-branes (extended fundamental objects in string 
theory) in type IIB string theory. 
As an example we take bound states of fundamental strings (F-strings) 
and D-strings. These two types of strings transform as a doublet
under the $SL(2,\Zvet)$ action of $S$-duality, so string duality 
predicts the existence of a whole tower of bound states of $p$ F-strings
and $q$ D-strings, with $p$ and $q$ relatively prime. 
It also suggests the following binding tension 
formula for a $(p,q)$ string \schwarz 
\eqn\noname{
\tau_{p,q} = {1 \over 2\pi\alpha'} \sqrt{p^2 + {q^2 \over g_s^2}}. 
}
This formula implies that for weak coupling the mass
difference between a single F-string plus a single D-string, and
a $(1,1)$ bound state is equal to the mass of a single fundamental string
up to zeroth order of the coupling constant.
Apparently the F-string dissolves almost entirely 
in the D-string when they form a bound state by minimizing 
their energy. This can be visualized 
by imagining the following procedure \wittenbound. 
Consider a fundamental string (F-string) in type IIB wrapped around 
a compact dimension. In the absence of D-strings,  
the string winding number is conserved. This winding number is equal
to the conserved charge of the gauge field that can be formed out of
the NS-NS two form $B$ by integrating it over the compact direction
\polstring\gsw. 
In the presence of a D-string however
the winding number of an F-string is no longer conserved. 
Instead the charge of the gauge field obtained from the $U(1)$ field
strength ${\cal F}_{\mu\nu} = F_{\mu\nu} + B_{\mu\nu}/2\pi\alpha'$ 
on the D-string is conserved.
Now the closed F-string can break in two parts and ``disappear'',  
thereby leaving electric flux on the world-volume of the D-string
behind, in such a way that the charge of the abelian gauge field 
on the D-string remains conserved. 

A D-string with electric flux on it is thus interpreted
as a bound state of a D-string and an F-string. 
It is a BPS state, because the configuration
can be mapped to a D-particle with non-zero momentum 
via T-duality, cf. \vsimpleaction\!\!-\surprise\!\!.

A proof that the bound states really exist has been given
by Witten. In \wittenbound\ he argued that 
the existence of a D-string/F-string bound state can be proven 
by considering the relevant low energy effective theory, perturbed
with a mass term. This mass term will break part of the supersymmetry
but it does not affect the BPS mass. 
By tuning the mass parameter of the 
extra term in the supersymmetric Yang-Mills theory, 
the effective coupling constant can be made small. 
For this theory then, it can be shown that 
the relevant supersymmetric ground state exists (and is unique)
\wittenbound. The existence and uniqueness of the 
ground state is not affected by putting the mass term to zero again,  
so that the whole tower of $(p,q)$ strings states 
with $p$ and $q$ relatively prime do exist in string theory. 

Analogous results hold for gauge theories that are associated with higher
dimensional branes. By turning on electric and magnetic fluxes 
not only the D-brane itself is described but also bound states with 
lower dimensional branes and strings. 
\\
\epsfysize=5cm
\htable{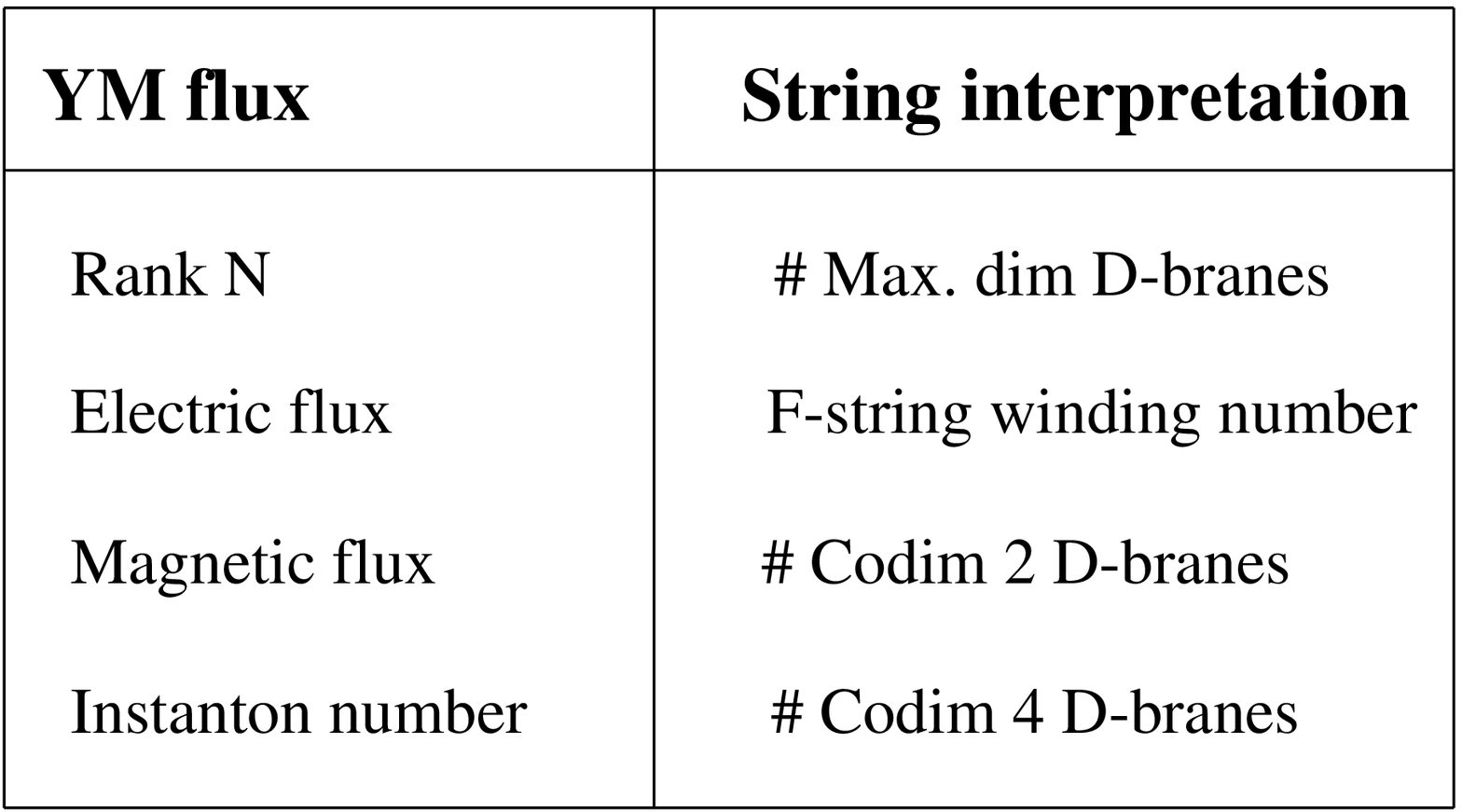}
{Bound states of a D-brane with lower dimensional branes and 
fundamental strings have a low energy description in gauge theory.
This table contains the translation code for the two theories.}
\xdef\boundedstates{\thechapter.\the\tablenumber\ } 
%%%
\\
This is illustrated in table \boundedstates\!\!. 
We will come back to this in section \thechapter.3 .

\newsubsection\nonamemth{M-theory and IIA string theory}  

M-theory is a conjectured model in eleven dimensions that should
unify all known string theories, and whose low energy limit is
eleven-dimensional supergravity. 

A concrete formulation of the model has not been established at 
present, but there are some proposals for particular sectors of the 
theory. A definition of M-theory could be the strong coupling limit 
of type IIA string theory. The motivation for this definition is, 
among other facts, the presence of D-particles in IIA theory whose
masses depend on the inverse of the string coupling constant $g_s$. 
When they are viewed as Kaluza Klein states of an eleven dimensional
theory, the radius of the extra eleventh dimension should be
proportional to $g_s$ 
\eqn\elevenra{
R_{11} = g_s l_s. 
}
Then in the strong coupling limit an extra dimension in
IIA string theory appears. Another motivation is the already long
known fact that the dimensional reduction
of 11D supergravity theory on a circle is IIA supergravity, the effective 
field theory of 10D type IIA string theory. 
The bosonic field content of 
11D Sugra \cremmer\ is the eleven dimensional metric $g_{MN}$ 
and an antisymmetric three-form gauge potential $C_{MNP}$. 
Upon dimensional reduction we therefore get a scalar, 
a gauge field $g_{\mu 11}$, a two form $C_{\mu\nu 11}$, 
a ten-dimensional metric and a three form, precisely the massless
bosonic fields of IIA supergravity theory. 
%
%\eqn\noname{
%\fbox{$
%\begin{array}{lcl}
%{\rm\large M\ theory} & \longleftrightarrow   &{\rm\large IIA\ string\ theory}
%\\[2mm] \hline\hline \\[-1mm]
%{\rm 11D \ graviton} &\leftrightarrow& {\rm D\ particle}
%\\[2mm] 
%{\rm wrapped\ membrane} &\leftrightarrow& {\rm fundamental\ string}
%\\[2mm]
%{\rm membrane} &\leftrightarrow& {\rm D2\ brane}
%\\[2mm] 
%{\rm wrapped\ five brane} &\leftrightarrow& {\rm D4\ brane}
%\\[2mm] 
%{\rm five brane} &\leftrightarrow& {\rm NS \ five brane}
%\\[2mm] 
%{\rm KK\ gauge\ field} &\leftrightarrow& {\rm RR\ gauge\ field}
%\\ 
%\end{array}
%$}
%}
%%

A graviton with momentum $N/R_{11}$ in the eleventh direction 
gets the interpretation as a bound state of $N$ $D$-particles, 
whose masses add up to $m = N/R_{11}$. The Kaluza Klein mode of the 
graviton 
$g_{\mu 11}$ couples to these D-particles and gets the role of the 
RR gauge field. The membrane, an extended solitonic object in 11D 
supergravity, wrapped along the eleventh direction gets 
associated to the IIA elementary string (as was realized for the
first time in \townsend.)
\\
\epsfysize=6.75cm
\htable{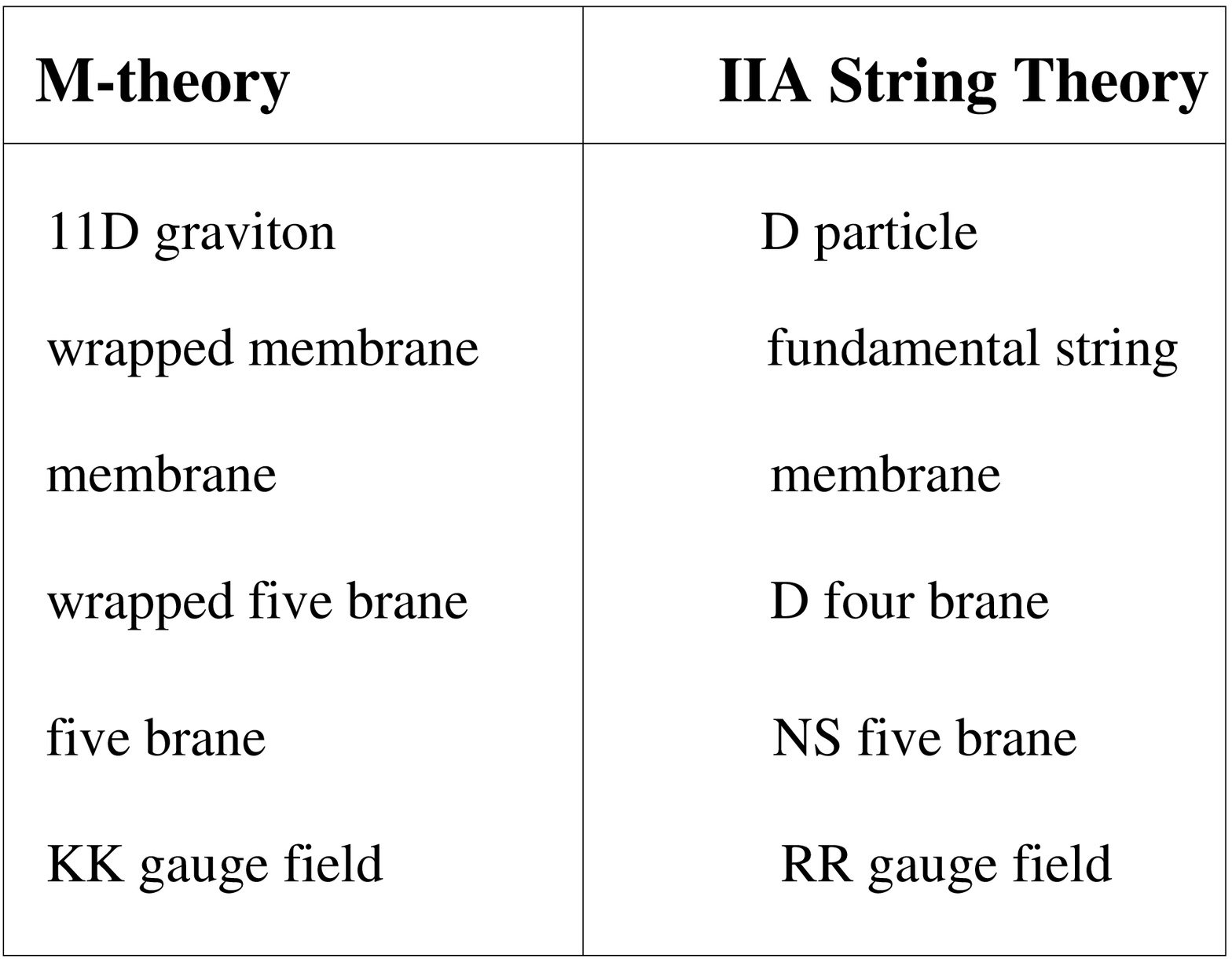}
{The dictionary between M theory (11D supergravity) compactified on 
a circle and IIA string theory.}
\xdef\mstringiden{\thechapter.\the\tablenumber\ } 
%%%
\\
The other solitonic object in supergravity, 
the five-brane, gives rise to the D four-brane and NS five brane. 
These identifications and others are summarized in the table  
\mstringiden\!\!. 

One can also read off the spectrum of 11D Sugra and M-theory 
from the eleven dimensional superalgebra \mmatrix 
\eqn\elevendsusy{
\left\{Q_\alpha , Q_\beta \right\} = 2p^\mu \gamma_{\mu\alpha\beta}
+ 2Z^{\mu_1\mu_2} \gamma_{\mu_1 \mu_2\alpha\beta} + 
2Z^{\mu_1 \cdots \mu_5}
\gamma_{\mu_1\cdots \mu_5 \alpha\beta}, 
}
where $\mu_i = 0,\cdots 10$ and $\alpha,\beta =1,\cdots 32$. The 
superalgebra \elevendsusy includes two central charges: a 
two brane charge $Z^{\mu_1\mu_2}$ and a five brane charge 
$Z^{\mu_1 \cdots \mu_5}$. These charges are infinite in non-compact
eleven dimensional space, but in compact space they can have finite
values.

The superalgebra \elevendsusy is well known to have $U$-duality invariance. 
The continuous versions of the $U$-duality groups are classical 
symmetries of the action of dimensional reduced eleven dimensional 
supergravity. The symmetries can be thought of as being generated 
by two groups, namely the $T$-duality group and the mapping class 
group of the torus on which 11D Sugra (M theory) should be
compactified to get supergravity theories (type II string theory) in
lower dimensions. The $U$-duality groups can thus be viewed as 
$SO(d-1,d-1,\Zvet) \bowtie SL(d,\Zvet)$. We list the groups in different
dimensions in table \stugroups\!\!. 

\newsection\noname{Matrix Theory} 
\newsubsection\nonamemst{The matrix theory proposal} 

In the previous section we saw that D-particles are the only objects
that carry $p_{11}$ momentum in M-theory. This was an important 
motivation for the following conjecture due to Banks, Fischler, 
Shenker and Susskind \bfss
\begin{list}{}{}
\item
{\bf Conjecture \bfss}:\ {\it M-theory in the infinite momentum frame
is exactly described by the $N \rightarrow \infty$ limit of 
0-brane quantum mechanics}
\eqn\matrixaction{
S = {1\over 2R_{11}} \int dt\ \! \tr \left(\dot{X}^2 + [X^I,X^J]^2 + 
\theta^T(i\dot{\theta} -\Gamma_I[X^I,\theta])\right), 
}
{\it where $N/R_{11}$ plays the role of the 11D momentum, and where 
$N/R_{11}$ and $R_{11}$ are both taken to $\infty$.}
\end{list} 
In writing down the D-particle action \matrixaction we used units
where $2\pi \alpha'=1$. 
The matrices $X$ are elements of the gauge group $U(N)$, and $\Rele$ is the 
radius on which M theory should be compactified to get type IIA string theory. 
When the fields $X$ are large, the finite energy configurations lie in
the flat directions. Along flat directions the fields $X^I$ are simultaneously 
diagonalizable and thus it is possible to interpret these diagonal matrix 
elements as coordinates of D-particles, with kinetic energy 
$M_{D0} \dot{X}^2 /2 = \dot{X}^2/2R_{11}$. For small distances the
fields no longer necessarily commute and the interpretation of the 
eigenvalues of the matrices as coordinates gets obscured. 
The off-diagonal components describe strings stretched between the 
branes, with characteristic energies $\sim |x_i-x_j| \Rele M_{Pl}^3$, 
where $M_{Pl}$ is the eleven dimensional Planck mass. 
The action neglects string oscillations and higher energy excitations 
(for example brane creation). 

Matrix theory has properties of a theory in the infinite
momentum frame. It is an effective theory of D-particles, so 
by construction it has only states with positive momentum in the 
eleventh direction. Furthermore the model \matrixaction has 
ten-dimensional (super)-Galilean invariance. 
Because of this it seems plausible that in 
the infinite momentum frame of M-theory the states are primarily 
composed of D-particles, whose effective low energy dynamics is 
determined by \matrixaction\!\!. 

Other important evidence for the model \matrixaction is the 
natural appearance of super\-mem\-branes in matrix quantum mechanics,  
as was originally observed in \dewiti. The matrix theory Hamiltonian 
is exactly the same as the light-cone Hamiltonian of the
super\-mem\-brane \dewiti. Due to supersymmetry this Hamiltonian has a 
continuous spectrum \dewitii, 
which from the mem\-brane point of view may be disappointing 
(as this would seem to rule out a generalization
of strings in terms of membranes). Matrix theory, however, 
gives a new interpretation of this result. The continuous 
supermembrane spectrum belongs to the collective dynamics of 
(many) D-particles. Then the continuity of the spectrum is 
exactly what we want. 

Additional evidence for the model \matrixaction came from calculations 
of D-particle scattering amplitudes, whose results are to be 
compared with supergravity \BeBe\BBPT\polpou. 
The first calculation, presented in \bfss\ was the remark
that in the Born approximation the scattering amplitude of two
gravitons in 11D supergravity corresponds with the leading potential
term between two D-particles. In \BeBe\ this agreement was checked, 
up to a two loop calculation in matrix theory. 

One can also investigate in how far the symmetries of M-theory are
present in matrix theory. These symmetries are Lorentz symmetry 
and U-duality symmetries. We will comment on this in the coming 
sections.

In calculations the rank $N$ is usually taken to be fixed 
and large. For finite $N$ the model \matrixaction has been 
conjectured to describe M-theory in the DLCQ formalism \dlcq. 
One might view this as a stronger conjecture than the 
original one in \bfss, as it tells us something about the matrix
quantum mechanics for any rank $N$. But the large $N$ limit needed in 
the conjecture of \bfss\ is a rather subtle issue and should be considered
with great care. 

We will go on with the proposal of \dlcq\ in the next section, and
adapt it to matrix string theory. In particular we show how the 
close relation between matrix string theory and DLCQ string theory 
can be made plausible.

\newsubsection\mstproposal{The matrix string theory proposal} 

We can now be more precise on the relation between string theory
and matrix string theory: the matrix string theory proposal at finite
$N$ is type IIA string theory in the DLCQ formalism, with rank $N$ and 
$p^+$ momentum identified. The D-particle number in string theory has 
the interpretation of (integer) electric flux in matrix string theory. 
\\
\epsfysize=6cm
\hfigure{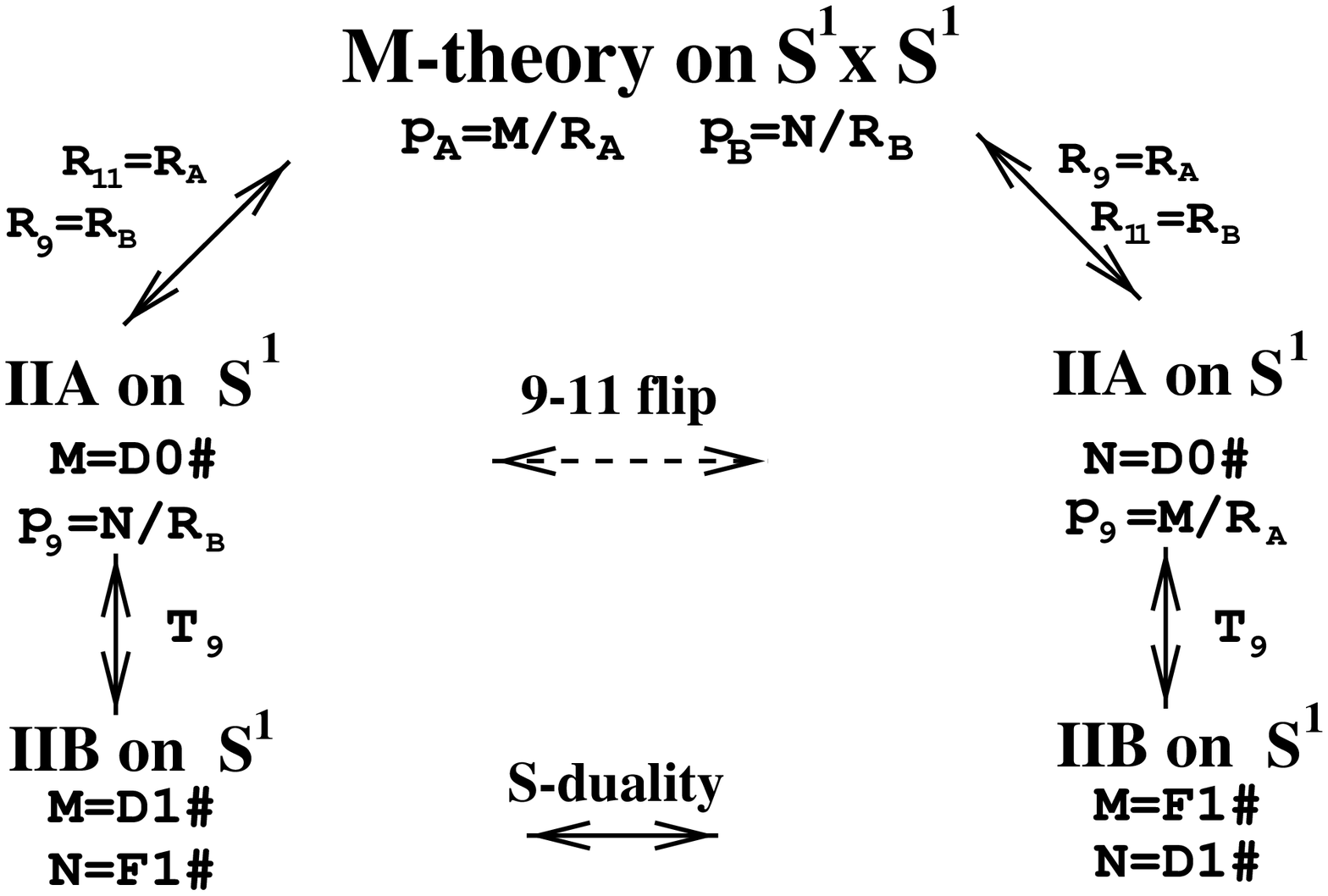}
{This diagram illustrates the 9-11 flip which is one of the 
essential features of matrix string theory. All identifications 
only depend on the M-theory conjecture \wittenstd\ and string duality. 
}
\xdef\idendual{\thechapter.\the\fignumber\ } 
%%%
%
%
\\
We can make this claim plausible by the following argument \seibergsen. 
In figure \idendual
we consider M-theory compactified on two circles 
with radii $R_A$ and $R_B$. Either of these radii can be chosen 
to be in the eleventh direction, so we can define two copies 
of type IIA string theories compactified on a circle of radius 
$R_A$ respectively $R_B$. 

The string coupling and string scale are determined by the 
eleventh radius and the 11-dimensional Planck scale $l_{Pl}$ through 
\eqn\noname{
R_{11} = g_s l_s = g_s^{2/3} l_{Pl}, 
}
so that the type IIA string theory obtained via compactification of 
the $A$ direction has string coupling and string scale given by 
\eqn\typeiiacouple{
g_s = (R_A/l_{Pl})^{3/2},  \qquad \alpha' = R^{-1}_A l^{3}_{Pl}. 
}
By a further $T$-duality in the nine direction on both sides of 
diagram \idendual we get two type
IIB theories that are related by an $S$-duality transformation. 
The type IIB theory in the lower-right corner in figure \idendual 
is defined on a circle of radius $R_A'$ and has string coupling 
and string scale (see \bergshoeff\ for a translation code between 
IIA and IIB)  
\eqn\typeiibcouple{
R'_A = l^3_{Pl}R^{-1}_A R^{-1}_B, \quad \quad 
\tilde{g}_s = R_B/R_A, \quad  \quad  \tilde{\alpha}' = 
R_B^{-1} l_{Pl}^{3}. 
}
The idea now is to obtain a relation between DLCQ IIA string theory and 
Yang-Mills theory by an infinite boost in the $R_B$ direction
of the equivalent IIA and IIB string 
theories \typeiiacouple and \typeiibcouple\!\!. 
This infinite boost is accompanied by a zero size limit of radius 
$R_B$ such that 
\eqn\noname{
R_B = e^{-\chi} R \quad ( \chi \rightarrow \infty), 
}
where $\chi$ is the boosting parameter which we will 
take to be $\cosh \chi  = (R^2+4R_B^2)^{1/2}/2R_B$.
This procedure has the effect that one of the light-like coordinates 
$x^\pm= x_B \pm t$ becomes compactified. Namely, for finite $\chi$ we have  
the simultaneous identifications 
\eqn\noname{
x^+ \simeq x^+ + e^{-\chi} R \quad , \quad 
x^- \simeq x^- + R \quad (\chi \rightarrow \infty ),
}
so that in the limit $R_B \rightarrow 0$ the light-cone coordinate 
$x^-$ becomes compact, while $x^+$ is identified with time. 
In this way we get on the left in figure \idendual\!\!,  
DLCQ IIA string theory in the sector
$p^+ = N/R$ and D-particle number $M$, with finite string coupling 
and string scale given by \typeiiacouple\!. 

On the right we have weakly coupled type IIB theory 
with $N$ D-strings and $M$ F-strings wrapped along a 
circle of large radius
\eqn\ymradius{
R'_A = l^3_p R^{-1}_A R^{-1}_B = e^{\chi} \ l^3_p R^{-1}_A R^{-1}  
\quad (\chi \rightarrow \infty).
}
At the scale of this compactification radius, the string length 
$\tilde{\alpha}'$ vanishes 
\eqn\noname{
{R'}^2_{A} / \tilde{\alpha}' = e^\chi \ l_p^3 R_A^{-2}R^{-1} 
\quad (\chi \rightarrow \infty).
}
This means that the IIB theory gets a Yang-Mills theory description, 
which can be derived from the low energy effective D-string action 
(Born-Infeld theory) via the zero slope limit $\alpha' \rightarrow 0 $. 
Using \yangmstring and \typeiibcouple we find the Yang-Mills coupling
constant of the gauge theory description 
\eqn\ymcoup{
g^2_{YM} = {g'_s \over \tilde{\alpha}'} = {R_B^2 \over R_A l_p^3}. 
}
In this formula $g'_s$ is the string coupling constant of the type
IIB theory on the right side of figure \idendual\!\!. 

There is one additional condition for
this reduction to make sense: namely finiteness of the Yang-Mills energy, 
as compared with the energy scale determined by the compactification
radius $R'_A$. 

The relevant Yang-Mills Hamiltonian is the energy of a
sector with integer electric flux $M$, 
\eqn\ymillenergy{
H_{YM} = g^2_{YM} R'_A {M^2 \over 2N}.   
}
This follows from the remarks made in section \boundstates where we
explained that a $(p,q)$ string bound state has a low energy
description in terms of Yang-Mills theory with electric flux turned 
on. 
The Yang-Mills energy \ymillenergy is finite with respect to the 
Yang Mills radius $R'_A$, as can be verified by using \ymradius and 
\ymcoup\!\!. 

When we include transversal momenta $p_\perp$ of the strings
the YM Hamiltonian becomes 
\eqn\noname{H_{YM} = {1 \over 2p^+}(p^2_\perp + {M^2 \over g_s^2}), 
}
which equals the DLCQ string theory Hamiltonian with D-particle charge 
$M$ \dlcq. 

%We rewrite the form of \ymillenergy by 
%using the string theory Hamiltonian 
%%
%\eqn\noname{
%p_- = {1 \over \alpha' g_s^2} {M^2\over 2p_+},  
%}
%%
%the relations in \typeiiacouple and \typeiibcouple and the 
%mass shell condition $p^2_\perp = 2p_+p_-$. We get 
%%
%\eqn\noname{
%R'_{A}H_{YM} = {l_p^3 \over R_A}{p_\perp^2 \over 2N}. 
%}
%%
We come to the conclusion that DLCQ IIA string theory 
is equivalent to super Yang-Mills theory defined on a cylinder, 
with dimensionless coupling constants identified via 
\eqn\noname{
g_{YM} = {1 \over g_s}.  
}
The equivalence is thus an example of a strong-weak coupling
duality as already emphasized in chapter \introintro\!\!. 

The relation between matrix string theory and M-theory in the discrete 
light-cone gauge is reflected at the level of the 
superalgebras too. 
The superalgebra of DLCQ 11-dimensional supergravity 
\elevendsusy on a $d$-dimensional torus is identical to 
the one of maximally supersymmetric Yang-Mills theory 
on the dual torus, after appropriate identifications of the fluxes 
\mmatrix. 

%\newsection\noname{N=4 SYM and DLCQ string Theory}

\newsubsection\nonamedecom{(De)compactification} 

In the previous section we argued why matrix string theory yields
a dual description of non-perturbative type IIA string theory. 
The gauge model describes type IIA strings in the DLCQ formalism, 
moving in eight transversal dimensions.

The way of reasoning of the previous section can be repeated for
type IIA string theory toroidally compactified to lower dimensions. 
The result is that the dual supersymmetric $U(N)$ gauge model is
defined on a higher dimensional torus.
This procedure, however, is of limited use, as Yang-Mills theories
are quantum mechanically well defined only up to 4 dimensions. 

This is one of the reasons why we will focus on the well known 
$\Nisf$ $U(N)$ SYM model in 3+1 dimensions, which, according to 
the reasoning above, describes three-branes or DLCQ string theory in 
8 dimensions.

Instead of repeating this reasoning again 
we will show that dimensional 
reduction of the gauge theory means decompactification of the
associated string theory, and vice versa. 
To begin with let us consider the Lagrangian of the $\Nisf$ model
\eqnar\fouraction{
S & = & -\frac{1}{g^2}\int \! d^4x\, {\rm tr}\Bigl(\frac{1}{4}
F_{\mu\nu}^2+\frac{1}{2}(D_{\mu}X^I)^2 +
\psi\Gamma^i D_i \psi \nonumber\\
& & \qquad \qquad +\psi\Gamma^I[X^I,\psi]+\frac{1}{4}[X^I,X^J]^2 \Bigr),
}
with $I = 1,\dots 6$. 
When we compactify this model on a three torus of the form 
$T^2 \times S^1$, and rescale the circle according to 
$x^3 \rightarrow \lambda x^3$, the bosonic part of the action 
\fouraction becomes
\eqn\actionscaled{
S = -{1 \over g^2} \int\! d^4x \, {\rm tr} \Bigl(
{1\over 2\lambda}F_{3i}^2
+{1 \over 4} \lambda F_{ij}^2 +{1\over 2 \lambda} (D_3 X^I)^2
+{1\over 2}\lambda (D_i X^I)^2 +\lambda [X^I,X^J]^2 
\Bigr). 
}
In the limit $\lambda \rightarrow \infty$ \actionscaled is conjectured 
to flow to an effective supersymmetric conformal field theory in the 
infrared \harvey\matrixstring. 
In this limit finite energy configurations satisfy the 
flatness conditions 
\eqn\flatness{
F_{ij} =0,\qquad
D_{i} X^I  =0 ,\qquad   
[X^I,X^J] = 0. 
}
These conditions are the same flatness conditions one gets in the strong  
coupling limit of matrix string theory (see also section 
\stringorbifold). We thus see that the limit 
$\lambda \rightarrow \infty$ (or equivalently shrinking the transversal
two torus to zero size) dimensionally reduces the model \fouraction 
to $1+1$ dimensions, i.e. to matrix string theory. The associated 
string theory is however decompactified from six to eight transversal
dimensions. 

Conversely compactifying string theory to lower dimensions
lead to higher 
dimensional gauge theories. Note that we have to define what we
actually mean by compactifying matrix string theory. Hereto we will
follow the original argument due to Taylor \taylor, 
who discussed toroidal compactifications of matrix theory.

Matrix string theory is equivalent to the low-energy description 
of a system of $N$ D-strings moving in flat space $\Rvet^9$. 
An obvious definition of compactified matrix string theory is therefore 
the low energy description of D-strings living on compact space. 

A collection of D-strings on a circle can be described by using orbifold 
techniques. The covering space of a circle $S^1$ is the real line $\Rvet$, 
so we can study the D-strings by considering the motion of an infinite family 
of D-strings on $\Rvet$ and then impose constraints that imply translation
invariance of the D-string configuration.  
The matrices $X^I$ get two type of indices: $X^I_{mi,nj}$ where 
$n \in \Zvet$ and $i$ runs over integers $1,\cdots,N$. 
The conditions the discrete symmetry $\Zvet$ 
imposes on these fields read \taylor\Tayll
\eqnar\restrictionsi{
X^I_{mn} & = &  X^I_{(m-1)(n-1)} \qquad \qquad I<9, 
\cr
X^9_{mn} & = & X^9_{(m-1)(n-1)} \qquad \qquad m \neq n, 
\\
X^9_{nn} & = & X^9_{(n-1)(n-1)} + 2\pi R_9 \one, 
\nonumber
}  
where we suppressed the indices $i,j$ of the matrices. We took 
the compactified direction in the ninth direction. 
As a result of the constraints \restrictionsi the matrix 
$X^9_{mn}$ can be written in the following form 
\Tayll
\eqn\nicematrix{
X^9 = 
\left(
\begin{matrix}
\ddots & X_1 & X_2 & X_3 & \ddots 
\cr
X_{-1} & X_0 -2\pi R_9\one & X_1 & X_2 & X_3 
\cr
X_{-2} & X_{-1} & X_0 & X_1 & X_2 
\cr
X_{-3} & X_{-2} & X_{-1} & X_0+2\pi R_9\one & X_1 
\cr
\ddots & X_{-3} & X_{-2} & X_{-1} & \ddots 
\cr
\end{matrix}
\right),
}
where we used the shorthand notation $X_k = X^9_{0k}$. 
One can interpret the matrix of the form \nicematrix as 
a representation of the covariant derivative of YM theory
defined on the dual circle
\eqn\covdev{
X^9 = i\hat\partial+ A(\hat{x}),
}
acting on the Fourier components of a periodic function defined on
the dual circle
\eqn\trivialsection{
\phi(\hat{x}) = \sum_n \hat{\phi}_n e^{in\hat{x}/\hat{R}_9}.
}
Decomposing the gauge field $A(\hat{x})$ into Fourier components
\eqn\noname{
A(\hat{x}) = \sum_n A_n e^{in\hat{x}/\hat{R}_9},
}
we see that the covariant derivative \covdev has precisely the
same action on the Fourier components of the function 
\trivialsection as the matrix \nicematrix\!\!. Thus we conclude that 
compactification of matrix string theory on a circle means 
adding an additional compact dimension to the YM theory, 
plus replacing a Higgs field by a gauge field. This procedure 
is the counterpart of dimensional reduction. 

Of course one can repeat this for tori of other dimensions, keeping
in mind that the gauge theory is well defined in low dimensions only.

\newsection\dualitymt{N=4 SYM and M-theory on $T^3$}

The $\Nisf$ SYM model \fouraction 
has been thoroughly studied in the past. 
Elegant semi-classical studies have revealed that the spectrum
of dyonic BPS-saturated states in the theory exhibits an exact
symmetry between electric and magnetic charges 
\montonen\osborn\wittenolive\vafawitten\newline \girardello. 
This $S$-duality is
expected to extend into the full quantum regime, thereby providing 
an exact mapping between the strong coupling and weak coupling
sectors. Although recent breakthroughs in non-perturbative
supersymmetric gauge theories and string theory have produced
substantial evidence for the duality conjecture, finding an explicit
construction of the duality mapping still seems as difficult as ever.  

The S-duality of the $\Nisf$ model \fouraction forms an important 
ingredient in the matrix theory \bfss\ formulation of 11-dimensional M-theory. 
As reviewed in section \mstproposal\!\!, matrix theory proposes a concrete 
identification between the $U(N)$ 
SYM model defined on a three-torus $T^3$ and DLCQ type IIA or IIB string
theory compactified on a two-torus $T^2$. A particularly striking 
consequence of this conjectured correspondence is that the $S$-duality 
of the gauge model gets mapped to a simple $T$-duality in the string 
theory language \susskind\ganor. 

Central in the correspondence with M-theory on $T^3$ is the following 
construction of the 11-dimensional supersymmetry algebra in terms of the 
SYM  degrees of freedom. The generators of the four-dimensional
$\Nisf$ supersymmetry algebra can be conveniently
combined into one single $SO(9,1)$ spinor supercharge $Q_\alpha$, by
considering the 4D SYM model as the 
dimensional reduction of ten-dimensional ${\cal N}=1$ SYM theory. 
In this ten dimensional notation   
the fermionic fields transform under the supersymmetry according to 
$\delta \psi= \Gamma^{\mu\nu}F_{\mu\nu}\epsilon$,
and the corresponding supercharge $Q_\alpha$ is equal to
\eqn\superchargeone{
Q=\int_{T^3} \tr \left[\Gamma^r\psi E_r
-\Gamma^0\Gamma^{rs}\psi \frac{1}{2}F_{rs}\right].
}
Here and in the following the indices $r,s$ run from 1 to 9.
As the spatial part of our space-time manifold is compact we have 
an additional global supersymmetry: the action is invariant under adding
a constant spinor to $\psi$ via $\delta \psi = \tilde{\epsilon}$.
We denote the corresponding supercharge by $\tilde{Q}$
\eqn\superchargetwo{
\tilde{Q}= \int_{T^3} \! \tr\, \psi,
}
where we put the volume of the three torus (and the length of 
all its sides) equal to one. The supersymmetry algebra is
\eqnar\susyalgebra{
\left\{\tilde{Q}_\alpha , \tilde{Q}_\beta \right\} &=&
N\delta_{\alpha\beta},
\nonumber\\
\left\{Q_\alpha , \tilde{Q}_\beta\right\} &=& Z_{\alpha\beta},\\
\left\{\bar{Q}_\alpha , Q_\beta \right\} &=& 2 \Gamma^0 H +
2 \Gamma^i P_i, \nonumber
}
where the central charge term
\eqn\centralcharge{
Z = \Gamma^0\Gamma^i e_i-\Gamma^{ij}m_{ij}
}
consists of the total electric and magnetic flux through the
three-torus, defined via 
\eqn\electroflux{
e_i = 
\int_{T^3} \! \tr\, E_i, \qquad 
m_{ij} = 
\int_{T^3} \! \tr \, F_{ij}.
}
$H$ in \susyalgebra denotes the supersymmetric Yang-Mills Hamiltonian
and the quantities $P_i$ are the integrated energy momentum fluxes, defined 
(in 3+1 notation) as 
\eqn\momenta{
P_i=\int_{T^3} \tr (E_jF_{ji}+\Pi^I D_i X_I+\frac{1}{2}i\psi^{\rm
T}D_i\psi),
}
where $\Pi_I$ is the conjugate momentum to the Higgs scalar field $X_I$.
In writing the above supersymmetry algebra 
we have assumed that the $U(1)$ zero mode 
part of $\Pi_I$ vanishes. 
From the eleven dimensional M-theory perspective, this
means that we assume to be in the rest-frame in the uncompactified 
space directions. 

The superalgebra \susyalgebra has the same form as the one of 
eleven dimensional supergravity (cf.\elevendsusy) 
in the DLCQ formalism:
\eqnar\susylcgsugra{
\Bigl\{\tilde{Q}_\alpha , \tilde{Q}_\beta \Bigr\} &=& 
p^+\delta_{\alpha\beta}, 
\nonumber \\[1mm]
\Bigl\{ Q_\alpha, \tilde{Q}_\beta \Bigr\} &=& 
2p^a\gamma_{a\alpha\beta}
+2 Z^{a_1a_2} \gamma_{a_1a_2\alpha\beta} + 
2Z^{a_1 \cdots a_5}\gamma_{a_1\cdots a_5 \alpha\beta}, 
\\[1mm]
\Bigl\{Q_\alpha, Q_\beta\Bigr\} &=& 
2p^- \delta_{\alpha\beta} + 2Z^a\gamma_{a\alpha\beta} + 
2Z^{a_1\cdots a_4} \gamma_{a_1 \cdots a_4\alpha\beta}. 
\cr\nonumber
}
Here we split the 32-component supercharge of 11D supergravity into
two sixteen-component supercharges $Q$ and $\tilde{Q}$. As prescribed 
by DLCQ, $x^-$ is defined on a circle and $x^+$ plays the role of time.
In writing up the supersymmetry algebra we have therefore set the charges 
with a $+$ component to zero, while the charges with a $-$ component do
not necessarily vanish. These charges are indicated by $Z^a$ and 
$Z^{a_1\cdots a_4}$ and they belong to membranes respectively 
five-branes wrapped around the light-like circle and transversal 
directions.  
Comparing the superalgebras \susylcgsugra and \susyalgebra 
we see that they have the same form, except for the five-brane 
central charge. But the five-brane is not included in M theory on 
low dimensional tori, because then it is infinitely massive, so 
the superalgebras are indeed identical. 

We thus get a dictionary for the various fluxes of $3+1$ dimensional
SYM theory in terms of charges of eleven-dimensional M-theory on 
$T^3$ and 10-dimensional IIA string theory compactified on $T^2$. 
%The 
%latter correspondence, with the IIA string, proceeds by writing 
%$T^3= T^2 \times S^1$ and interpreting the third direction (along the
%$S^1$) as the eleventh direction of M-theory. 
This leads to the 
following list of correspondences (here $i,j$ run from 1 to 2):  
\\[-3mm]
%
%%%
\begin{eqnarray}
\begin{array}{ccccc}
  \begin{array}{|l|l|}
  \hline
  N & H \\[2mm]
  \hline
  e_3 & e_i\\[2mm]
  \hline
  m_3 & m_i\\[2mm]
  \hline
  p_3 & p_i\\[2mm]
  \hline
  \end{array}
 &\quad \leftrightarrow \quad &
  \begin{array}{|l|l|}
  \hline
  p_{+} & p_-\\[2mm]
  \hline
  p_9 & p_i\\[2mm]
  \hline
  m_{ij} & m_{9 j}\\[2mm]
  \hline
  m_{9-} & m_{i-}\\[2mm]
  \hline
  \end{array}
 &\quad \leftrightarrow \quad &
  \begin{array}{|l|l|}
  \hline
  p_{+} & p_- \\[2mm]
  \hline
  q_0 & p_i\\[2mm]
  \hline
  m_{ij} & w_j\\[2mm]
  \hline
  w_- & m_{i-}\\[2mm]
  \hline
  \end{array}\\[1.7cm]
\mbox{4D SYM} && \mbox{M-theory on $T^3$ } 
&& \mbox{IIA string on $T^2$} 
\end{array}%\nonumber
\end{eqnarray}
\xdef\itable{{\theequation}\ }
%%%
%
\\[0mm]
Here $q_0$ denotes the D-particle number in the IIA string theory. 
In the second and third table, the integers $m_{ij}$ - corresponding 
to the SYM magnetic fluxes -  denote the wrapping numbers of the 
M-theory membrane around the $T^3$ and the D2-brane around 
the $T^2$ respectively. The M-theory membranes wrapped $m_{9j}$ 
times around the compact light-like direction turn into IIA strings 
with winding number $w_i$. 

States with non-zero momentum flux $P_i = p_i$ in the SYM theory
correspond
to membranes that are wrapped around the longitudinal $x^-$ light-like 
direction. In the original proposal of \bfss , M-theory arises in the 
limit $N\rightarrow \infty$ while restricting the spectrum of the 
$\Nisf$ SYM model to the subspace of states that have energy 
of order $1/N$. This limit amounts
to a decompactification of the longitudinal direction. States with
non-zero $p_i$ will thus correspond to infinite energy configurations 
containing membranes stretched along the light-cone direction. All
finite energy states therefore must have $p_i=0$. 

In the DLCQ setup, the longitudinal membranes are no longer 
infinitely massive, and must be naturally included in the spectrum. 

\newsubsection\nonameudual{$U$-Duality in matrix theory on $T^2$}

The correspondence with string theory and M-theory gives a number 
of predictions concerning the duality properties of the gauge theory.
These predictions in particular concern the BPS spectrum of the 
SYM model at finite $N$ as well the behavior of the model at large $N$.
For example, from the above table \itable 
it is seen immediately that 
electric-magnetic duality in the SYM theory follows from the 
$T$-duality on the two torus, as first noted in \harvey. 
The $T$-duality that exchanges the D-particles with the D2-branes and the
KK momenta with the NS winding numbers in type IIA string theory, 
when translated to the first table indeed gives rise to the $S$-duality
that interchanges all the electric and magnetic fluxes. Via this 
correspondence, the complete $U$-duality symmetry 
$SL(2,\Zvet)\times SL(3,\Zvet)$ of M-theory on $T^3$ is expected to be
realized as an exact symmetry of the matrix formalism
\susskind\ganor. 
\footnote{For a review of $U$-duality in matrix theory we refer to \obers.}
A discussion about the expected $SL(5,\Zvet)$ duality symmetry group of 
matrix theory on $T^4$ can be found in \rozali. This case is more 
subtle as the theory is not simply described by $4+1$ dimensional 
gauge theory, because it is not renormalizable.  

At finite $N$ there are reasons to suspect that the duality
group that acts on the BPS sector is in fact enlarged. 
As we have discussed in section \mstproposal\!\!, for finite $N$
matrix theory describes M-theory in the DLCQ formalism, and 
correspondingly one no longer needs to restrict to states with 
vanishing momentum flux $p_i$. BPS states at finite $N$ can therefore
carry a total of 10 charges, labeled by $(N,e_i,m_i,p_i)$. 

The gauge theory descends from Born Infeld theory 
in the zero slope limit $\alpha' \rightarrow 0$ 
(cf. section \dbranedual). 
Therefore we can interpret the $U(N)$ SYM model as a 
low energy description of all possible bound states of $N$ D3 branes
of IIB string theory on $T^3$. In this correspondence, $e_i$ denotes 
the NS string and $m_i$ the D-string winding number, while $p_i$ is 
the KK momentum. The SYM theory should thus encompass all BPS bound
states of this system. 

It should be emphasized, however, that from
this reasoning we should only expect the BPS {\it degeneracy} formula 
to be $SL(5,\Zvet)$ symmetric, while of course the energy spectrum is
not, since the Yang-Mills model only represents a particular
limit of the $N$ D3 brane system. 

\xdef\voetnootpagina{\thepage}

\newsection\bpsmtheory{BPS spectrum }

The goal of our study in this section 
is to determine the explicit form of the BPS degeneracies for finite 
$N$ as a function of the charges $(N,e_i,m_i,p_i)$, 
and thereby exhibit its full duality symmetry.
We will approach this problem in two ways: first from
M-theory and then directly from the $\Nisf$ SYM model. 
Our main finding is that the spectral degeneracy of individual bound 
states from both points of view
is identical, and furthermore exhibits a full $SL(5,\Zvet)$
duality symmetry. Part of this large duality group also acts
on the rank $N$ of the gauge group, an example of so called Nahm type
transformations \wnahm. 
%In fact, once we include a Nahm type transformation 
%the manifest $SL(2,\Zvet)\times SL(3,\Zvet)$ group is extended to 
%$SL(5,\Zvet)$. 

The BPS-states that we will consider respect 1/4 of all supersymmetries.
Every such BPS-state in a fixed multiplet satisfies
\eqn\multipletcon{
\left(\epsilon^\alpha Q_\alpha+\tilde{\epsilon}^\alpha 
\tilde{Q}_{\alpha}\right)
\left|{\rm BPS}\right>=0,
}
for a certain fixed collection of $SO(9,1)$ spinors $\epsilon$ and 
$\tilde{\epsilon}$. These spinors are, up to an overall factor, 
completely determined by the set of charges.

By taking the commutator with $\bar{Q}$ and $\tilde{Q}$ 
in the preceding equation
we get two conditions for the spinors $\epsilon$ and $\tilde{\epsilon}$
\eqnar\tepsilon{
2\Gamma^0H\epsilon
+2\Gamma^iP_i\epsilon
+\Gamma^0 Z \tilde{\epsilon}= 0,\nonumber \\
N \tilde{\epsilon}+Z^\dagger\epsilon = 0.
}
Plugging $\tilde{\epsilon}$ 
in the first equation gives the following
equation for $\epsilon$
\eqn\conditie{
\left(\Gamma^0H'+\Gamma^i P'_i\right)\epsilon=0,
}
where $H'$ and $P_i'$ denote the Hamiltonian and momentum fluxes with
the zero-mode contributions removed. Explicitly,
\eqnar\hammom{
H =  \frac{1}{2N}(e_i^2 + m_{i}^2) + H', \nonumber \\ 
P_i = (e \wedge m)_i/N + P_i'.
}
From the last equation, 
we see that the eigenvalues of $P'_i$ are equal
to $p'_i = \kappa_i/N$ with 
\eqn\threekappa{
\kappa_i = Np_i - (e\wedge m)_i.
}
The operator in equation \conditie should have eigenvalues equal to zero.
This is only the case when the magnitude of $H'$ and the length of $P'_i$
are equal, $H' = |P_i'|$. 
Using this relation, we can write equation \conditie 
in the following way
\eqn\gamcon{
\left(\Gamma_0+\Gamma^i\hat{\kappa}_i\right)\epsilon=0,
}
where 
the vector $\hat{\kappa}_i$ denotes the unit vector in 
the direction of $\kappa_i$.

\newsubsection\nonamebps{BPS spectrum from M-theory}

Now we are ready to determine the detailed BPS spectrum 
from discrete light-cone
M-theory. To this end it is useful to introduce the notion of an 
{\it irreducible} BPS state, as a state that contains only one 
{\it single} BPS bound state.
Indeed, general BPS states can in principle combine more 
than one such bound state into one second quantized BPS state. 
The combined state will still be BPS, provided each of the 
irreducible constituent bound states is left invariant by the {\it same}
set of supersymmetries, {\it i.e.} with the same spinors 
$\epsilon$ and $\tilde{\epsilon}$.

The degeneracy of irreducible BPS states 
is determined almost uniquely from U-duality invariance, and from 
the known BPS spectrum of perturbative string states.
The relevant U-duality group for our case turns out to be as large as the
complete $SL(5,\Zvet)$ duality group of string theory on $T^3$, 
{\it i.e.} M-theory on $T^4$. 
%At first sight, the appearance of this large symmetry may seem 
%surprising, in particular since the supersymmetry algebra 
%\susyalgebra itself is not (manifestly) $SL(5,\Zvet)$ invariant. 
%The origin of this large duality group can be understood, 
%however, as explained in section \mstproposal, 
%by noting that the discrete light-cone version
%of M-theory on $T^3$ can be obtained from M-theory on $T^4$ 
%via a suitable limit, essentially by applying an infinite boost to one
%of the compactified directions of the $T^4$ \seibergsen. 
%We expect the BPS spectrum to be invariant under such boosts. 

The 10 quantum numbers $(N,e_i,m_i,p_i)$ can be combined into a
$5\times 5$ anti-symmetric matrix \gysbert 
\eqn\fivematrix{
\left(
\begin{matrix}
0  & p_3& -p_2 & e_1 & m_1 \\
-p_3& 0  & p_1 & e_2 & m_2 \\
p_2 &-p_1&  0   & e_3 & m_3 \\
-e_1&-e_2& -e_3 & 0   & N \\
-m_1& -m_2& -m_3& -N   & 0\\
\end{matrix}
\right), 
}
on which $SL(5,\Zvet)$ acts by simultaneous 
left- and right-multiplication. 
\footnote{The $SL(5,\Zvet)$ transformations also acts on the 
metric of the torus and the coupling constant in a non-trivial way. 
We will discuss this in the last section of this chapter.} 
The smaller $SL(3,\Zvet) \times SL(2,\Zvet) $ symmetry group
is included in the group $SL(5,\Zvet)$ and is represented
by block-diagonal matrices, with an $SL(3,\Zvet)$ group element in the 
upper left corner and an $SL(2,\Zvet)$ element in the lower right corner. 
Together with an $SL(5,\Zvet)$ matrix that is not in this block-diagonal form
the elements of $SL(3,\Zvet) \times SL(2,\Zvet)$ generate the complete
extended duality group. We will call the extra transformations 
Nahm type transformations, because they mix rank and
electro-magnetic fluxes with each other. They resemble the well known 
Nahm duality mapping that exchanges instanton number and rank of gauge
group in Euclidean $U(N)$ Yang-Mills theory \braam. 
We come back to this issue in section \thechapter.5.

The following bilinear combinations of the fluxes 
\eqn\kfive{
K_i = (Np_i - (e\wedge m)_i, \, p\! \cdot\! m\, , \, p\!\cdot\! e\, ),
}
transform as a 5 vector under $SL(5,\Zvet)$ duality. 
It can be formed out of the fluxes by contracting the matrix 
\fivematrix with its Hodge dual, a three tensor.  
From this we deduce that the relevant $SL(5,\Zvet)$ invariant scalar 
combination we can make out of the ten charges is the integral 
length of this five-vector 
\footnote{Another $SL(5,\Zvet)$ scalar is $\gcd(N,p_i,e_i,m_i)$ which
is not relevant for the degeneracies of BPS states in Yang-Mills
theory. This scalar will be important however 
when one considers the BPS spectrum of supersymmetric 
Born Infeld theory on $T^3$ \gysbert .}  
\eqn\kscalar{
|K| = \gcd(Np_i - (e\wedge m)_i, \, p\! \cdot\! m\, , \, p\! \cdot\! e\, )
.}
Duality invariance thus predicts that the degeneracy of irreducible
BPS bound states should be expressible in terms of the quantity $|K|$ only.

To determine the explicit degeneracy formula, we note that by using
the U-duality symmetry, any irreducible bound state can be rotated into a
state that carries only KK momentum and NS string winding (in the SYM
language these states have zero momenta $p_i=0$). 
Such a state must
necessarily be made up from a single fundamental IIA string. The invariant
$|K|$ for this perturbative string state simply reduces to the bilinear
combination of momenta and winding numbers that determines (via the BPS
restriction) the oscillator level of the string. Single BPS states of
toroidal compactified II string theory have only left-moving 
(or right-moving) excitations. The number of irreducible
BPS bound states is therefore counted by means of the chiral
string partition function \polstring\ 
\eqn\generator{
\sum_K c(K)q^K=(16)^2 \prod_n \left(\frac{1+q^n}{1-q^n}\right)^8, 
}
with $K=|K|$ as defined in \kscalar.

General BPS states of discrete light-cone gauge M-theory may consist of more
than one irreducible bound state. The total BPS condition requires that the
charges of these separate bound states must be compatible.
The complete second quantized partition sum is obtained by taking into
account all possible such ways of combining individual bound states into
a second quantized configuration with a given total charge. More detailed
comments on the combinatorial structure of the second quantized BPS
partition sum will be given in section \thesection.3.

\newsubsection\bpsnisf{BPS spectrum from N=4 SYM on $T^3$ }

The above description of the BPS spectrum can be reproduced directly
from the $U(N)$ gauge theory on $T^3$ as follows. First let us recall
the definition of the electro-magnetic flux quantum numbers.
To this end it is useful to decompose
the $U(N)$ gauge field into a trace and a traceless part
\eqn\traceless{
A_\mu %=\frac{\tr A_\mu}{N} \one + (A_\mu-\frac{\tr A_\mu}{N}\one)
=A^{U(1)}_\mu\one+A^{SU(N)}_\mu
}
and to allow the fields on $T^3$ to be periodic up to gauge
transformations of the form
\eqnar\boundary{
A_i^{SU(N)}(x+a_j)\is \Omega_jA^{SU(N)}_i(x)\Omega_j^{-1},
\nonumber \\ \\[-\the\baselinetruc] \cr
A^{U(1)}_i(x+a_j)\is A^{U(1)}_i(x)-2\pi m_{ij}/N, \nonumber
}
with $m_{ij}$ integer. 
Here $a_i$ with $i=1,2,3$ denote the translation vectors
that define the three torus $T^3$.
\footnote{We repeat here that for simplicity
we mostly take $T^3$ to be cubic with sides of length 1.}
The $SU(N)$ rotations $\Omega_i$ must satisfy the $\Zvet_N$ cocycle conditions
\eqn\cocycle{
\Omega_j\Omega_k = \Omega_k\Omega_j
e^{2\pi i \mu_{jk}/N},
}
for integer $\mu_{jk}$.
The quantities $\mu_{ij}/N$ define the 't Hooft $\Zvet_N$ 
magnetic fluxes \hooft\ and $m_{ij}/N$ can be identified with the $U(1)$
magnetic flux defined in \electroflux\!\!. 
The integers $\mu_{ij}$ and $m_{ij}$ are restricted 
via the condition that the combined gauge transformation \boundary
defines a proper $U(N)$ rotation. This requirement translates into the 
Dirac quantization condition that the total flux $(\mu_{ij}-m_{ij})/N$
must be an integer. Hence $m_{ij} = \mu_{ij} (\mbox{mod $N$})$.

Electric flux carried by a given state is defined via the action of
quasi-periodic gauge rotations $\Omega[{\bf n}]$ defined via

\eqn\quasiper{
\Omega_j\Omega[{\bf n}] = \Omega[{\bf n}]\Omega_j
e^{2\pi i n_{j}/N}, 
}
with integer $n_j$. Such gauge rotations will preserve the boundary
condition \boundary on the gauge fields. A state $|\psi\rangle$ is defined 
to carry $SU(N)$ flux $\epsilon_j$ if it satisfies the eigenvalue condition
\eqn\egvalue{
\widehat{\Omega}[{\bf n}] | \psi \rangle = 
e^{{2\pi i \over N} n^j \epsilon_j} | \psi\rangle, 
}
where $\widehat{\Omega}[{\bf n}]$ denotes the quantum operator 
that implements the
gauge rotation $\Omega[{\bf n}]$ on the state $|\psi\rangle$. 
Similarly as for the magnetic flux, the electric
flux receives an overall $U(1)$ contribution $e_i$ defined in 
\electroflux\!\!. The abelian
and non-abelian parts of the flux must again be related via 
$e_{i} = \epsilon_{i} (\mbox{mod $N$})$.

To determine the supersymmetric spectrum for a given set of charges, 
the idea is to first reduce the phase space of the $U(N)$ SYM model 
to the space of classical supersymmetric configurations and then to 
quantize this BPS reduced phase space.
The justification for this procedure should come from the high degree
of supersymmetry in the problem, while furthermore the degeneracy of 
BPS states is known to be a very robust quantity.

To obtain the reduced phase space, we recall that the SUSY
transformation for the fermionic partners of the Yang-Mills fields 
reads (again using ten-dimensional notation)
\eqn\susytrafo{
\delta\psi=\left(E_r\Gamma^{0r}+\frac{1}{2}F_{rs}\Gamma^{rs}\right)
\epsilon+\tilde{\epsilon}.
}
The BPS restriction requires that the right-hand side vanishes 
for those $\epsilon$
and $\tilde{\epsilon}$ determined in the previous section. 
Thus in particular we can 
use the equation \tepsilon to express $\tilde{\epsilon}$ in terms of
$\epsilon$ and the $U(1)$ zero modes. The result is
\begin{equation}
\delta\psi=\left(E'_r\Gamma^{0r}+\frac{1}{2}F'_{rs}\Gamma^{rs}\right)
\epsilon=0.
\end{equation}
where the primed quantities are equal to the un-primed ones with the constant
$U(1)$ parts removed. For BPS-states in a fixed multiplet, supersymmetry is 
unbroken for $\epsilon$ satisfying the equation \conditie above.
Hence for these BPS-states the following must hold for all spinors 
\begin{equation}
\left(E'_r\Gamma^{0r}+\frac{1}{2}F'_{rs}\Gamma^{rs}\right)
\left(\Gamma^0-\Gamma^k\hat{\kappa}_k\right)\epsilon=0.
\end{equation}
Note that $\Gamma^0-\Gamma^k\hat{\kappa}_k$ acts like a projection
operator on the space of spinors satisfying equation \conditie\!\!. 
We conclude that the matrix in spinor space in the last equation
has to vanish. This is the case when $E'$ and $F'$ satisfy the following
two conditions
\eqnar\condition{
E'_i \hat{\kappa}^i &=& 0, \\  \nonumber
E'_{[r}\hat{\kappa}^{\phantom{'}}_{s]} &=& F'_{rs}.
}
From now one we shall omit the prime, 
and simply denote by $E$ and $F_{ij}$
the $U(N)$ fields without the $U(1)$ constant mode. We will also return to 
a 3+1-dimensional notation, and for additional notational convenience,
use the $SL(3,\Zvet)$
symmetry to rotate the three-vector $\kappa_i$ defined in \threekappa
in the 3 direction. So we will choose coordinates such that
\eqnar\kkk{
\qquad \qquad \qquad \kappa_3 &=& Np_3 - e_i m_{i3}, \\[2.5mm]
\qquad 
\qquad \qquad \kappa_j &=& Np_j - e_i m_{ij} - e_3 m_{3j} \, = \, 0 \qquad 
\quad i,j = 1,2.
\nonumber
}
Here
and from now on the indices $i,j$ run from 1 to 2.
From the second condition in \condition we read off that the gauge and
Higgs
fields are flat on the plane perpendicular to $\hat{\kappa}$, meaning
\eqnar\flat{
F_{ij} &=& 0, \nonumber \\ 
D_i X_J &=& 0, \\[0mm] %[2mm]
[X_I,X_J] &=& 0. \nonumber
}
In addition we have 
\begin{eqnarray}
\qquad E_i &=& F_{3i},
\\
\nonumber
\Pi_I &=& D_3X_I.
\end{eqnarray}
Finally, the first condition in \condition simply becomes
\begin{equation}
E_3 = 0.
\end{equation}
We can interpret this constraint equation as a gauge invariance condition under
arbitrary local shifts in the longitudinal gauge field $A_3$. We can therefore
exploit this gauge invariance by putting $A_3 = 0$. The relations 
\flat then simplify to the statement that the transversal
gauge fields $A_i$ and Higgs scalars $X_I$ satisfy the chiral 2D free field
equations
\eqnar\chiral{
\partial_0 A_i&=&\partial_3A_i,\\ \nonumber
\partial_0 X_I&=&\partial_3 X_I.
}
Furthermore, the equations of motion imply that the fields do not 
depend on the coordinates of the transverse torus. 
Thus  
we conclude that the BPS reduced theory is described by the left-moving
chiral sector of a two-dimensional sigma model with target space given by the
space of solutions to the flatness conditions \flat subject to the 
twisted boundary conditions specified by the electro-magnetic flux 
quantum numbers.

To determine the detailed properties of this sigma model, let us first consider
the case with all electro-magnetic fluxes equal to zero.
The only non-zero quantum numbers are therefore $N$ and $p_3$. 
In this case we can parameterize the space of solutions to \flat
by means of the orbifold sigma model on the $N$-fold symmetric product space
\eqn\symorbifold{
{(\Rvet^6 \times T^2)^N\over S_N},  
}
where $S_N$ denotes the permutation group of $N$ elements, acting on the
$N$ copies of the transversal space $T^2 \times \Rvet^6$.
\footnote{This orbifold sigma model was first considered in relation 
with $\Nisf$ SYM theory on the 
three torus in \harvey.} To see that this is the right space, we observe that 
for solutions to \flat one can always choose a gauge in which
all $U(N)$ valued fields take the form of diagonal matrices. Each such
matrix field thus combines $N$ separate scalar fields, corresponding to
the $N$ eigenvalues. The $S_N$ permutation symmetry arises as a
remnant of $U(N)$ gauge invariance, acting via its Weyl subgroup on 
the space of diagonal matrices.
Finally, the flat transversal gauge fields $A_i$ with $i=1,2$ give rise to
{\it periodic} 2D scalar fields, since constant shifts in $A_i$ by multiples
of $2\pi$ are pure gauge rotations.

The model thus reduces to the free limit of type IIA matrix string theory
\motl\tomnati\matrixstring\ in the discrete light-cone gauge \dlcq.
As explained in section \stringorbifold  
the Hilbert space of the model decomposes into
twisted sectors labeled by the partitions of $N$, 
in which the eigenvalue fields combine
into a collection of `long strings' of individual 
length $n_k$ such that the total length adds up to $\sum_k n_k = N$.
Each such string is made up from, say, $n_k$ eigenvalues that, 
by their periodicity condition around the 3-direction are connected 
via a cyclic permutation of order $n_k$. In the M-theory
interpretation, all these separate strings will indeed correspond to 
separate bound states, {\it i.e.} particles that each can move 
independently in the uncompactified space directions. 
The general form of the BPS partition function of symmetric product 
sigma models of the form \symorbifold has been described in detail
in \dmvv.

%\be \sum_{k,l}  p^k q^l D(k,l) =\prod_{n>0, m\geq 0} \frac{1}{(1-p^n q^m)}{}_{c(nm)} \ee

In the following we will mainly concentrate on the irreducible states,
describing one single BPS particle. These necessarily consist of one 
single string of maximal length. In the present case, with zero total 
electro-magnetic flux, this maximal string has total winding number
$N$ around the 3-direction.
Correspondingly, its oscillation modes have energies that are quantized in
units of $1/N$. Thus the degeneracy of states as a function of $N$ and $p_3$
is obtained by evaluating the chiral superstring partition function,
as given in \generator\!\!, at oscillator level $Np_3$.

Next let us turn on the magnetic flux $m_3$. The space of solutions to
\flat with twisted boundary conditions \boundary 
around the transverse torus again takes the form of a symmetric product
\eqn\symp{
{(\Rvet^6 \times T^2)^{N'}\over S_{N'}},
}
but where now $N' = \gcd(N,m_3)$. In order to visualize this reduction,
we note that gauge rotations $\Omega_i$ with 
$\Omega_1 \Omega_2 = \Omega_2\Omega_1 \exp(2\pi i m_3/N)$ that define
the twisted boundary conditions, can be chosen to lie within an
$SU(k)$ subfactor of $U(N)$ where $k= N/\gcd(N,m_3)$. By decomposing
the matrix valued fields according to the action of $SU(k) \otimes U(N')$
with $N' = \gcd(N,m_3)$, we can thus factor out a sector of $U(N')$ valued
field variables that are unaffected by the twisted boundary conditions.
Now following the same reasoning as before, these fields parameterize
the symmetric product space of the above form. Note that in the
particular case that $m_3 =1$, the whole $SU(N)$ part of the moduli
space of solution to \flat collapses to a point, so that only the 
$U(1)$ part survives.

By a very similar reasoning we can also include the electric flux
$e_3$ in our 
description. Like with the magnetic flux, an electric flux $e_3=1$ has
the effect of reducing the $SU(N)$ part of the vacuum moduli space
\flat to a point, or rather, it projects out just one single
supersymmetric state in the $SU(N)$
sector \wittenbound. More generally, however, it can be seen that one can
again factor out a $U(N')$ subfactor of the model, that is unaffected by
both the electric and magnetic flux $e_3$ and $m_3$, where now
\eqn\nprime{
N' = \gcd(N,m_3,e_3).
}
The BPS sector for non-zero $m_3$ and $e_3$ is thus obtained by quantizing the
supersymmetric orbifold sigma model on \symp\!\!, 
with $N'$ equal to \nprime\!\!.

The spectrum of irreducible BPS bound states is obtained as before, 
by considering the Hilbert space sector defined by the eigenvalue
string of maximal length. The maximal winding number is now equal to
$N'$. The total momentum along this string is determined by the
remaining quantum numbers of the BPS state, and should be equal to
\eqn\pthree{
p_3 - e_im_{i3}/N,
}
which is the total momentum of the BPS state minus the contribution from the
$U(1)$ electro-magnetic fluxes. Notice that the latter contribution is
in general fractional. However, since the oscillation modes of the
long string states also have fractional oscillation number quantized
in units of $1/N'$, this fractional 
total momentum \pthree can in fact be obtained via integer string oscillation
levels. The total oscillation level corresponding to this momentum flux is
\eqn\kprime{
K = N' \times (p_3 - e_im_{i3}/N) 
}
and it is easy to check using \kkk that this is an integer. In fact, after 
taking the direction of $\kappa_i$ again arbitrary, we find that the
integer quantity
\kprime becomes equal to the $SL(5,\Zvet)$ invariant length
$|K|$ defined in \kscalar of the five-vector \kfive\!\!. In this way we 
reproduce the description of the BPS spectrum given in the previous section.
In particular we find confirmation that the degeneracy of supersymmetric 
bound states has a discrete $SL(5,\Zvet)$ duality invariance.

\newsubsection\nonamecomplete{Complete BPS partition function} 

As mentioned earlier, the complete supersymmetric spectrum of the $\Nisf$
SYM model contains many more sectors, that in M-theory describe configurations 
of multiple BPS bound states. These correspond to the other twisted sectors
in the orbifold model \hbox{\symp\!\!,}  
describing multiple strings with separate
lengths $n_k$ with $\sum_k n_k = N'$. Via their zero modes these strings
each separately carry all the possible flux quantum numbers. The degeneracy 
of these separate states as a function of these charges is identical to the 
$SL(5,\Zvet)$ invariant result just described. However, it turns out that the 
combinatorics by which many such states can be combined into one second 
quantized BPS state no longer respects the full $SL(5,\Zvet)$ symmetry. For
this it would be necessary that, via the compatibility of the individual 
BPS conditions, the 10 dimensional charge vectors of all constituent states 
must align in the same direction. It can be seen, however, from the
second condition in \tepsilon and condition 
\conditie that this alignment 
is not entirely implied: in the above notation, all charges must indeed align,
{\it except} for the individual $p_3$ momenta. 

In case of nonzero $p_3$ momentum (and other fluxes equal to zero) the 
multiple states are in one-to-one relation with 
partitions of rank $N$. 
One can distribute the $p_3$ momentum freely among the 
different sectors. We conclude that the degeneracies are 
simply determined by a second quantized string partition function
\eqn\noname{
\sum_{N,p_3} p^N q^{p_3} = \prod_{r,s} {1 \over (1-p^rq^s)}{}_{c(rs)} 
,}
where the integers $c(K)$ are the degeneracies of long string 
states defined by equation \generator\!\!. 

In the situation where we have general fluxes, 
the multiple BPS states are, as explained above due to supersymmetry, 
related to partitions $\sum_{k} n'_k = n'$, with $n'$ defined by 
the greatest common divisor of all charges except $p_3$
\eqn\littlen{
n' := \gcd(N,e_i,m_i,p_1,p_2).  
}
A single short string has length $n_k= n'_kN'/n'$ and oscillator level
$\tilde{p}^3_k= p^3_k - {n'_k \over n'}{(e\times m)_3 \over N}$. 
%n_k(e\times m)_3/(NN')$. 
Its degeneracy is 
equal to $d(n_k\tilde{p}^3_k)$. 
Again the way the $p_3$ momentum is divided among the 
different short string sectors is not limited by supersymmetry. 

Hence the total BPS partition function is again
generated by a second quantized string partition function
\eqn\degterrible{
\sum_{n',p_3} p^{n'} q^{p_3} = \prod_{r,s} {1 \over (1- p^r q^s)}
{}_{d(s_r)},
}
where 
\eqn\noname{
s_r = r{N' \over n'} 
\left(s - {r \over n'}{(e\times m)_3 \over N} 
\right).
}
The degeneracy formula \degterrible is invariant under the Nahm 
type transformation that exchanges $m_3$ and $N$. We therefore 
conclude that the BPS partition function of the $U(N)$ SYM model
does not exhibit the $SL(5,\Zvet)$ symmetry, but still has a 
symmetry essentially larger than the manifest 
$SL(2,\Zvet) \times SL(3,\Zvet)$. In the next section we will 
discuss the geometric origin of this  extra symmetry in some 
detail. 

\newsection\nahmduality{Nahm duality} 

An especially interesting class of duality transformations that act on 
BPS sectors of $U(N)$ SYM theory are those that interchange the rank $N$
of the gauge group with the magnetic flux. 

In the notation used in section \bpsnisf (with
$\hat\kappa$ rotated in the 3-direction) the BPS reduced quantum
phase space exhibits a manifest symmetry under the interchange
\eqn\symmetryi{
N \rightarrow m_3 \qquad m_3 \rightarrow -N \qquad 
m_i \rightarrow \epsilon^{ij} m_j \qquad
e_i \leftrightarrow p_i, 
}
as well as under its electro-magnetic dual counterpart
\eqn\symmetryii{
N \rightarrow e_3 \qquad e_3 \rightarrow -N \qquad 
e_i \rightarrow \epsilon^{ij} e_j \qquad 
m_i \leftrightarrow p_i.  
}
In particular one can verify that the last two relations and 
$\kappa_3/N$ in \kkk as well as the
integers $N'$ in \nprime and $n$ defined in \littlen 
are invariant under these two mappings. 

The second type of duality symmetry \symmetryii
is particularly interesting, because from the M-theory perspective, 
this symmetry \symmetryii must be some manifestation of
11-di\-men\-sional covariance, as it exchanges the eleventh direction
(more precisely a light direction) with a transversal direction.   

Because of the possible relevance for Lorentz invariance of matrix
theory, it is of interest to know whether 
the symmetry extends to the full $\Nisf$ model.
Although we do not expect that this is the case, \footnote{
In the context of gauge theories on non-commutative tori, Nahm-type 
transformations are part of a larger manifest (classical) duality 
symmetry \hofmani\hofmanii. 
We will briefly comment on this at the end of this section.} 
we cannot resist reviewing the geometrical origin of this duality
mapping, which can in fact be defined for arbitrary non-BPS gauge 
configurations.

The type of transformations \symmetryi and \symmetryii 
are very similar to the Nahm-type transformations,
considered e.g. in \braam. 
There it was proven that a $k$-instanton solution of $U(n)$ 
Euclidean gauge theory on a four torus can be mapped to an 
$n$-instanton solution of $U(k)$ gauge theory on the dual
four torus. This mapping was shown to induce a bijection between 
the moduli spaces of the two instantons. 
In two dimensions it is possible to formulate an analogous mapping, 
where now rank and magnetic flux of an arbitrary gauge 
field configuration are exchanged.

\newcommand{\Aa}{A}

\newcommand{\DAa}{{D}}% \hspace{-8pt} \slash}}
\newcommand{\hhat}{\widehat}

Consider an arbitrary $U(N)$ gauge field 
$A=A_x+iA_y$ on $T^2$ with magnetic flux $M$.
This configuration can be related to a dual $U(M)$ gauge field  with 
flux $N$ defined on the dual torus $\hat{T}^2$ as follows. The key idea of 
the construction is to consider the parameter family of connections 
on $T^2$ of the form (here $a,b$ denote the $U(N)$ color indices, we
use the fundamental representation of the gauge group $U(N)$) 
\begin{equation}
\Aa^{ab}(x;z) =  A^{ab} (x)  +  2\pi {\bf 1}^{ab} z
\end{equation}
with $z=z_1+iz_2$ a complex coordinate 
on the dual torus $\hhat{T}^2$. 
The mapping proceeds by considering the space of zero modes of the 
Dirac-Weyl equation defined by $A$. Let $\DAa$ and $\DAa^*$ denote the
corresponding Dirac operator acting on right- and left-moving spinors,
respectively. We will assume that $\DAa^*$ has no zero-modes (we can 
make this assumption because, depending on the sign of the magnetic
flux, either $DD^*$ or $D^*D$ is a strictly negative definite operator 
on the two torus). 
According to a Dirac index theorem \gsw, the number of zero-modes of
$D$ then equals the magnetic flux $M$. We have  
\eqn\diracone{
\DAa\psi_i(x;z) = 0, 
}
with $i=1,\ldots,M.$ 
We can choose an orthonormal basis of zero modes, satisfying 
the orthogonality relations 
\begin{equation}
\int \! d^2 x \; \overline{\psi}_i(x;z)\psi_j(x;z) = \delta_{ij}. 
\end{equation}
The dual $U(M)$ gauge field $\hhat{\Aa}$ on $\hhat{T}^2$ is now defined as 
\begin{equation}
\hhat{A}_{ij}(z)  =  
\int \! d^2 x \; 
\overline{\psi}^a_i(x;z) 
\hhat\partial \psi_{aj}(x;z), 
\end{equation}
with $\hhat\partial  = {\partial\over\partial z }$.
This definition is equivalent to the following formula
for the dual covariant derivative $\hhat \DAa$
\eqn\definingproj{
\hhat{\DAa} \psi = 
({\bf 1}- {\bf P}) \hhat\partial  \psi, 
}
where ${\bf P}$ denotes the projection on the space of zero 
modes $\psi_i$ of the operator $D$. 
It can be shown that this
dual gauge field has magnetic flux equal to $N$, and moreover
that the mapping from $A$ to $\hhat{A}$ is a true duality
(it squares to the identity).

To make these properties more manifest, it is useful to obtain a somewhat 
more explicit form of the dual covariant derivative $\hhat{D}$.
To this end, define the Green function
\begin{equation}
\Delta G(x,y) = \delta^{(2)}(x-y)
\end{equation}
of the Laplacian $\Delta = D D^* $,  and introduce the notation
\begin{equation}
({\bf G}\psi)(x;z) = \int \! d^2y \; G(x,y) \psi(y;z).
\end{equation}
Then the projection operator ${\bf P}$ satisfies the relation 
\eqn\pgreen{
{\bf 1} - {\bf P}=  D^* {\bf G} D. 
}
Inserting this identity into the analogous formula for 
$\hhat{\DAa}^* \psi$ of the definition \definingproj\!\!, we find that 
\eqn\diractwo{
\hhat{\DAa}^* {\psi} = D^* {\bf G} D \hat{\partial}^* {\psi}  = 0, 
}
since $[D,\hat{\partial}^*] = 0$.
Hence the zero modes of the Dirac-Weyl operator on the dual torus are
equal to the ones on the original torus, with opposite
chirality. By an explicit construction \christiaan\ one can indeed 
verify that the dual $U(M)$ gauge field has magnetic flux equal to $N$,
as predicted by the index theorem. 
Thus the Nahm transformation can be summarized in an elegant way 
by means of the two equations \diracone and \diractwo\!\!,
which together specify the map from $A$ to $\hhat{A}$.
Note in particular that in this form, one of the `magical' properties
of the Nahm transformation has become manifest, namely that it is a
map of order 2, {\it i.e.} it squares to the identity:
\begin{equation}
\hhat{\hhat A} = A,
\end{equation}
up to gauge equivalence. Notice further that the mapping
is defined for arbitrary connections $A$.

When we translate the above construction back to our 3+1-dimensional setting,
the resulting mapping indeed interchanges $N$ and $m_3$ as advocated. 
In addition, since it is a mapping from $T^2$ to the dual 
torus $\hhat{T}^2$, the definition
of the momentum flux in the direction of the $T^2$ gets interchanged with that
of the electric flux along this direction, as indicated in \symmetryii\!\!. 
Naturally, the electric flux represents the conjugate momentum to the constant
mode of the gauge fields $A_i$, and thus defines the total momentum on
the dual torus $\hhat{T}^2$. Further inspection also shows that the magnetic
fluxes $m_i$ get reflected, as predicted.

Finally, we should of course note that in the interpretation of the $U(N)$
SYM model as describing $N$ D3 branes, this Nahm duality is nothing other 
than a double T-duality along the 1-2 directions of the three torus.
\footnote{To recognize this interpretation of the mapping \symmetryi\!\!, 
see the translation code summarized on page \voetnootpagina. 
This interpretation of the Nahm transformation, as related 
to T-duality in string theory, was first suggested in \mikegreg.} 
From this perspective, it seems somewhat 
surprising that the Yang-Mills theory (via the Nahm transformation) 
still knows about this T-duality symmetry, despite arising
from string theory via the zero-slope limit. We will give 
an explanation for this in the next section. 

\newsubsection\nonamenond
{Nahm duality for gauge theory on noncommutative tori} 

We end this section with some remarks on Nahm-type duality 
transformations of gauge bundles defined on non-commutative tori. 
These gauge theories are relevant for compactifications
of matrix theory with non-zero $B$ fields \connes\douglashull. It is 
therefore of interest to investigate their duality symmetries. 
 
As we have seen above the Nahm transformation on commutative 
tori is somewhat involved. It requires knowledge about the 
zero modes of the Dirac equation. 
On non-commutative tori however, Nahm type transformations
can be made manifest \hofmani\hofmanii.
These transformations are part of a larger $SO(d,d)$ 
symmetry, that is known in the mathematics literature as Morita 
equivalence. 

The most simple example of a non-commutative space 
is the non-commutative two torus $\Tvet^2_\theta$. 
The coordinates of this torus do not commute, but they 
satisfy the commutation relation 
\eqn\noncomtorus{
[x^1,x^2] = {\theta \over 2\pi i} \quad {\rm or} \quad
U_1U_2 = e^{2\pi i\theta} U_2U_1. 
}
Here $\theta$ is a real parameter and $U_i = e^{2\pi ix^i}$ are  
exponentials that generate the Fourier modes 
of the functions which can be defined on the non-commutative torus. 

Trivial gauge bundles can be constructed by defining the 
gauge connection by $\nabla_j = \partial_j + iA_j(\tilde{x})$ 
where $\tilde{x}^i = x^i+{i\theta \over 2\pi} \epsilon^{ij}\partial_j$
are modified coordinates that commute with the original coordinates. 
The connection depends on the modified coordinates and not on the
coordinates $x^i$, because this guarantees that the connection 
commutes with the full algebra of functions and therefore obeys 
the correct Leibniz rule \hofmani\hofmanii.  

To construct a non-trivial gauge bundle with magnetic flux $M$ 
modified Fourier modes have to be used to construct a gauge field that
has the right properties 
(for example the translation operators should act like 
gauge transformation on the fields, as in \boundary\!\!). 
Now the interesting thing is that these modified Fourier modes themselves 
generate an algebra of an abelian gauge theory on a non-commutative 
torus, with modified $\hat\theta$-parameter that is 
related to the original $\theta$ 
via an fractional $SL(2,\Zvet)$ transformation \hofmani\hofmanii. 
The gauge fields and fluxes also transform in a representation of
$SL(2,\Zvet)$. We can turn this around, by starting 
with an abelian theory defined on a particular non-commutative torus, 
apply an arbitrary $SL(2,\Zvet)$ transformation on the gauge fields 
and the non-commutative parameter, and thus get a $SL(2,\Zvet)$ 
family of equivalent gauge bundles. A Nahm transformation
interchanging rank and magnetic flux of the gauge bundle is included 
in this $SL(2,\Zvet)$ group. 
Thus Nahm duality can be made manifest in gauge theories on non-commutative
spaces. 

Analogous results hold for gauge bundles 
on 4 dimensional non-commutative tori. 

The degeneracies of $1/4$ BPS states of $U(N)$ gauge theory on a
non-commutative three-torus have been calculated in \aschwarz. The 
result is the same as \generator in this chapter. This confirms the 
idea that the degeneracies of BPS states are rather robust quantities.

\newsection\noname{Relation to Born Infeld theory}

In this chapter we have shown that the multiplicities of 
single BPS-states of $\Nisf$ supersymmetric Yang Mills theory 
on $T^3$ has the extended $SL(5,\Zvet)$ $U$-duality invariance. 
This result may be remarkable, but it is clear that a possible 
explanation comes from the relation between Yang-Mills theory and 
Born-Infeld theory. Yang-Mills theory is a truncated theory of 
D-branes, whereas Born-Infeld is a more accurate low energy 
effective description.  

This is most clearly reflected in the BPS mass spectrum of the two 
theories: the BPS mass spectrum of BI theory compactified on a
three torus has the expected $SL(5,\Zvet)$ invariance, as has been 
shown in \hvz\gysbert. This duality symmetry is broken however in 
the limit $\alpha' \rightarrow 0$, that is needed to get 
SYM theory. Yet the symmetry in the degeneracies of irreducible BPS 
states remains preserved in the zero slope limit.
It is the aim of this last section to explain how this can happen.  

\newsubsection\nonamebi{Born Infeld BPS mass spectrum}

Let us first summarize the analysis that was done in \hvz\ 
and \gysbert\ where the BI-theory BPS mass spectrum was calculated 
explicitly. The spectrum was derived by starting from the bosonic 
action with abelian gauge group, (cf. \borninfeldaction\!\!)
\eqn\biaction{
S_{BI} = T_3 \int d^3x dt \ e^{-\phi} 
\sqrt{\det{(G_{\alpha\beta} + 2\pi\alpha' F_{\alpha\beta})}}.
}
For given electro-magnetic fluxes and momenta, the energies of the 
BPS states can be obtained from the BI Hamiltonian by a Bogomolny type
of argument \hvz. These masses are in agreement with the BPS 
masses that were obtained from the NS five-brane in \fivebraneo\fivebrane. 
In units where $2\pi\alpha'=1$ the BPS mass spectrum is 
\hvz\gysbert
\eqnar\BImasses{
E^2_{\rm\small BPS} = 
g_s^{4/5} (\det G) ^{-1/5}  \left(
{1 \over g_s^2}( N^2 \det G + G_{ij} m^im^j) 
+e^i G_{ij} e^j + p_i G^{ij} p_j\right) + \qquad
\nonumber \\  \\[-\the\baselinetruc] \cr \nonumber
2g_s^{-1/5}(\det G)^{-1/5} 
\sqrt{
\det G \ \kappa_i G^{ij} \kappa_j +
(p \cdot m)^2 + g_s^{2} (p \cdot e)^2 
}.\qquad \qquad
%\biggr)^{1/2}
}
Here $G_{ij}$ is the string metric.
Note that we have put the rank $N$ in the mass formula by hand.
The mass spectrum \BImasses 
is invariant under the complete $SL(5,\Zvet)$ symmetry. 
Part of this $U$-duality symmetry is the electric-magnetic $S$-duality
under which the coupling constant is inverted and the string metric
gets transformed \bergshoeff\ in the following way 
\eqn\noname{
g_s \rightarrow g_s^{-1}, \qquad 
G_{ij} \rightarrow {1 \over g_s} G_{ij}.  
}
The other non-trivial symmetry of the mass spectrum \BImasses is
a double $T$-duality (e.g. in the 1 and 2 direction), that 
as explained in section \nahmduality\!\!, can be interpreted in gauge 
theory, as a Nahm transformation.  
Under this duality the string metric and the string
coupling constant transform according to \bergshoeff
\eqn\trafonormal{
G_{11} \rightarrow G_{11}^{-1}, \qquad 
G_{22} \rightarrow G_{22}^{-1}, \qquad 
g_s \rightarrow g_s (G_{11}G_{22})^{-1/2}. 
}
When combined with the transformations of the fluxes it can be
easily checked that the Born Infeld BPS mass spectrum is invariant
under $S$-duality and Nahm transformations. Together with the manifest
$SL(3,\Zvet)$ symmetry of the torus 
these duality transformations generate
the complete $SL(5,\Zvet)$ symmetry, and therefore the mass spectrum 
\BImasses is indeed $U$-duality invariant. 
Analogous duality symmetries hold for Born Infeld theory
on 4 dimensional and 2 dimensional tori. For example in four
dimensions the symmetry group of the BPS mass spectrum is 
$SO(5,5,\Zvet)$, the $U$-duality group of string theory in 
6 dimensions \hvz.

The invariance can be made manifest by writing 
the BPS masses in the following way \gysbert
\eqn\noname{
E^2_{\rm\small BPS} = -{1\over 2}\tr({\cal G F G F}) +
2\sqrt{K_\rho {\cal G}^{\rho\sigma} K_\sigma}.
}
Here ${\cal F}$ is the $5\times 5$ matrix \fivematrix 
containing the fluxes, 
$K_{\sigma} = (Np_i - (e\wedge m)_i, p\cdot m, p\cdot e)$ 
is the five vector \kfive and ${\cal G}$ is defined by 
\eqn\metriccoupleii{
{\cal G} = 
\left(
\begin{matrix}
{(\det{g})^{-1/2}g_{ij} }
\cr\cr
& g_s (\det{g})^{1/4}
\cr\cr
&& g_s^{-1}(\det{g})^{1/4}
\end{matrix}
\right),
}
The metric $g_{ij}$ is related to the string metric $G_{ij}$ 
via 
\eqn\smallgbigg{
G_{ij} = g_s^{1/2} (\det g)^{-1/8} \ g_{ij}. 
}
On both matrices ${\cal F}$ and ${\cal G}$ the U-duality group 
$SL(5,\Zvet)$ acts by conjugation, and hence the $SL(5,\Zvet)$
invariance becomes manifest. 

It is instructive to extract the SYM BPS mass spectrum from 
Born Infeld theory. To this end we reintroduce the string scale
$\alpha'$ and expand the expression \BImasses 
\eqnar\noname{
E_{\rm\small BPS} = (2\pi\alpha')^{-5/4} NV^{1/2} + 
{1 \over 2NV}\biggl(
{1 \over g_s} m^i g_{ij} m^j + 
g_s e^i g_{ij} e^j + 
{2\pi\alpha'}^{5/4} V^{1/2} p_i g^{ij} p_j \biggr) + 
%\qquad \qquad \qquad 
\nonumber \\ \\[-\the\baselinetruc] \cr \nonumber
{1 \over N}\biggl( \kappa_i g^{ij} \kappa_j +
{2\pi\alpha'}^{5/4} g_s^{-1} V^{-3/2} (p\cdot m)^2 + 
{2\pi\alpha'}^{5/4}g_s V^{-3/2}(p\cdot e)^2
\biggr)^{1/2}
+ \cdots , 
}
where the volume is defined in terms of the metric, $V= \sqrt{\det g}$.  
This expansion shows that in the large $N$ limit (or large 
volume limit) and the string decoupling limit 
$\alpha' \rightarrow 0$ we recover the BPS mass spectrum of SYM, 
%
%
%
%\eqnar\BImassesi{
%E_{\rm\small BPS}^2 = 
%{1\over \alpha'}
%\left(
%(\det g)^{1/2} N^2  + 
%g_s^{-1}{g_{ij}\over (\det{g})^{1/4} } m^im^j +
%g_s{ g_{ij}\over (\det{g})^{1/4} } e^i e^j + 
%g^{ij} p_i p_j 
%\right) + \qquad \qquad 
%\nonumber \\ \\[-\the\baselinetruc] \cr \nonumber
%{2\over \alpha'} 
%(\det g)^{-1/8}
%\sqrt{ (\det g)^{3/4} \kappa_i g^{ij} \kappa_j +
%g_s^{-1} (p \cdot m)^2 + g_s (p \cdot e)^2
%} \qquad \qquad 
%}
%
%
\eqn\symmasses{
E_{\rm BPS} =
{1 \over 2NV}\left(
g^2_{\rm YM} e^i g_{ij} e^j  + 
{1\over g^2_{\rm YM}}m^i g_{ij}m^j\right) + 
\sqrt{p'_i g_{ij} p'_j},
}
where $p'_i = p_i -(e\times m)_i/N=\kappa_i/N$. 
We identified the string coupling constant $g_s$ with the 
Yang-Mills coupling constant squared 
\eqn\noname{
g_{YM}^2 = g_s.  
}
Obviously the SYM mass spectrum \symmasses is invariant under 
$SL(2,\Zvet)\times SL(3,\Zvet)$ but not under 
the complete $SL(5,\Zvet)$ $U$-duality group. 
For example the Nahm type transformation that exchanges rank $N$ and
electric flux $e_3$ does not leave the mass spectrum invariant. 
From the matrix theory point of view this is no surprise 
(at least for finite $N$) as for that
theory the flux $N$ is interpreted as a light-cone momentum, 
whereas $e_3$ is momentum in a spatial direction. M(atrix) theory 
is not likely to be invariant under the exchange 
of a light-like direction and a space-like direction, so we do
not expect the BPS mass spectrum to be invariant under Nahm-type
transformations. 

\newsubsection\nonamedeg{Degeneracies}

What about the degeneracies of the $1/4$ 
BPS states in Born-Infeld theory? 
For this theory the same philosophy that was used in the SYM case 
can be applied: first the phase space of the model is reduced to 
the space of classical supersymmetric configurations, and then 
this BPS reduced phase space is quantized. 

When doing this analysis an immediate important difference emerges
between the BI case and the YM case: 
As emphasized in section \mstproposal the superalgebra of SYM is equal
to the one of 11D supergravity in the DLCQ formalism, whereas the 
superalgebra of Born Infeld theory 
\footnote{We repeat here that the action for supersymmetric abelian 
BI theory has been constructed \aganagic. We make the assumption that
results for the non-abelian case follow from the analysis of the 
abelian model.}
is just the 11D superalgebra with 4 spatial directions 
compactified \gysbert. 
The last mentioned superalgebra has an $SL(5,\Zvet)$-invariance and 
therefore we expect that the degeneracies of the BI states exhibit 
this duality symmetry as well. Another difference is of course the 
square root in the Born Infeld action, that makes explicit
calculations more complicated. 

As in section \bpsmtheory the zero modes and fluctuations of the 
BPS states are restricted by the equation 
\eqn\biconi{
\left\{\bar {Q}, Q \right\} \epsilon =0, 
}
where $Q$ are the supercharges of supersymmetric Born-Infeld 
theory, and $\epsilon$ is a 16 component spinor. 
The restriction on the zero modes can be rewritten with the help 
of a projection operator \gysbert
\eqn\biconii{
(|K| + K_{i} \tilde{\gamma}_i ) \epsilon = 0, 
}
for some suitable $\gamma$ matrices. 
Here $K_i$ is again the five vector \kfive\!\!. 
In the $\alpha' \rightarrow 0$ limit this condition reduces to the 
condition \gamcon\!\!, which can be seen from the inverse 
of the 5 dimensional metric \metriccoupleii\!\!, which collapses to 
a 3 dimensional metric in the $\alpha' \rightarrow 0$ limit. 

Unlike the Yang-Mills case the BPS equations of Born-Infeld 
theory implied by conditions \biconi and \biconii\!\!, are in general 
quite complicated. However, in the special case that the flux 
configuration consists of only non-zero rank $N$ and momentum $p_3$, 
they lead to the simple chiral 2D free field equations \chiral
\gysbert. 
Any configuration of electric and magnetic 
fluxes can be mapped to this simple case by an appropriate 
$SL(5,\Zvet)$ transformation (see reference [17] in \gysbert). 
This transformation will also act on the gauge fields and the 
rank $N$, that can be combined into an antisymmetric matrix of the 
form \hbox{\fivematrix\!\!.} The component of the $5 \times 5$ matrix that is
supposed to represent the rank of a gauge group may therefore be
non-constant in the transformed matrix. Then the relation to a 
gauge theory seems to be obscured. 
This apparent problem can be solved, however, by noting that the BPS 
equations only depend on the quotients of the gauge fields and rank 
$N$. As argued in \gysbert\ this observation makes it possible to 
use $SL(5,\Zvet)$ transformations to generate BPS solutions for
general fluxes. 

The procedure just sketched extends to the BPS quantum theory
\gysbert. The degeneracy formula for the irreducible states 
(the long string states) thus obtained is equal to the one 
in \generator and therefore agrees with the M-theory result.

For the irreducible states (the long string states) 
both theories give the same counting formula \generator
for the multiplicities. 
A difference appears when we consider the second quantized BPS spectrum. 
Single BPS states can be combined into a second quantized BPS state, 
provided these single states preserve the same supersymmetries. 
In BI theory the condition \biconi implies 
that all 10 charges of the single BPS states align; 
in SYM only 9 charges have to align. 
An alternative way to arrive at this conclusion is by considering
the BPS energy (mass) spectra. The energies of BPS states add up when 
they are combined into a new BPS state. For SYM theory this precisely 
means that 9 charges align (not $p_3$, when $\kappa_i$ is turned in the 
three direction); for BI theory all 10 charges have to align, because
all fluxes come in quadratic form in the BPS energies. 

We thus see that Born Infeld gives more restrictions 
on the charges of single BPS states
to combine into a second quantized BPS state, than SYM does. 
As expected the BI conditions are relaxed and become equal to 
conditions in SYM, in the $\alpha' \rightarrow 0$ limit.
We therefore conclude that SYM 
contains more second quantized states than BI does. 

Despite the success in finding appropriate string $U$-duality 
symmetries in supersymmetric gauge theories, it is not clear whether 
these symmetries are actually genuine duality symmetries of the Born 
Infeld theories. The problem one has in proving electro-magnetic duality
for $\Nisf$ SYM theory appears here as well:
it is difficult to find an explicit duality mapping for the fields. 
Another problem arises when one investigates Nahm duality, namely
the rank $N$ is not the zero mode of any field in the gauge theory.  
It may be possible to embed the 3+1 BI theory into a 
six dimensional string theory in the light-cone gauge. 
One could then think of $N$ as the $p^+$-momentum of this theory. 
For a discussion of ideas in this direction see \gysbert.

\xdef\thechapter{3}
\newchapter\hesmst{High energy scattering \break in matrix string
theory}
{High energy scattering in matrix string theory}

\lhead[\fancyplain{}{\sc High energy scattering in matrix string
theory}]{\fancyplain{}{}}
\thispagestyle{empty} 
\newsection\introfive{Introduction} 

High energy processes in string theory were first considered 
from the point of view of conventional string perturbation theory 
by Gross and Mende \grossmende\ in the regime of fixed angle scattering
and in the near forward regime by Amati et al in \Ven.   
The recent insights from M-theory, however, have
provided a large number of new non-perturbative tools which can
now be used to put these works into a new perspective, and extend
the results into new directions. For instance,
it was long believed that the string length $\ell_s$ marks the minimal 
distance that can be probed via scattering processes in string theory. 
This belief was based on the fact that fundamental strings tend to 
increase in size when boosted to high energies, and thus appear to be
incapable of penetrating substringy distance scales. Since the discovery 
of D-particles as non-perturbative solitons of the IIA theory, however, 
we know that there exists small scale structure that, at least for weak 
string coupling, extends well below the string length
\Shen\danielsson\kabat\dkps. 
This particular realization provided important motivation for
the matrix theory conjecture of \bfss\ 
% \footnote{For  reviews see 
% \BiSu\Tayll\ and references therein.}
that all localized excitations 
of M-theory  (including the fundamental strings) are representable as 
multi-D-particle bound states \Polc\wittenbound.

In this chapter we begin a study of high energy processes in type IIA 
string theory, by making use of this matrix theory formalism.
We focus on the four graviton scattering amplitude, and in particular 
we will present a detailed calculation of the pair production rate of 
D-particles via this process. Our aim is to probe in this way the 
transition region between the conventional perturbative string regime 
and the strong coupling regime described by 11-dimensional M-theory
(see figure \thechapter.1). 

From the ten dimensional perspective of IIA string 
theory, D-pair production is an inelastic scattering process, in which 
two strings exchange one unit of D-particle charge. It is inherently 
nonperturbative and thus inaccessible to conventional perturbative methods.
It is also inaccessible in the traditional matrix theory approach since
the anti-D particles are boosted to infinite energy.

From the eleven dimensional perspective, on the other hand, 
the D-pair creation process can simply be thought 
of as the elastic scattering of two particles in which one unit of 
Kaluza-Klein momentum in the 11 direction is exchanged. Via this 
interpretation, one can rather straightforwardly obtain a tree level 
estimate of the probability amplitude. This estimate should be reliable
for large values for the $S^1$ compactification radius $R_{11}$ and for 
collision energies sufficiently below the 11-dimensional Planck energy. 
At high energies and/or small values for $R_{11}$, on the other hand,
we expect the physics of the scattering process to be quite different 
from (semi-)classical supergravity. In the following we will attempt 
to gain more insight into this regime via the matrix string approach.

%%%%%%
\epsfysize=8cm
\hfigure{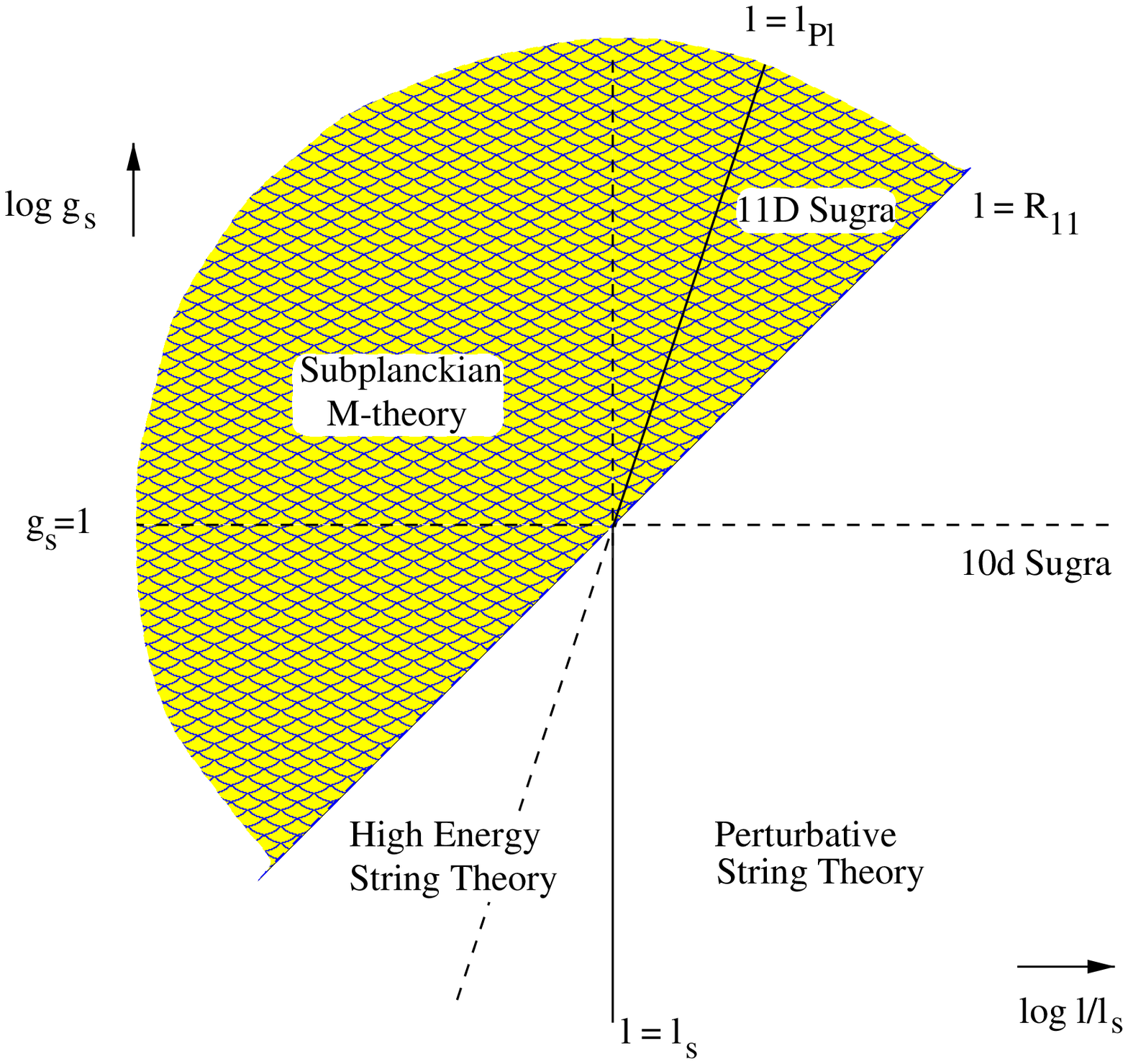}
{The phase diagram of the $S^1$ compactification 
of M-theory, with horizontal axis the log of the length scale and 
vertical axis the log of the string coupling. The various perturbative
and low energy limits are indicated. The shaded region marks the
regime where D-pair production is expected to be the dominant high
energy process.} 
\xdef\domifig{\thechapter.\the\fignumber\ }
%%%%%%

\noindent
Matrix string theory arises from the original matrix theory proposal 
\bfss\ via compactification on a circle, and starts from the action 
of 1+1-dimensional maximally supersymmetric Yang-Mills theory 
with gauge group $U(N)$. Via the identification of the eigenvalues of 
the matrices $X^I$
with the transverse location of type IIA supersymmetric strings, this SYM
model can be reinterpreted as a non-perturbative formulation of light-cone
gauge IIA string \motl\tomnati\matrixstring.  In this correspondence, the
string coupling constant $g_s$ is inversely proportional to the Yang-Mills
coupling $g_{{}_{\! YM}}$ (cf. \gaugestringc\!\!) 
and the free string limit therefore arises in the
strong coupling limit of the Yang-Mills model. 
This correspondence has been worked out in some detail in 
section \mstproposal\!\!. 
 
More generally, however, all regimes of the $S^1$ compactification 
of M-theory, as indicated in figure \domifig
should according to the matrix string
conjecture of \bfss\motl\tomnati\matrixstring\ via the above 
identifications be described by particular regimes of 
(the large $N$ limit) of the 1+1D supersymmetric gauge theory.

In particular, it is expected that in the weak coupling, moderate energy
limit 
of the SYM theory it effectively reduces to the matrix quantum
mechanics description of 11-dimensional supergravity. Indeed, a new feature
of matrix string theory (relative to standard light-cone string theory) is
that via the electric flux of the gauge field, the string states can be
adorned with an extra quantum number, identified with the D-particle charge
\matrixstring.  In a small $g_s$ expansion, these flux sectors 
energetically decouple, corresponding to the fact that
D-particles can not be produced via perturbative string interactions.
Nonetheless, electric flux can get created in the gauge theory:
it is easy to see that electric flux creation is a simple one-loop
effect that takes place whenever a virtual pair of charged particles gets
created and annihilated, after forming a loop that winds one or more times 
around the $\sigma$ cylinder.

In the following we will develop a new method for studying high energy
scattering and D-pair production in matrix string theory, which
will be based on a semi-classical expansion from the SYM perspective.
An important novelty of this method is that it applies to processes with 
arbitrary longitudinal momentum exchange. 
In the gauge theory language, this means that 
the transitions between the initial 
and final states that we will consider will involve a non-perturbative 
tunneling process in which an arbitrary number 
of eigenvalues get transferred between the two scattering states. 
Most previous calculations in matrix theory relied on perturbative
SYM corrections and thus were necessarily restricted to zero $p^+$ 
transfer. \footnote{In \polpou\ Polchinski and Pouliot analyzed graviton
scattering with non-zero M-momentum transfer in matrix theory.  
In their case,  the M-momentum was identified with the magnetic 
flux of the SYM gauge theory, and the corresponding instanton was 
a magnetic monopole. Here we will consider different kind of momentum
transfer, namely of longitudinal momentum represented 
by the size  $N$ of the matrix bound states, {\it i.e.} the number of 
D-particles in the original matrix dictionary of \bfss. 
This will require a different, less familiar type of instanton
process.}

Concretely, we will construct SYM saddle point configurations that
will allow us to interpolate between in-going matrix configurations 
of the form
\eqn\instate{{\vec X}_{in}(\tau) = \hf
\left(
\begin{matrix}
(\frac{\vec p_1}{N_1}\tau + {\vec b})I_1 &0\cr
0&({{\vec p_2}\over N_2}\tau -{\vec b})I_2\cr
\end{matrix}
\right)\ }
and outgoing configurations of the form
\eqn\outstate{{\vec X}_{out}(\tau) =
\left(
\begin{matrix}
({{\vec p_3}\over N_3}\tau + {\vec b})I_3 &0\cr
0&({{\vec p_4}\over N_4}\tau -{\vec b})I_4 \cr
\end{matrix}
\right), }
where $I_i$ are $N_i {\times} N_i$ identity matrices, where all $N_i$'s
are {\it different} (but subject to the momentum constraint 
$N_1\! +\! N_2 = N_3\! +\! N_4$).  
These {\it in} and {\it out} configurations each
describe two widely separated gravitons with different light-cone 
momenta
\eqn\noname{
p^+_{(i)} = N_{(i)}/R 
}
and transverse momenta $\vec{p}_{(i)}$, 
and with relative impact parameter ${\vec b}$. 

The interpolating solutions that we will construct, essentially look
like an appropriate matrix generalization of perturbative string
world-sheets.  The importance of these solutions is not entirely obvious,
however, since a priori one would expect that the range of validity of the
semi-classical Yang-Mills approximation has no overlap with that of
perturbative string theory. Indeed, as emphasized in section 
\mstproposal 
the two regimes appear related via a strong/weak coupling duality. 
However, as we will argue in the following, even at small or moderate 
string coupling $g_s$, at sufficiently high collision energies and/or 
impact parameters one enters a regime in which the semi-classical SYM 
methods may provide an accurate description of the scattering process.

Just like string/M-theory, the 1+1 SYM model contains various length
scales: (i) the circumference of the cylinder,
(ii) the scale set by the Yang-Mills coupling 
$\ell_{{}_{YM}} = 1/g_{{}_{YM}}$, in units where the circumference of
the cylinder is set equal to $1$
\eqn\ymcoup{
\ell_{{}_{YM}}\simeq g_s,
}
(iii) 
the typical mass scale set by the Higgs expectation values of the SYM model. 
The latter length scale is inversely proportional to the
impact parameter $b$ of the string/M-theory scattering
process:
\eqn\bscale{
\ell_{b} \simeq {g_s\over b}\ ,
}
in units where $2\pi\alpha' =1 $. 
%There is also another scale arising from the instanton analysis that we
%will perform later in the paper. 
Finally, (iv) there is also the length scale $\ell_E$ 
determined by the typical size of the SYM energy $E$, 
which is related to the relative space-time momenta via $E\simeq p^2/N$. 
%\eqn\Escale{
%\ell_{E} \simeq {N\over p^2}.}

The existence of these scales allows us to find small
dimensionless ratios that may parameterize the strength of the SYM
processes taking place at that scale.  For example, 
while
$g_{{}_{YM}}=1/g_s$ defines the effective coupling of SYM processes 
that take place at the scale of the YM cylinder, we also have
\eqn\gYMb{
g^{eff}_{{}_{\! YM}}(b)\, \simeq \,
 \ell_b/\ell_{{}_{YM}} \, \simeq \, 1/ b,}
as the dimensionless coupling at the scale $\ell_b$.  
%
%Corresponding to these various length scales, there are an equal number
%of dimensionless ratios that parameterize the effective strength of the
%SYM interactions at these length scales. Specifically, while
%$g_{{}_{YM}}=1/g_s$ defines the effective coupling of SYM processes 
%that take place at the scale of the YM cylinder, we also have
%%
%\eqn\gYMb{
%g^{eff}_{{}_{\! YM}}(b)\, \simeq \,
% \ell_b/\ell_{{}_{YM}} \, \simeq \, 1/b,}
%%
%and
%%
Similarly, we can also associate an effective coupling
$g^{eff}_{{}_{\! YM}}(E)$ with the scale set by the
SYM energy $E$.
%\eqn\gYME{
% \, \simeq \,
%\ell_E/\ell_{{}_{YM}}\, \simeq \, {N\over g_s p^2}. 
%}
%%
This suggests the possibility that even if $g_s$ is small or of order 1,
processes at these other 2D length scales can be accurately described by
perturbative and/or semi-classical SYM methods. This will require however
that we consider the limit of high collision energies and sufficiently
large impact parameters.\footnote{In this context it may be of
relevance that 
in classical 10-dimensional DLCQ supergravity, the impact parameter 
$b$ scales with the transverse relative momentum $p$ via $b 
\simeq \left(g_s^2 p^2/ N \sin \theta \right)^{1/6}$ with $N$ the DLCQ 
$p_+$-momentum.
Hence, at least in this classical context, and for fixed scattering angles 
$\theta$ and $g_s$ of order 1, the condition that $b$ is large is 
automatically satisfied in limit of large $p^2\gg N$.}

The two types of processes that we will consider, high
energy scattering with non-zero $\Delta p_+$ and the 
D-pair production, may at first sight seem quite unrelated.
% The first is a matrix string generalization of
%Gross and Mende's treatment of fixed angle scattering.  
%This generalization arises from an instanton in the two-dimensional 
%SYM theory.  The second is the pair creation of
%D-charge.  From the string point of view this is an intrinsically
%non-perturbative process.  In the SYM theory on the other hand it is
%described by a perturbative process
%in which one of the massive charged states winds around the cylinder,
%thereby creating an electric flux.  
However, there are several connections between these two types of 
processes. First of all, it is worth pointing out that in both 
cases the scattering process involves (depending on which duality frame
one chooses) the transfer of D-particle charge and/or momentum between 
the two scattering particles. Indeed, the rank $N$ started out as 
identified with D-particle charge, and only after the duality 
it and the electric flux $E$ are mapped onto each other under an 11-9 flip: 
i.e. the interchange of the 11-th and 9-th direction (a detailed 
explanation on this flip was given in section \mstproposal\!\!).
Hence quantitative understanding both types of processes will
have a direct bearing on the Lorentz invariance of the matrix
formalism.

%%%%%%
\epsfysize=4.2cm
\hfigure{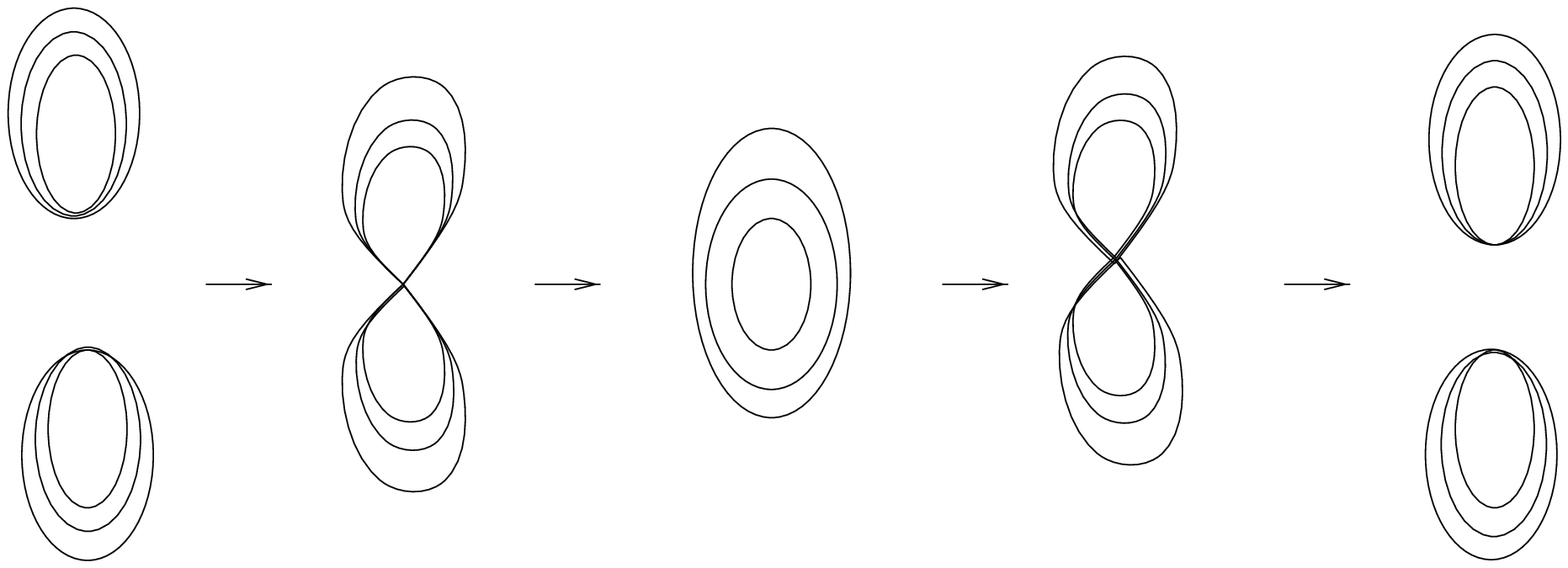}
{This figure depicts a typical saddle point trajectory
that contributes to the high energy scattering amplitude of
fundamental strings, according to the perturbative physical picture 
proposed in \grossmende.}
\xdef\nstringfig{\thechapter.\the\fignumber\ }
%%%%%%

Another, more speculative 
connection between the two calculations is related to
a fundamental puzzle in the original calculation of \grossmende, 
namely the apparently dominant contribution of arbitrarily high genus 
to scattering amplitudes.  The saddle point trajectory at loop 
order $G$ typically describes a process as depicted in 
figure \nstringfig two 
incoming strings, that are wound $N=G+1$ times, interact and then 
propagate as $N$ intermediate short strings. The $N$ strings then  
join together again, producing a final state of two different 
$N$ times wound outgoing strings (see fig. \nstringfig\!\!). 
It was found in \grossmende\ that the contributions of these higher 
order interactions 
grows larger with the genus $G$. This instability appears to signal a 
fundamental breakdown of conventional string  perturbation theory in the 
high energy regime.

On the other hand, the fragmented form of the intermediate state 
in figure \nstringfig
gives a strong hint of some underlying non-perturbative 
structure that looks quite similar to that of the multi-D-particle 
bound state dynamics of matrix theory.  This suggests that the
matrix treatment may provide a rather natural stabilizing mechanism 
for a cutoff on the genus. Furthermore, our study will show that
D-particle pair production becomes relevant at this cutoff -- when 
the strings become maximally fragmented. This leads us to
suspect a deeper relation between Gross and Mende's high energy, fixed
angle scattering and the non-perturbative process of D-pair creation.

We will begin this chapter with a quick review of the kinematics 
of fixed angle scattering in the traditional string framework, 
with particular emphasis on its description in the light-cone gauge.
This will be followed by a discussion of the Gross Mende saddle
points, first at tree level and then generalized to arbitrary 
genus $g$ world-sheets.  
Then we turn to string interactions in the matrix 
string formulation. In particular, we will find a local 
instanton solution in the two-dimensional theory that describes the 
splitting/joining interaction.
Furthermore, the condition for this instanton to be matched to the
incoming/outgoing states is precisely that we be working at the 
world-sheet moduli corresponding to the saddle-point surfaces of 
reference \grossmende. 

After this we will briefly discuss recent results of \wynteriii, 
where by a detailed zero mode analysis of the interaction instanton, 
the four tree-level string scattering amplitude was reproduced
exactly, up to a numerical factor. 

We then turn to D-pair production, which we consider both from 
the supergravity perspective as well as via a one loop
Yang Mills calculation valid for arbitrary $N$.  This is followed
by a discussion of the ranges of validity of our calculations. In
particular, we suggest that the ranges of validity of the two calculations
overlap, and allow the picture we have mentioned, in which they complement
each other.  This connection is further discussed in the final section,
together with some other observations and speculations.

\newsection\fixedangles{Fixed angle scattering of strings}

High energy, fixed angle processes in superstring theory were first 
studied in detail from the point of view of conventional string 
perturbation theory by 
Gross and Mende \grossmende. Central to their approach is the observation 
that in the limit of large external momenta, the Polyakov path integral
at each given perturbative order is dominated by a finite number
of saddle point configurations. Furthermore, it was proposed 
that all these saddle points essentially describe the same preferred 
world-sheet trajectory, up to an overall factor depending on the 
loop order.

In the subsequent sections we will find independent evidence 
from the point of view of matrix string theory that supports this 
physical picture. In addition we will give a useful characterization 
of the Gross-Mende saddle points in terms of the light-cone gauge 
formulation of string perturbation theory. 

\newsubsection\nonamekin{Kinematic relations for four string scattering}

It will be useful to first establish a few kinematic relations 
of the tree level diagram that describes the scattering of four 
external massless particles with light-cone momenta 
$p^+_i = N_i/R$ and transverse momenta ${\vec p}_i$. 
\\
%%%%%%
\epsfysize=6cm
\hfigure{kinema2.eps}
{This figure indicates the kinematics of the transverse momenta $p_i$.}
%%%%%%
\\
For definiteness, we will describe this process in the center of mass 
frame in the transversal direction  
\eqnar\cmframe{
{\vec p}_1 + {\vec p}_2 &=& 0,
 \nonumber \\ \\[-\the\baselinetruc] \cr \nonumber
{\vec p}_3 + {\vec p}_4 &=& 0. 
}
The four transversal momenta $p_i$ 
can all be chosen to lie within one given plane. We can thus write
all $p_i$ as complex numbers. In addition,
longitudinal momentum and energy conservation imply that 
\eqn\sumn{
N_1+N_2 = N_3 + N_4,
}
and
\eqn\sumpm{
{|p_1|^2\over N_1}+{|p_2|^2\over N_2}={|p_3|^2\over N_3}+{|p_4|^2\over
N_4}.
}
For a given set of locations $z_i$ of the corresponding vertex operators,
the classical location of the world-sheet is described by the equations
\xplusn and \xtrans
\eqn\xplusntrans{
X^+(z,\bar{z}) = {1 \over 2} \sum_i \epsilon_i N_i \log| z-z_i|^2 
,\qquad 
X(z,\bar{z}) = {1 \over 2} \sum_i \epsilon_i p_i \log|z-z_i|^2. 
}
In the light-cone gauge, one chooses a fixed 
world-sheet parameterization by identifying $X^+$ with the world-sheet
time $\tau$, which via \xplusntrans amounts to setting (after a 
rescaling) 
\eqn\lcgauge{
w\equiv \tau +i \sigma = {1\over 2\pi}  \sum_{i} \epsilon_i N_i \log(z-z_i)\ .
}
The differential $\omega=dw$ (see \localcoord\!\!) 
is a specific globally defined holomorphic 
differential on the world-surface; existence and uniqueness of such a 
differential at arbitrary genus \GiWo\SBGTasi\  
generalizes the construction to higher loop amplitudes.  
Notice that (due to the branch cuts in the logarithm) 
the coordinate $\sigma$ in \lcgauge is defined on an interval
$0\leq \sigma <  (N_1+N_2)$, in accord with the rescaling of the 
world-sheet coordinates with total light-cone momentum $p^+$. 

The light-cone coordinate system \lcgauge specifies a particular 
time-slicing of the string world-sheet. 
As argued in section \lightconebasics\!\!, in this coordinate frame there 
are specific points on the world-sheet at which strings split or join. 
Namely, these interactions take place at zeros of $\omega$, that is 
critical points $z=z_0$ of the light-cone coordinate $X^+$. 
Inserting the explicit form \xplusn for $X^+$ then gives the condition
\zsolve that can be reduced in the specific case of the four-point 
scattering amplitude, to an equation relating the interaction point and the
M\"obius invariant cross ratio 
\eqn\moebius{
\lambda = {(z_1-z_3)(z_2-z_4) \over (z_1-z_2)(z_3-z_4)}, 
}
via
\eqn\znsolve{
{N_1\over z_0} + {N_2\over z_0-1} = {N_3\over z_0-\lambda}\ .
}
For given $\lambda$, this is a quadratic equation for $z_0$ with in 
general two solutions $z_0^+$ and $z_0^-$, representing the simple
splitting and joining interaction respectively.

\newsubsection\nonamegm{Gross Mende saddle points}

Now we are ready to discuss the Gross-Mende saddle point. 
Already in chapter one we wrote down the typical (bosonic) 
$g$-loop amplitude of $n$ scattering particles in string theory 
\expectation\!\!. The only place where the external momenta $p_i$ 
enter in the amplitude is the exponential of   
\eqn\coulomb{
{\cal E}_C(p_i,z_i,m_I) = {1 \over 2} 
\sum_{i < j} p_i \cdot p_j G_m(z_i,z_j).
}
Here $z_i$ are the insertion points of the vertex operators and 
$G$ is Green's function associated to the $g$-loop 
world-sheet with moduli $m$. In the limit of large external
momenta, the Polyakov path integral is dominated by saddle point 
configurations that minimize the exponential \coulomb\!\!. 
This minimizing problem has an elegant physical interpretation: 
it is equivalent to finding a semi-stable
configuration of two-dimensional Minkowski charges $p_i$ 
placed at positions $z_i$ on a Riemann surface of genus $g$ and 
moduli $m$ \grossmende.

In the simplest case of four string scattering at tree level, 
the Green's function in \coulomb is simply given by 
the logarithm $G(z_1,z_2) = \log|z_{12}|$ on the plane, so that 
the Coulomb energy \coulomb becomes 
\eqn\coutreelevel{
{\cal E}_C = {1 \over 2} \sum_{i <j} p_i \cdot p_j \log |z_i -z_j|^2.
}
Due to conformal invariance, the energy ${\cal E}_C$ depends on 
the locations $z_i$ of the vertex functions only by means of 
the cross ratio $\lambda$ defined by \moebius\!\!. 
The variation of ${\cal E}_C$ with respect to this variable 
$\lambda$ reads 
\eqn\ecvar{
\partial_\lambda {\cal E}_C(\lambda) = 
{p_1\! \cdot\! p_3\over \lambda} + 
{p_2\! \cdot\! p_3\over \lambda-1}\ .}
The saddle point equation $\partial_\lambda {\cal E}_C=0$ 
is solved by
\eqn\saddle{
\lambda = {p_1\! \cdot\! p_3\over {p_1\! \cdot\! p_2}} = {t\over s}\ ,}
where $s,t$ are the usual Mandelstam variables. 
The saddle point 
corresponds to a particular classical world-sheet
trajectory which at high energies gives the dominant contribution to
the scattering amplitude.

For later reference, it will be useful to translate the above description
of the GM saddle point into the light-cone gauge language.
To begin with, in the complex parameterization for the $p_i$, 
the Mandelstam parameters $s$ and $t$ are
expressed as 
\eqn\mandelsti{
s = - 2 p_1\! \cdot\! p_2 = (N_1 + N_2)^2{|p_1|^2\over N_1 N_2}, 
} 
\eqn\mandelstii{
t = - 2 p_1\! \cdot\! p_3 = {|N_3 p_1 - N_1 p_3|^2\over N_1 N_3}, 
}
so that \saddle takes the form
\eqn\saddl{
\lambda =  {N_2\over N_3} \, {|N_3 p_1 \!- N_1 p_3|^2\over  
(N_1+N_2)^2 |p_1|^2} .
}
Together with \znsolve\!\!, this saddle point specifies a particular
set of locations for the two interaction points $z_0^\pm$ of the light-cone
string diagram. We now claim that this preferred location of the 
interaction points $z=z_0^\pm$ is singled out by the requirement that,
in the immediate neighborhood of $z=z_0^\pm$,
the transverse coordinate fields $X(z)$ are 
\hbox{(anti-)analytic} functions of $z$ 
%
%\eqnar\bpscond{
%\partial_z X\mid_{z=z_0}
%%{\strut \Bigl{|}{\textstyle z=z^+_0}} 
%\is 0 \cr
%\partial_\zbar \Xbar\mid_{z=z_0}
%%{\strut \Bigl{|}{\textstyle z=z^+_0}} 
%\is 0\ .}
%%
%%
\eqn\bpscond{
\partial_z X\mid_{z=z_0} =0, \qquad 
%{\strut \Bigl{|}{\textstyle z=z^+_0}} 
\partial_\zbar \Xbar\mid_{z=z_0} =0 .
%{\strut \Bigl{|}{\textstyle z=z^+_0}} 
}
To verify this claim, let us compute the cross ratio $\lambda$ from 
\bpscond\!\!. The result should be equal to \saddle\!\!.  
Inserting the solution \xtrans into \bpscond gives
\eqn\bpssum{
\sum_i {\epsilon_i p_i\over z^+_0-z_i} = 0.
}
In terms of the cross-ratio $\lambda$ defined in \moebius this reads
\eqn\bpscross{
{p_1\over z^+_0(z^+_0-1)} =- {p_3 \over z^+_0 - \lambda},
}
where we used that $p_1+p_2 = 0$.
When combined with the equation \znsolve\!\!, which relates $\lambda$ with 
the location of the interaction points $z_0$, this equation can  
be used to compute $\lambda$ in terms of the scattering data. 
If we subtract $N_3$ times \bpscross from $p_3$ times \znsolve\!\!, 
we obtain a linear equation for $z_0$, solved by
\eqn\znot{
z^+_0 = {N_1p_3 - N_3 p_1 \over (N_1+N_2) p_3}\ .}
%{N_1 \over N_1+N_2} + {p_1\over p_3} {N_3\over N_1+N_2}}
%
Further, from \bpscross\ we find that
\eqn\zznot{
\lambda  = z^+_0 \Bigl(1+ {p_3 \over p_1} (z^+_0-1)\Bigr)\ .}
After inserting \znot into \zznot\!\!, it is a simple calculation to
verify that the resulting expression for $\lambda$ coincides with the
high energy saddle point \saddle\!\!.
Note that for the saddle-point configuration, $\lambda$ is in fact 
real. The interaction points $z^+_0$ and $z^-_0$ are in this case
each others complex conjugate.

We conclude that for Gross-Mende saddle point world-sheets the
transverse coordinate fields behave holomorphically near the interaction
points. This light-cone characterization of the GM saddle point
in terms of the holomorphicity conditions \bpscond will be critical 
in establishing the implementation of high energy scattering in the 
matrix string context.

\newsubsection\nonamehigh{Higher genus contributions} 

In general extending the results to higher genus is not straightforward. 
Especially finding the dominating saddle points will not be 
so easy. The one-loop case is still relatively simple. Using the
explicit form of the Green function on the torus, one can again 
extremize the electrostatic energy \coulomb of four charges placed on 
the torus. The resulting extremized energy is a factor two smaller
than the tree level energy \grossmende, 
and surprisingly the saddle point is exactly the
same as in the tree level case. 

There is little hope to do successful analogous  
calculations for higher genus diagrams. 
However in \grossmende, Gross and Mende proposed 
the following attractive 
generalization of the saddle-point to higher orders in the string 
perturbation expansion. They assume that the dominant saddle points 
at genus $g$ take the form of an $N=g+1$ fold cover of the same 
four-punctured sphere as described above, branched over the four 
locations $z_i$ of the vertex operators. The resulting surfaces are 
known as $\Zvet_N$ curves. 
Such a $\Zvet_N$ curve is defined by the polynomial expression
\eqn\zncurve{
y^N = \prod_{i=1}^L(z-z_i)^{L_i}. 
}
This curve represents an $N$ sheeted covering of the complex plane, 
with $L$ branch points of order $N-1$ 
(provided $\sum_i L_i =0 \mod N$ and the $L_i$'s relatively prime
with respect to $N$). In case of four string scattering $L=4$. 

When we place electric charges $p_i$ at the branch points the total 
electric field takes the same value on all $N$ sheets, 
and is given by 
\eqn\electricfield{
E_N(z) = E^x_N-iE^y_N = {1 \over N} \sum_i {p_i \over z-z_i}. 
}
It is straightforward to show that \electricfield is the right 
expression for the electric field. 
It is a conservative field and it moreover satisfies Gauss's law. 
The last property can be easily checked by considering a cycle 
that encloses a region that contains a branch point. This 
cycle will encircle the branch point $N$ times, one time on each 
sheet. Thus the contribution of ``surface'' integral of the
electric field will be equal to the charge enclosed. So the 
field defined in \electricfield obeys Gauss's law.  

The electrostatic energy is given by 
\eqn\energyn{
{\cal E}_N = -{1 \over 2N} \sum_{i < j} 
p_i\cdot p_j \log|z_i-z_j|^2,
}
that is $1/N$ times the energy of the tree level case.
Obviously 
extremizing \energyn with respect to the M\"obius invariant $\lambda$
will yield the same saddle point \saddle we found before. This
confirms the earlier mentioned claim that at all orders the scattering 
is dominated by the same saddle point \saddle\!\!. 

For the $\Zvet_N$ curves the classical spacetime surface 
swept out by the strings at loop level $N$ is  
\eqn\spacetimesurface{
X^\mu  = {1 \over 2N} \sum_i \epsilon_i p^{\mu}_i \log|z-z_i|^2, 
}
where $\epsilon_i =\pm 1$ for incoming respectively outgoing strings. 
We see that the coordinate fields grow with increasing energy, so 
that strings cannot be used as probes to explore string theory
at small distances (high momenta).
The classical trajectory of a higher order saddle point
\spacetimesurface has the same shape as the tree level 
trajectory, but its size is $N$ times smaller. 
The intuitive reason is that they describe multiple wound strings,  
so that the effective string tension is $N$ times bigger than usual. 
Correspondingly, since the different trajectories are 
weighted by the world-sheet area, the higher order 
trajectories give contributions proportional 
to $e^{-{\cal E}_C/N}$ (with ${\cal E}_C$ 
defined by \coutreelevel\!\!). The higher genus contributions are 
thus quite strongly enhanced at high energy. 

It is worth pointing out that the structure of the $\Zvet_N$ curves and 
the corresponding space-time trajectories, as depicted in 
figure \nstringfig\!\!,  are quite reminiscent of the description of 
the ``long string'' boundary 
conditions in section \stringorbifold\!\!. 
In our view, this (proposed) structure 
of the higher order interactions is one of several indications 
that the Gross-Mende approach to high energy string scattering may 
have a natural implementation in the matrix string context.

\newsection\mstfirst{Matrix string interactions}

In this section we will prepare the ingredients for the semi-classical 
study of high energy scattering in the matrix string framework. 
To begin with, we notice that the above light-cone gauge description 
of the dominant string world-sheet trajectories can rather easily be 
put into a matrix form, by the procedure introduced in section 
\towardsmst\!\!. Starting from equations \xplusn 
and \xtrans\!\!, we represent the classical string trajectory by means 
of a diagonal $N\times N$ matrix (with $N=N_1+N_2$) by first writing 
the transversal coordinates $\vec{X}$ as a function of $w$ defined 
in \lcgauge\!\!, and then ``roll up'' the spatial interval 
$0\!\leq \!\sigma\!<\! N$ onto the short interval 
$0\!\leq\! \sigma\! <\! 1$.
Concretely, we define the diagonal matrix elements of $X(\sigma)$
via $X_{kk}(\sigma) = X(\sigma+ k)$, and  in this way we indeed 
create matrix configurations that, away from the interaction times, 
satisfy the long string boundary condition \perio and \cycle\!\!.
\\
%%%%%%
\epsfysize=3cm
\hfigure{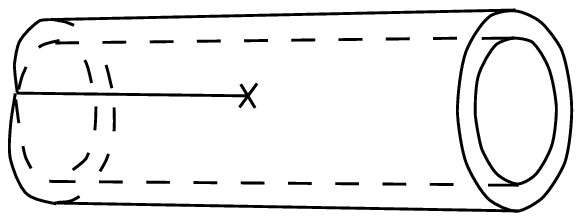}
{The string interaction relating a one string to a two string state.
This interaction occurs when two eigenvalues $X_I$ and $X_J$ coincide,
we enter a phase where an unbroken $U(2)$ symmetry is restored.
Figure taken from \wynteri.}
%%%%%%
\\
These diagonal matrix configurations represent particular solutions
to the SYM equations of motion, that are regular everywhere {\it except}
at the interaction points.  If at some point in the $(\sigma,\tau)$ 
plane two eigenvalues $X_I$  and $X_J$ coincide, we enter a phase where 
locally the gauge symmetry is restored to $U(2)$. In general we should 
thus expect that in this local region the semi-classical
SYM solution will need to become truly non-abelian. 

It is readily seen that the diagonal matrix configurations constructed
via the above procedure from the CFT solution \xtrans is not single valued
around the interaction points. Instead, as explained in section 
\mstinteraction\!\!, in going around the interaction point, the matrix $X$
undergoes a simple transposition of the two degenerating eigenvalues. 
This transposition changes the long string boundary conditions and 
therefore corresponds to an elementary string interaction. 

In the gauge theory language, the diagonal CFT 
solution \xtrans in fact hides a delta-function Yang-Mills curvature
at the interaction point, such that the infinitesimal Wilson line 
around it coincides with this permutation group element \wynteri. 
In this section we will describe how the Yang-Mills dynamics 
smoothes out this singularity.
\\
%%%%%%
\epsfysize=5cm
\hfigure{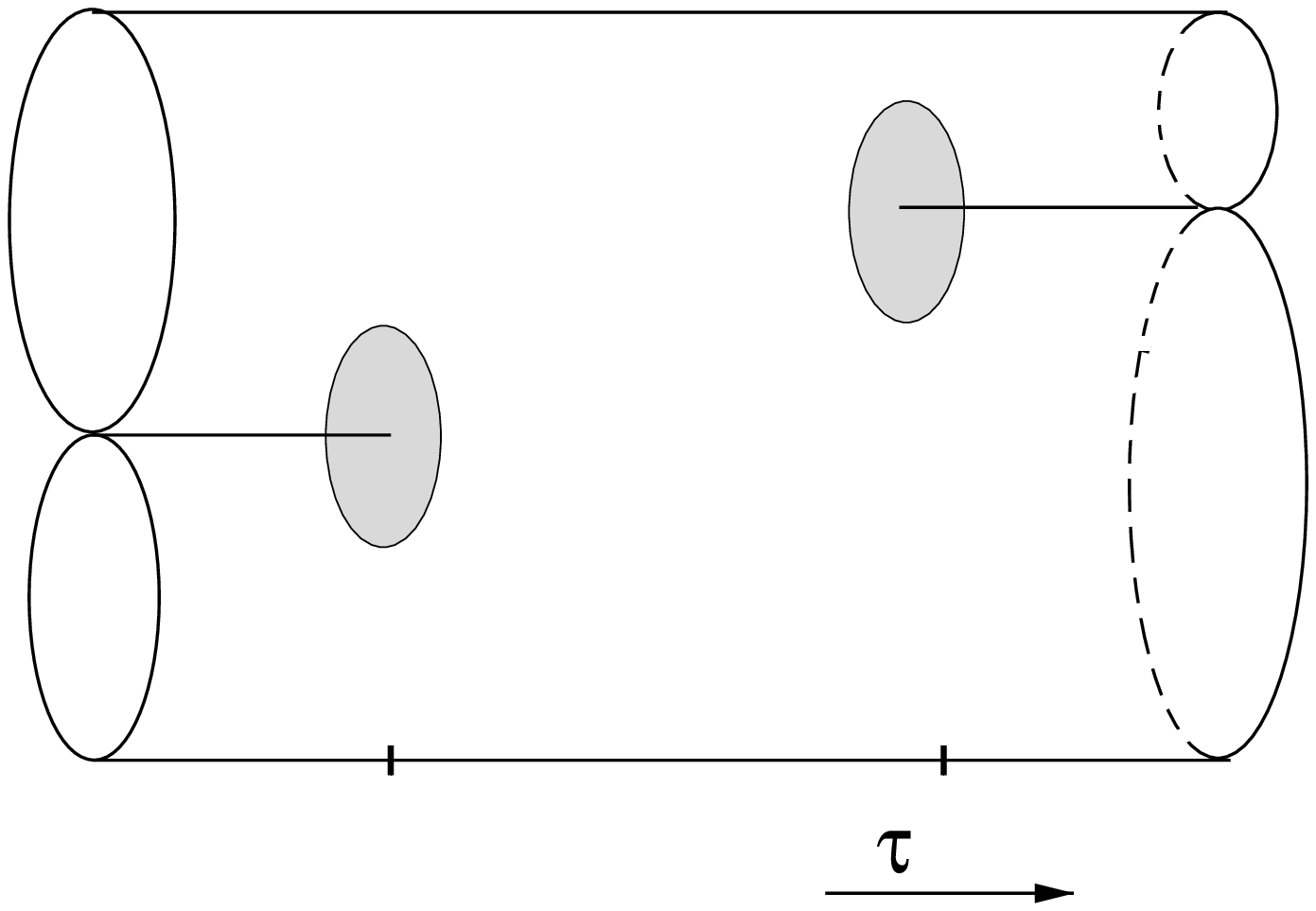}
{The shaded areas are the interaction regions where the 
singularity of the abelian string configuration is smeared 
out by an instanton-like solution. The diagram corresponds to
a four string scattering process at tree level.}
\xdef\shadedinstanton{\thechapter.\the\fignumber\ }
%%%%%%
\\
Concretely, we will now exhibit a smooth and single-valued Yang-Mills 
configuration that describes the local splitting or joining of one or
two matrix strings. The idea is to glue this solution into the abelian
CFT configuration in a neighborhood of the interaction points
(see figure \shadedinstanton\!\!). The conditions we have to impose for 
this gluing procedure will help us to find the Yang-Mills
configuration. 

Ultimately, we will be interested in obtaining global
classical solutions to the SYM equations of motion that minimize the
Yang-Mills action for given asymptotic conditions on the matrix fields $X$,
as written in equations \instate and \outstate in the introduction.

\newsubsection\nonamesym{SYM Solution near interaction point}

It seems reasonable to assume that, at least in the immediate
neighborhood of the interaction point, these minimal action
configurations of the SYM model are described by supersymmetric
configurations. Hence, instead of trying to solve the full Yang-Mills
equations, we will restrict ourselves to the special class
of solutions satisfying a dimensionally reduced version of the 
self-duality equations from four to two dimensions.  
We will choose to work with complex variables 
\eqn\cplxvar{X=\hf(X^1+iX^2)\ ,\ {\overline X} = \hf (X^1 -i X^2)\ ,}
setting the remaining $X^i$'s to zero.  The self-duality conditions then
become
\eqnar\selfdual{
 F_{w\wbar} = - {i\over g^2_s}[\, X,\overline{X}\, ], \quad \cr
\qquad \qquad \quad D_wX = 0, \quad 
D_\wbar\Xbar = 0.    \nonumber
}
The above equations are most conveniently analyzed by writing 
\eqnar\Aglc{
A_w(w,\wbar) &=& -iG \partial_w G^{-1}, 
\nonumber \\ \\[-\the\baselinetruc] \cr
A_\wbar(w,\wbar) &=& i   (\partial_\wbar \Gbar^{-1}) \Gbar, 
\cr\nonumber 
}
where $G(w,\wbar)$ denotes an element of the {\it complexified} (${\bar
G}\neq G^{-1}$) gauge group.
This parameterization of $A_\alpha$ allows one to solve the second and third
equation of \selfdual\!\!, via
\eqn\xsolve{
X(w,\wbar) = G \widehat{X}(\wbar) G^{-1}\ .  }
The first equation in \selfdual then takes the following form
\eqn\sinhG{
\partial_\wbar (\Omega \partial_w \Omega^{-1}) = -{1\over g_s^2}[ \Omega 
\widehat{X}(\wbar) \Omega^{-1}, 
\widehat{\Xbar}(w)]}
with 
\eqn\Omeg
{\Omega = \Gbar G.
}
Let us now look at the local neighborhood of an interaction point. 
For convenience, we choose coordinates such that it is located at $w=0$.
Since the interaction involves only two eigenvalues, it is sufficient
to consider only the corresponding $SU(2)$ part of the matrices.
The matrix $\widehat{X}$, which parameterizes the local
coordinate distance between the two interacting strings, can be 
chosen of the following form
\eqn\xparam
{\widehat{X}(\wbar) \simeq \pm B \; \sqrt{\, \wbar} \, \tau_3, 
}
for some constant $B$. The $\pm$ indicates that the interaction point $w=0$
represents a square root branch point for the diagonal matrix 
$\widehat{X}$ in \xparam\!\!, which therefore is multi-valued.

%In fact, if it where not for this multi-valuedness, the diagonal matrix
%$\widehat{X}(\wbar)$ would represent a perfectly valid solution 
%to the SYM equations near the interaction point. Instead, however,
%we must require that the physical matrix variable $X(w,\wbar)$ in \xsolve
%is single-valued around the interaction point, or alternatively,
%compensate its multi-valuedness by means of an appropriate boundary
%condition on the Yang-Mills potential $A_\alpha$ at $w=0$. We'll first 
%choose the latter option.
%

The diagonal matrix $\widehat X(\wbar)$, together with $A=0$, represents a
valid solution of the SYM equations \selfdual except at the interaction
point, where analyticity fails.  Therefore we will look for a true solution
of the form \xsolve\!\!, where $G\rightarrow1$ asymptotically far from $w=0$.
A helpful {\it Ansatz} for $G(w,\wbar)$ is
\eqn\gparam{
G= e^{{1\over 2} \alpha \, \tau_1},  
}
%
%\Gbar \is e^{-{1\over 2} \alpha \, \tau_1} }}
%
where for $\alpha(w,\wbar)$ we choose a real function (so 
that $G=\Gbar$ and $\Omega = \exp(\alpha \tau_1)$) 
that tends to zero far away from the interaction point.
We now compute
\eqn\comp{%\eqalign{
\Omega \widehat{X}\Omega^{-1} = B \sqrt{\wbar} e^{\alpha \tau_1}
\tau_3 e^{-\alpha\tau_1}
= B \sqrt{\wbar} 
\left(
\begin{matrix}
\cosh 2\alpha & -\sinh 2\alpha\ \cr \sinh 2\alpha & 
-\cosh 2\alpha 
\end{matrix}
\right).}
Hence
\eqn\comp{
\Bigl[ \Omega \widehat{X}\Omega^{-1}, \widehat{\Xbar}\Bigr] 
= 2|B|^2 |w| \, \sinh 2\alpha \; \tau_1}
and thus we find that under the present Ansatz the equation of motion 
\sinhG reduces to 
\eqn\sinhg{
\partial_w \partial_\wbar \alpha = {2\over g^2_s} |B|^2 |w| 
\sinh2\alpha, 
}
%+ i\pi \delta(w)}
%
which is essentially the familiar sinh-Gordon equation.  (It can
be transformed to the exact sinh-Gordon equation after a (multi-valued) 
coordinate transformation $w \rightarrow \tilde{w} = w^{3/2}$.)

The boundary condition that we must impose on $\alpha(w,\wbar)$ at $w=0$ 
follows from the requirement that the Yang-Mills configuration be regular. 
This condition is most easily understood in the gauge where $X$ is {\it
single-valued} near $w=0$; in this gauge the YM curvature
$F_{w\wbar}$ should be a regular function at $w=0$.  
The configuration
\xsolve\!\!--\gparam\!\!, however, is (for single-valued and real $\alpha$)
multi-valued. We can make $X$ single-valued by applying the
singular gauge transformation
\eqnar\gaugetr{
X & \rightarrow U X U^{-1},\cr
A_w & \rightarrow - i U D_w U^{-1},
}
with gauge parameter
\eqn\utheta{
U = e^{\pm i\theta \tau_1/ 4},
}
with $\theta = {1\over 2i} \log(w/ \wbar)$ the azimuthal angle around
$w=0$. In this gauge
\eqnar\asing{
A_w &=& i \Bigl[{1\over 2}\partial_w\alpha \pm {1\over 8w}\Bigr] 
\; \tau_1, \cr
A_\wbar &=& \; -i \Bigl[{1\over 2}\partial_\wbar\alpha \pm {1\over 8 \wbar}
\Bigr] \; \tau_1.
}
Using that $\partial_w {1\over \wbar} = \pi \delta^{(2)}(w)$, this gives
\eqn\fww{
F_{w\wbar} = -i\tau_1 \Bigl(\partial_w \partial_\wbar \alpha \pm {\pi \over 4} 
\delta^{(2)}(w)\Bigr).}
The regularity requirement at $w=0$ is therefore that
$\partial_w \partial_\wbar \alpha \simeq \mp {\pi \over 4}
\delta^{(2)}(w)$. We thus deduce that the solution to equation 
\sinhg\ that we want must satisfy the following asymptotic condition
\eqn\alphlimit{\qquad \qquad 
{\alpha}(w,\wbar)  \; \simeq \;  \mp{1\over 2} \log | w| + {const.}  
\qquad \quad 
w \rightarrow 0,
}
while at large distances from the interaction point ${\alpha}$ must tend to 
zero. 
%
%Note that the interaction point gets mapped to $r \rightarrow - \infty$.

Now let us write $\alpha = \alpha(r)$ with $r = |w|$. The equation
of motion \sinhg reduces to the ordinary non-linear differential
equation 
\eqn\sing{
(\partial_r^2 + {1\over r} \partial_r) {\alpha} = 
{8\over g^2_s} |B|^2 r \sinh 2\alpha\ .}
\\
%%%
\noindent
\epsfysize=7cm 
\hfigure{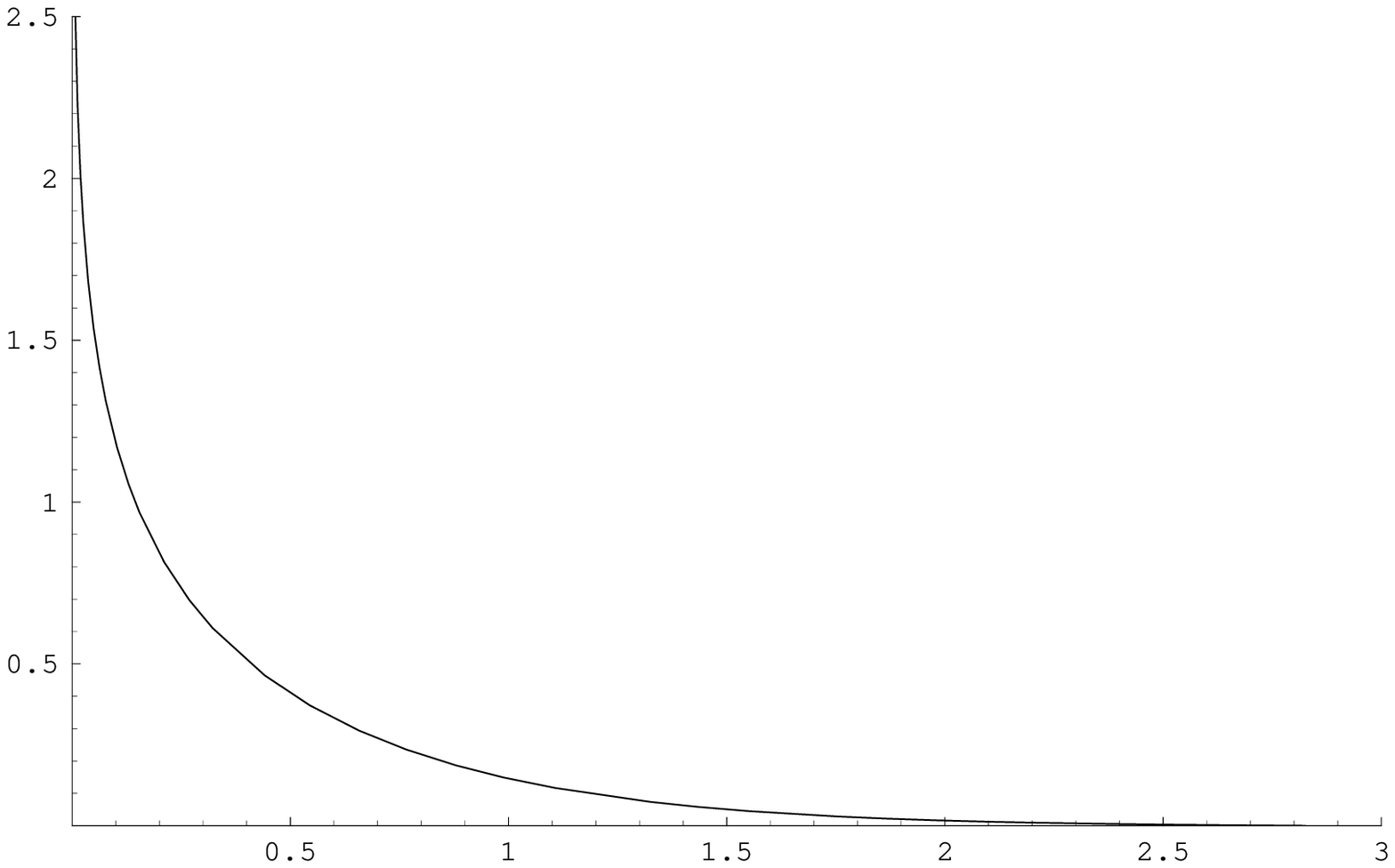}{
The numerical solution $\alpha(s)$ to $(\partial_s^2 + 
{1\over s} \partial_s) {\alpha} = s \sinh 2\alpha$, as 
a function of the distance $s$ from the interaction 
point. The initial condition for small $s$, that after integrating 
leads to the correct asymptotic behavior for large $s$, reads 
$\alpha(s) = -{1\over 2} \log s + 0.0305070(1)$.
}
\xdef\numericalfig{\thechapter.\the\fignumber\ }  
%%%
\\
A numerical solution to this equation is depicted in figure 
\numericalfig\!\!. 
\footnote{We thank Pierre van Baal for pointing out an error in the 
original figure in \ghv.}
For large $r=|w|$ the solution looks like 
\eqn\rlimitp{%\eqalign{
\alpha(r)  \sim  %J_0({|b|\over 3 g_s} (2r)^{3/2}) \cr
{\mp 1\over |w|^{3/4}} \exp\Bigl(-{8|B|\over 3 g_s} 
|w|^{3/2} \Bigr)\ .}
From this asymptotic behavior we read off 
the typical instanton size 
\eqn\instantonsize{
\ell_{\rm inst} \sim g_s^{2/3}|B|^{-2/3}.
}
%

%%%%%%%%%%%%%%%%%%%%%%%%%%%%%%%%

\newsection\noname{High energy scattering of matrix strings}

The matrix solution of the string interaction constructed in the 
previous subsection, should be
viewed as a local description in the immediate neighborhood
of the interaction point. In general, it must therefore be glued
via an appropriate patching procedure into a complementary
CFT type solution ({\it e.g.} as described 
in section \fixedangles\!\!) that matches 
with the asymptotic scattering data at the far past and future. 
The idea here is that (as we will see shortly) at sufficiently 
high collision energies, the size of the interaction regions are 
small compared to the rest of the matrix string world-sheet. 
Hence, while the behavior \xparam provides the asymptotics 
for large $|w|$ at the UV scale of the matrix solution, it also provides 
the local boundary condition near the interaction point for the
CFT solution for $X$ that describes the IR part of the saddle point.

The solution \selfdual that we described is  not the most
general SYM description of a string splitting and joining event,
but rather the most symmetric one, with smallest action.
This means therefore that there is a non-trivial matching
condition on the corresponding matrix string world-sheet: 
from \xparam we see that we must require that the transverse
string coordinates $X$ behave holomorphically near the 
interaction point. Remarkably, this is exactly the same  
condition  as \bpscond\!\!, which tells us that the shape of the
string world-sheet must be precisely that of the 
Gross-Mende saddle point!  Therefore these solutions  
seem appropriate to a YM generalization of 
the high-energy scattering of \grossmende. In this section, we fill in 
a few more details of this connection.

\newsubsection\nonameeva{Evaluation of the classical action}

In order to estimate scattering amplitudes via the instanton processes, one
must calculate the instanton action.  The bosonic part of the SYM action 
(with only two $X$-fields non-vanishing) 
can be written as
\eqnar\bpsact{
S = \int d^2 w \left\{-{g_s^2} 
\Bigl(F_{w\wbar} + {i\over g^2_s}[\, X,\overline{X}\, ]\Bigr)^2
+ 4 D_w X   D_\wbar\Xbar\right\}\cr 
+\oint (\Xbar {\overline D} X + X D \Xbar - \Xbar DX - X {\overline D}\Xbar )
} 
and thus for the supersymmetric configuration that satisfy
\selfdual\!\!, the total classical action reduces to a boundary term
\eqn\bounds{
S_{cl} = 
\oint (\Xbar {\overline \partial} X + X \partial \Xbar)}
identical to the boundary term needed to glue the non-abelian
matrix solution described in the previous section into the
CFT type solution.
Hence we claim that, in the limit that the matrix interaction points
become sufficiently small, the SYM action for the above saddle point
configurations coincides with the CFT action, {\it i.e.} for the case of
a tree level string diagram it equals the ``Coulomb energy'' 
\coutreelevel\!\!,
where me must insert the saddle-point value for locations $z_i$.
It is perhaps worth pointing out that this saddle point actions is 
fully Lorentz invariant, as it should be. While this is not 
surprising once we have established the connection with the GM saddle 
point, it does seem to represent a rather non-trivial statement from 
the SYM point of view!

More generally we see that from \bpsact we can derive
(as usual) an inequality, which suggests that whenever the interaction
does not take place at a holomorphic point for the $X$-fields,
the SYM action is always larger than the corresponding CFT
action. This provides additional evidence for the conjecture that 
the above type of configurations represent dominant saddle-points,
that minimize the SYM action. 

Obviously, there exist a large number of CFT-type solutions for which
$X$ varies (anti-) holomorphically near all interaction points. In
particular, there are the higher genus $\Zvet_N$-curves of \grossmende.
In addition it is also possible to write down SYM solutions that 
describe multiple string world-sheets, but nonetheless still satisfy
the appropriate boundary conditions, as specified in equations 
\instate and \outstate in the introduction. Ideally, one would
like to know which (sub-class) of these solutions provide the truly
dominant contribution to the scattering amplitude.

\newsubsection\nonamemd{Minimal distance}

The parameter $|B|$ that governs the size
of the interaction vertex, as seen in \rlimitp\!\!, can be straightforwardly
determined in terms of the momenta of the external states.
The coordinate system $(w,\wbar)$ on the Yang-Mills cylinder that we used
in the analysis of the self-dual Yang-Mills equation \selfdual\!\!, 
coincides with the light-cone coordinates defined in \lcgauge\!\!. 
From this and \xparam we immediately find 
\eqn\bexpr{
|B|^2 \, = \,2 
{\;\; |\partial_\zbar X(z_0^+)|^2\over |\partial^2_z X^+(z_0^+)|}\ .}
A straightforward calculation then gives 
\eqn\bans{
|B|^2= {|p_1\overline{p}_3-
\overline{p}_1 p_3| (N_1+N_2)\over \sqrt{N_1N_2N_3N_4}}\ .}
It is interesting to note that for this
solution, even though the eigenvalues of the complex coordinate matrix
$X$ vanish at the 
interaction point, the full matrix coordinate $X$ in fact does not! 
Instead, near $w=0$ it approaches the constant non-diagonalizable
matrix 
\eqn\xlimit{
\qquad \qquad 
X(w,\wbar)  \; \simeq \;  const. \ g_s^{1/3} B^{2/3}  
\left(
\begin{matrix}
1 & \mp 1 \cr \pm 1 & -1 
\end{matrix}
\right)
 \qquad \quad 
w \rightarrow 0.
}
The value of the overall constant can be determined numerically.
From this we read off that the minimal ``distance'' between the 
two interacting strings is in fact non-zero! Instead, we have
\eqn\dlimit{
d_{min} = \sqrt{\tr(X(0) \Xbar(0))} \sim  g_s^{1/3} |B|^{2/3}.
}
Although it is tempting to speculate 
(as indeed we will do in the concluding section), 
the precise physical significance of this result is as yet 
unclear to us.  We do notice, however, that the typical world-sheet 
size $\ell_{inst}$ of the matrix interaction region, 
as can be read off from \rlimitp\!\!, 
is naturally expressed in terms of this minimal relative distance as 
$\ell_{inst} = (g_s/|B|)^{2/3} = g_s / d_{min}$.

\newsection\noname{One loop fluctuation analysis}

In principle it is now possible to do a computation of the 
one-loop determinant of the quantum fluctuations around the
semi-classical saddle-point configurations.
An important motivation for performing such an analysis is 
to obtain a semi-classical estimate for the absolute strength
of the splitting and joining interactions in matrix string theory.
Duality symmetries of M-theory give the precise prediction that 
this strength should be governed by the string coupling $g_s$. 
To verify this, one needs to compare the SYM one-loop determinant
with the Gross-Mende fluctuation determinant, coming from the
Gaussian integration over the Riemann surface moduli around the
saddle-point. 

This comparison has been done for tree level fixed angle scattering 
in the high energy limit \wynteriii. 
The result is that all of the expected structure of the high energy 
scattering amplitude is reproduced, including the power of the string
coupling constant $g_s$ and the kinematical factor. 

When one does an expansion around a classical background one has
to do two things: a calculation of the fluctuation determinant 
and a separate treatment of the zero modes. 
In our case the fluctuation determinant is approximately equal to one,
because the background field is taken to be self-dual in a
neighborhood of the interaction points. Then the fermionic and bosonic
contributions cancel each other. Note however that non self-dual 
corrections to the background will change this simple result.

In the approximation where the fluctuation term is equal to one, the  
zero modes determine the form of the scattering amplitude. 
Below we will discuss them briefly. For an extensive and detailed
treatment we refer to the work of Wynter in \wynteriii.

\newsubsection\nonamezerom{Zero modes of the instanton}

The strategy of the zero mode analysis of \wynteriii\ is based 
on the philosophy we sketched in section \mstfirst\!: We have no 
explicit solution of the full YM saddle-points, but we know 
their form near the interaction points and their asymptotics far away 
from the interaction points. The instanton type solutions near the
interaction points satisfy boundary conditions that are compatible
with the abelian matrix string configurations. These configurations 
consist of commuting matrix fields, together with an abelian gauge 
field that generates the appropriate monodromies around the
interaction points \wynteri. Their singular behavior at the 
interaction points is smoothened out by the instanton solutions. 
The same is true for the non-trivial
zero modes: the abelian matrices have non-trivial zero modes with 
singularities at the interaction points. These singularities are, 
however, smoothened out by the instanton zero modes.  

To make this more explicit, we will now review the zero modes of the 
instanton solutions. For this, it is 
convenient to write them in the multi-valued gauge
\eqnar\noname{
A_w = {1\over 2}i \partial_w \alpha \ \! \tau_1, 
\cr
X = B\sqrt{\bar{w}}\left[\cosh\!\alpha \ \!\tau_3 -i\sinh\!\alpha \ \! 
\tau_2 \right]. 
}
We distinguish between four types of bosonic zero modes \wynteriii, 
namely those associated to translations and deformations of the instanton 
solution (the zero modes of the fields $X$ and $\bar{X}$), 
gauge field zero modes, the zero modes of the six transversal coordinates
$X^I$ and those of the ghost fields.  

All bosonic zero modes satisfy the eigenvalue equation for quadratic 
fluctuations around the classical instanton background which reads 
(after inclusion of the background gauge fixing term 
$(\bar{D}_\mu A_\mu)^2$ to the action) 
\eqn\zerocondition{
D^2 V_\mu + F_{\mu\nu} V_\nu =0. 
}
Here we are using a ten dimensional notation; $D^2$ and $F_{\mu\nu}$ 
are taken with respect to the instanton background field. 

The gauge field zero modes originate from the left-over gauge 
degree of freedom; they can be written as pure gauges, 
$V_w = D_w \Lambda$. They will prove to be of particular 
interest in the following, because they determine the $g_s$ coupling 
constant dependence of the contribution to the scattering amplitude 
at $g$-loop order \bonelli\wynteriii. 
Plugging the expression for 
the instanton and the pure gauge $V_w$ in equation \zerocondition\!\!, 
we get the following condition for $\Lambda$ 
\eqn\zmodeq{
D_wD_{\bar{w}}\Lambda + D_{\bar{w}}D_w \Lambda - 
{1 \over g^2}(
[X,[\bar{X},\Lambda]]-[\bar{X},[X,\Lambda]])    
= 0.  
}
Solutions to this equation are $\Lambda_n$ and $(\Lambda_n)^*$ 
\wynteriii, given by 
\eqn\zeromodes{
\Lambda_n = \left\{ 
Bw^n(\cosh{\alpha}\ \tau_3 - i\sinh\alpha\ \tau_2) \qquad
n=-{ 1 \over 2}, {1 \over 2}, {3 \over 2}, \dots  
\atop w^n \one \qquad n=1,2,3 \dots 
\right.  
}
Of particular relevance is the zero mode for which $\Lambda$ 
coincides with the $X$-field of the instanton solution 
\eqn\zeromodeinst{
V_{\bar{w}} = {1 \over 2}B\bar{w}^{-1/2} 
(\cosh\alpha\ \tau_3 - i\sinh\alpha\ \tau_2). 
}
Though this zero mode has a branch starting at the origin, it is 
single-valued when we go back to the gauge \utheta\!\!. 
Away from the interaction region it behaves as an 
{\it abelian} zero mode with $\bar{w}^{-1/2}$ asymptotic behavior
\eqn\noname{
V_{\bar{w}}  \rightarrow {1\over 2}B\bar{w}^{-1/2}\tau_3 \qquad
(\bar{w} \rightarrow \infty ). 
}
Hence the zero mode \zeromodeinst can be glued into the zero 
mode of an abelian gauge field that is defined on the light-cone
cylinder, except for an interaction point, where it has a 
singularity. One can remove this singularity by going to the double cover. 
As explained in section \lightconebasics the light-cone cylinder 
can be lifted to a double cover in a neighborhood of an interaction 
point via $w=z^2$. On the Riemann surface the abelian zero mode 
becomes 
\eqn\noname{
V = \bar{w}^{-1/2}d\bar{w} = d\sqrt{\bar{w}} = d\bar{z}.   
}
Together with the other solutions \zeromodes we get a complete 
set of abelian gauge field zero modes in the complex $z$-plane 
\eqn\noname{
V_{\bar{z}} = 1, \bar{z}, \bar{z}^2,\bar{z}^3, \dots
}
The zero modes for the bosonic fields $X^1$ and $X^2$ correspond to 
translations of the interaction points (branch points of the 
light-cone cylinder). 
An obvious guess for the translation mode is 
\eqn\zmodetrans{
V_\mu = \partial_{\alpha} A_\mu - D_\mu A_\alpha = F_{\alpha\mu}, 
}
which can be easily verified to be a solution of equation 
\zerocondition ($\alpha$ runs over the light-cone coordinates).  
The first term in \zmodetrans corresponds to a translation in the 
$\alpha$-direction, the second term is a gauge transformation which 
forces the zero mode to satisfy the background gauge condition. 

Its asymptotic behavior is 
\eqn\noname{
V_w  \rightarrow \bar{w}^{-1/2} \tau_3, 
}
which is again smoothened out at the origin of the instanton solution
in the regular, single-valued gauge. 

The remaining bosonic zero modes are those of the Higgs fields $X^I$ 
(where now $I= 3 \dots 8$), 
which correspond to translations in the (six) transverse directions; 
and the ghost zero modes. Both types of zero modes satisfy 
the equation \zmodeq \wynteriii, so the functions \zeromodes 
form a basis for them. However there is no solution with asymptotic
$\bar{w}^{-1/2}$ behavior that is finite at the origin (in the regular
gauge). Hence we have no non-trivial zero modes for the ghost 
fields and the Higgs fields. 

Finally we have fermionic zero modes. They are associated to the
unbroken supersymmetry of the instanton configuration.  
There is one zero mode with abelian asymptotic $\bar{w}^{-1/2}$ 
behavior, which is finite at the origin \wynteriii.  

Now we know the local behavior of the zero modes in a 
neighborhood of the interaction points. They are smooth
at the interaction points, and behave like abelian modes
away from these points. This last property makes it possible 
to glue them into the zero modes 
of the matrix string configurations. These were constructed in
\wynteriii. We will not discuss them here; instead we refer 
to reference \wynteriii. 

We conclude that not only the field configurations 
of the instantons can be glued into the abelian background field, 
but their non-trivial zero modes as well. 

\newsubsection\nonametree{Tree level high energy scattering}

For the four string scattering process illustrated in figure 
\shadedinstanton the calculation of the tree level amplitude
amounts to a careful treatment of the bosonic and fermionic 
zero modes. 

Integration of the bosonic zero modes corresponding to simple 
translations of the string configuration in the six transverse 
directions $X^I$, gives rise to transverse momentum conservation. 
The non-trivial zero modes which are associated to simultaneous 
translations of the two branch points, in the 
$\sigma$ respectively $\tau$ direction, lead to 
invariance under shifts in $\sigma$ respectively 
conservation of light-cone energy $p^-$. Relative displacements 
of the two branch points in figure \shadedinstanton are somewhat 
more involved. A careful integration over these zero modes leads to 
a kinematic factor in the scattering amplitude. This contribution
has the form \wynteriii
\eqn\expresfam{
{c \over sut} e^{-{1\over 4}(s\log s + t\log t + u \log u)}, 
}
where $c$ is a constant. The expression \expresfam is 
a familiar term from the calculation by Gross and Mende 
of tree level high energy scattering \grossmende. 

The integration over the ghost zero modes and the gauge field 
zero modes together yield a factor $g_s^2$, the correct string 
coupling constant dependence for a four string scattering amplitude at
tree level. 

Finally the fermionic zero modes give rise to the right kinematical
factor. Thus the high energy limit of the string theory scattering 
amplitude is reproduced by matrix string theory, up to a 
numerical factor. This is an encouraging result; it gives clear
evidence for the matrix string theory conjecture of section 
\mstproposal\!\!. 
As a further check one should compare loop contributions
with the results of Gross and Mende as well. In principle this 
calculation can be performed, but the large $N$ limit should
be taken with great care (see also \wynterii).

We end this section with by remarking that, as was originally argued 
in \bonelli, the $g_s$ coupling dependence is reproduced by matrix
string theory for any light-cone Riemann surface. 
The power of the string coupling is determined by integration over the
gauge field zero modes and the ghost field zero modes. 
Each (abelian) gauge field zero mode contributes a positive power 
of $g_s$. The constant ghost zero mode yields a factor $g_s^{-1}$
\bonelli. 
This reasoning is justified by the analysis of Wynter, who verified
the local existence of the zero modes around the interaction points. 

Than it is just a 
matter of counting the number of independent abelian gauge zero 
modes or equivalently the number of independent cycles
on the Riemann surface. When we start with a tree level diagram with
$n$ external states, we have $n-1$ independent cycles. For each 
loop (handle) we glue to the diagram, we have to add two cycles, so
that we have $2g+n-1$ abelian gauge field zero modes for a $g$ genus
light-cone diagram with $n$ external states. There is only one 
constant ghost field zero mode, thus we come to the conclusion 
that the string coupling dependence is 
$g_s^{2g+n-2} = g_s^{-\chi}$ where $\chi$ is the Euler number of the
Riemann surface. This is the right power of $g_s$, known 
from string theory \grossmende.

This concludes our discussion of the one-loop fluctuation analysis.

It seems even more worthwhile to look for true new physical
effects that might arise from the one-loop corrections. Compared to
the conventional perturbative string description, the new degrees
of freedom in matrix string theory are the charged components of 
the $X$-fields, as well as the extra gauge potential $A_\alpha$. 
These new degrees of freedom are non-perturbative from the string 
perspective, and their quantum fluctuations could thus potentially 
lead to new physics. As we will show in the next section, there
is indeed such a new effect: the pair creation of D-particles.

\newsection\dpairprod{D-Particle pair production} 

In this section we turn to the process of pair creation of D charge, which
is in our description $x^9$ momentum or equivalently (under the matrix
string duality) electric flux.  This can be viewed as a contribution to the
fluctuations about the high-energy scattering processes of the preceding
sections, or as a process worthy of interest in its own right in the
context of graviton scattering.
There are several viewpoints from which
this can be investigated.  In the limit where $x^9$ decompactifies, this
simply matches onto the standard supergravity calculation 
\BeCo.  In fact, we can
work backwards from this, using the method of images, and compute the
amplitude at large finite $R_9$, in the special situation with source and probe
particles, $N_1\gg N_2$.  We will discuss this calculation first.  
Alternately, one can study this process directly
in the matrix string approach, and derive the pair-production rate via a
one-loop Yang Mills calculation.  This latter approach gives a leading
order result valid for arbitrary $N_1$ and $N_2$, and also more readily
makes connection with the other results of this chapter.  Furthermore, the
Yang-Mills calculation also apparently extends beyond the region
where supergravity is a valid approximation.
\\
\epsfysize=5.3cm
\hfigure{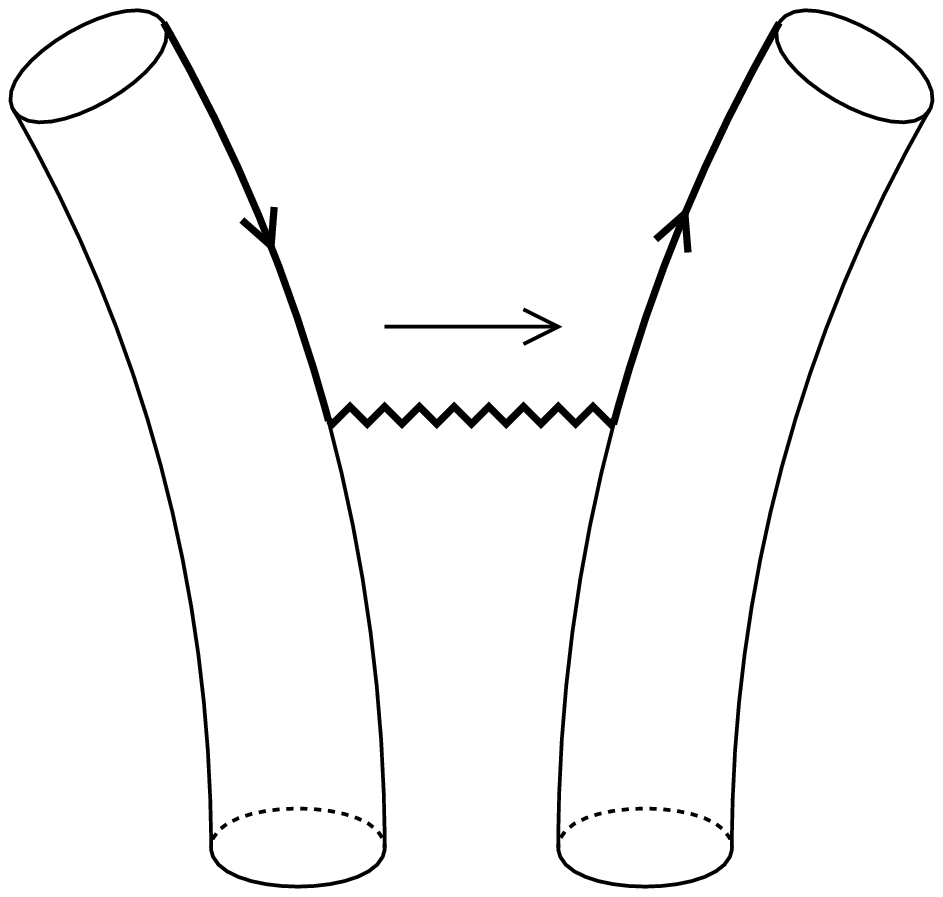}
{In matrix string theory strings can be adorned with an extra quantum number,
the D-particle number. This enables one to compute non-perturbative 
corrections to perturbative string scattering amplitudes. 
The figure illustrates the creation/annihilation of a D-particle pair,
a process that is reminiscent of electron-positron creation/annihilation.} 
\xdef\alain{\thechapter.\the\fignumber}

\newsubsection\nonamesugra{Supergravity calculation}

Consider 11-dimensional supergravity compactified on $S^1\times S^1$,
where one $S^1$ a light-like circle of radius $R$
\eqn\dlcq{
x^- \equiv x^- + 4\pi R
}
and the other $S^1$ denotes a space-like circle of radius $R_9$. 
As we have seen in section \mstproposal\!\!, in the M-theory/matrix string
correspondence this second radius
$R_9$ is expressed as  $R_9 = g_s $ in string units 
(cf. \typeiiacouple\!\!). 

Consider in this set-up the scattering process of two massless
particles of light-cone momenta $p^+_i = N_i/R$ and transverse momenta $p_i$. 
Let us first consider the probe situation $N_2\ll N_1$. Then one
can already get quite useful information about the scattering process
from considering the classical gravitational force between the two
particles.
The boosted particle with $p_+=N_1/R$ produces via its stress-energy 
a non-trivial gravitational background, described by the generalized 
Aichelburg-Sexl shock-wave geometry of the form \AiSe\BBPT\ 
\eqn\ASm{ds^{2}=-dx^{-}( dx^{+}+f(r,a)dx^{-}) +dx_{\bot }^{2} +g_s^2
da^2,
}
with
\eqn\fr
{f(r,a) = \sum_{k} {-15 N_1 g_s^3 \over 2 R^2 
(r^2 +g_s^2 (2\pi k + a)^2)^{7/2}}.
}
Here $a$ denotes the coordinate distance from the gravitational source in the
compact $x^9$ direction, and the sum over $k$ arises from the image points 
in this direction.
 
The momentum four vector
of the second massless particle moving in this background geometry 
will satisfy a dispersion relation of the form
\eqn\dispersion{
2 p^-(p^+ + f(r,a) p^-) = p^2 + e^2/g_s^2, 
}
where $e$ denotes the quantized momentum in the $x_9$ direction.
We can solve for the light-cone Hamiltonian $p^-$ of the
particle and obtain
\eqn\pminus
{p^- = 
{p^+\over 2 f(r,a)}\left\{\sqrt{\Bigl(1 + {2 f(r,a)\over (p^+)^2} 
(p^2+ e^2 /g_s^2)\Bigr)} -1\right\} . 
}
Substituting $p^+ = N_2/R$, (and rescaling the light-cone time by a factor 
of $R$) we can write this as 
\eqn\htwo{H = H_0 + H_{int}, 
}
where 
\eqn\hnot{
H_0 = {1\over 2N_2} (p^2+ e^2/ g_s^2),
}
and
\eqn\Hint{H_{int}
\simeq  - {15 N_1 g_s^3 \over 8 N_2^3} \sum_{k} {(p^2+e^2/g_s^2)^2 \over 
\Bigl(r^2 +g_s^2 ( 2\pi k  +  a)^2\Bigr)^{7/2}} + \ldots \ . 
}
Hence the motion of the second particle in terms of the light-cone
time $x^+$ looks like that of a particle with mass $N_2$ moving
in $R^8 \times S^1$ under the influence of an interaction potential
given by \Hint\!\!. 

From this description we can now quite easily extract a low
energy prediction for the D-pair production rate. To this end,
it is useful to rewrite the interaction Hamiltonian via a
Poisson resummation as
\eqn\hint{
H_{int}\simeq
-{1 \over 2\pi} {N_1 \over N_2^3}
g_s^2(p^2+e^2/g_s^2)^2
\sum_n \exp(ina) \int dT T^2 \exp(-Tr^2) \exp(- n^2/4g_s^2T).
}
The $n=1$ term in this series is the term that corresponds to changing the
compact momentum by one unit, {\it i.e.} to D-charge production.  Working to
first order in perturbation theory, we can then compute the corresponding
phase shift, using 
\eqn\phham
{
\delta=-\int d\tau H_{int}(b^2 +p^2\tau^2)\ ,
}
where $b$ is the impact parameter and $\tau$ the light-cone time.

\newsubsection\nonamedpair{D-pair production via electric flux creation} 

We now study this problem of D-charge creation in the matrix string
framework. More generally, we consider scattering states which 
asymptotically have momenta of the form 
\eqn\momdef{p^\mu=(p^-,\vec{p},p_9=n/R_9,p^+=N/R)\ .}
These include both gravitons ($n=0$) and D0-branes -- or anti-branes --
($n=\pm1$). The case of current interest begins with an initial state of
two gravitons, and pair produces a D particle pair.  This process is
intrinsically non-perturbative from the point of view of string
theory. It is also a process not accessible in the standard matrix 
theory approach, where the anti-branes are boosted away to infinite energy.  

In principle (for example on a sufficiently large computer) 
it appears possible to calculate such amplitudes to arbitrary
order in the coupling $g=g_{YM}=1/g_s$, and calculate the D-pair production
rate even for small $g_s$.
In the coming  sections we will work to leading
non-trivial order (one-loop), and leave further calculations to other
work.  Similar calculations have been performed in the context of matrix
theory in \BeBe.

\newsubsection\nonamedmst{D-particles in matrix string theory}

As we have indicated in chapter 2, an important new feature
of the matrix string theory formalism 
(relative to standard light-cone string theory) is that via the
electric flux, string states can be adorned with a non-vanishing
D-particle charge. In this subsection we will describe this in 
somewhat more detail. 

To add to this interpretation, let us first show
that each separate string can carry only one type of electric flux.
Consider a single long string in matrix string theory with length $N$.
Define the $U(N)$ matrix $U$ such that
\eqn\uvcommutation{
UV = VU e^{{2\pi i \over N}},
}
with $V$ the usual cyclic permutation matrix 
on the $N$ eigenvalues defined in \cycle\!\!. 
% 
%\eqn\cyclicper{  
%V= \left(
%\begin{matrix}
%&~1~&&& \emptyset \ \ \cr &&~1~&&\cr & %\emptyset 
%&&\ddots &\cr &\emptyset &&&1 \ \cr ~1~&&&&
%\end{matrix}
%\right).
%}
%%
Hence we can take
\eqn\formu{
U =\left(
\begin{matrix}
1~&&& \emptyset \  \cr &e^{{2\pi i \over N}}&&\cr
&&\ddots &\cr \ \emptyset &&&e^{{2(N-1)\pi i \over N}}
\end{matrix}\right).
}
The $SU(N)$ part of the electric flux in this sector
is defined as
\eqn\hatu{
\hat{U} |\, \psi_e\, \rangle = \exp\Bigl({2\pi i e \over N}\Bigr) 
|\,\psi_e\, \rangle,
}
with $e \in {\bf Z}_{N}$ and
$\hat{U}$ the quantum operator that implements the constant
gauge rotation 
\eqn\aix{
(A,X) \rightarrow (U\!AU^{-1}, U\!XU^{-1}).}
Since diagonal matrices are inert under this gauge rotation, 
we conclude that the $SU(N)$ part of the electric flux 
dynamically decouples from the diagonal matrix string configurations 
\matrixcoo that describe the separate freely propagating strings.
Now recall that in $U(N)$ SYM theory, the overall $U(1)$ part of the 
electric flux is related to the $SU(N)$ part $e$ via
\eqn\sunflux{\tr E = e \quad ( {\rm mod} \ N )\ .}
Supersymmetry ensures that the ground state in the $SU(N)$ sector
has zero energy even for $e \neq 0$. Hence the total ground state 
energy receives only a contribution from the overall $U(1)$ flux.
In the following we will thus identify $e$ with the total 
$U(1)$ electric flux. From the above description it is 
further clear that we can turn on
only one electric flux per long string, as is appropriate for
its identification with D-particle charge.

The energy of the ground state in this electric flux sector is equal to
\eqn\energy{
H_{0} =  {e^2 \over 2 Ng_s^2}.}
General ground state configurations 
\eqn\grstate{
|N^{(i)},p^{(i)}_\pperp, e^{(i)} \rangle }
of $s$ separate strings of individual length $N^{(i)}$, 
transverse momenta $p_\pperp^{(i)}$, and D-particle charge 
$e^{(i)}$ have a SYM energy equal to
\eqn\hnul
{H_0 = \sum_{i=1}^s {1 \over 2 N^{(i)}}
\Bigl[(p_\pperp^{(i)})^2 +  (e^{(i)}/g_s)^2\Bigr], 
} 
% 
%(this is defined with respect to the time $\tau=tR$)
which, when rescaled by $R$, is 
the sum of the $p_-$ light-cone momenta of the corresponding
collection of string ground states
\eqn\pminus{\sum_{i=1}^s {1 \over 2 p_+^{(i)}}
\Bigl[(p_\pperp^{(i)})^2 +  (M^{(i)})^2\Bigr].}
In particular, we read off from 
\hnul that the states with D-particle charge $e^{(i)}$ each have 
mass  
\eqn\mass{
M^{(i)} = {e^{(i)}\over g_s}= {e^{(i)}\over R_9}, 
}
in accordance with their identification as graviton states with non-zero 
KK momentum in the compact direction.

\newpage 

\newsubsection\nonameonel{One-loop calculation, $N=2$}

For simplicity we begin with the case where the incoming and
outgoing particles all have $N=1$.  The next subsection will
generalize to arbitrary $N$.  The asymptotic states take the form
\eqn\initstate
{{\bar X}^1=\hf\left(
\begin{matrix}p\tau&0\cr0&-p\tau
\end{matrix}\right)\ ,\
{\bar X}^2 =
\hf \left(
\begin{matrix}
b    &0\cr0&-b
\end{matrix}
\right)\ ,}
corresponding to two particles with center of mass momentum $p$ and impact
parameter $b$ (measured in string units).
It will also be 
useful work with a non-trivial gauge background 
\eqn\clfield{
{\bar A}_\sigma=\left(
\begin{matrix}
a/2+e\tau/g_s^2&0\cr 0&-a/2 -
e\tau/g_s^2\cr
\end{matrix}
\right) ,\ {\bar E }= \left(
\begin{matrix}
e&0\cr 0&-e\cr
\end{matrix}
\right)\ .}
The constant electric field corresponds to a non-zero D-charge
for the incoming and outgoing particles, with quantization
\eqn\equant{e \in \Zvet\ .}
The prototypical example of production of D-charge is 
in processes where this
changes by one unit,
\eqn\elflux{\Delta E=\left(
\begin{matrix}1&0\cr0&-1
\end{matrix}
\right)\ .}
Introducing the constant background potential $a$ will help keep track of
such changes.

In the large string coupling/small Yang-Mills coupling limit, the
leading contribution to D-charge producing processes is easily computed 
via a one-loop super-Yang Mills calculation.
Calculations at higher loop order then give subleading corrections
in $g = 1/{g_s}$.

Our starting point is the Yang-Mills action \mstaction\!\!, although it will
be simpler to begin with it written in its un-dimensionally reduced
form in terms of the gauge field
\eqn\Adecomp{
A^M=(A^\mu,gX^i),}
with ${M=0}, \cdots ,{9}$, ${\mu} = {0}, {9}$ and ${i} = {1},
{\cdots}, {8}$.  We will decompose the gauge field into background and
fluctuation pieces,
\eqn\backdecomp{
{A_M} = {\bar A}_M + {g} {\tilde A}_M\ .}
The Feynman background gauge-fixed Lagrangian is
\eqn\gflang{
{\cal L} = {-}{\rm Tr} \left\{ {{1}\over{4g^2}} ({F_{MN}^2}) {+}
{{1}\over2{g^2}}
({\bar D}^M{A_M})^2 {-} {i}{\bar\psi}{\qi}{\psi}\right\} ,    
}
\noindent 
where ${\bar D}_M = {\partial_M} + {i}{\bar A}_M$.
Using the decomposition \backdecomp
\noindent 
we find
\eqnar\sdecomp{{\cal L}=-\Tr \Biggl\lbrace{{{\bar F}^2}\over{4g^2}} +\hf ({\bar
D}_M {\tilde A}_N)^2 + {i}{\bar F}^{MN} [{\tilde A}_M,{\tilde A}_N] {-}
{i}{\bar\psi}{\qi}{\psi} + \nonumber\\ \\[-\the\baselinetruc] \cr
{g} {\bar\psi} {\qA}{\psi}+{ig}{\bar
D}_M{\tilde A}_N [{\tilde A}^M,{\tilde A}^N] {-} {{g^2}\over{4}} [{\tilde
A}_M, {\tilde A}_N]^2\Biggr\rbrace. \cr\nonumber 
}
The amplitude in question is given by
\eqn\fetint{
{\cal A} (a,e) =  \int {\cal D}
{\tilde A}_\mu{\cal D}{\tilde X}^i{\cal D}{\psi}{e^{iS}},
}
where the boundary conditions on the functional integral are
chosen to correspond to the asymptotic behavior given in
\initstate\!\!,\clfield\!\!.   

If we write
\eqn\Asum{
{A_M} = {{1}\over{2}} {A_{Ma}}{\sigma}^a = {{1}\over{2}} {A_M}_+
{\sigma}^+ + {{1}\over{2}} {A_M}_- {\sigma}^- {+}\hf {A_{M3}}
{\sigma}^3,}
\noindent with $\sigma^{\pm}$ defined in terms of the usual 
Pauli matrices via 
\eqn\sigmadef{
{\sigma}^{\pm} = {{{\sigma}^1 \pm {i}{\sigma}^2}\over{\sqrt{2}}},}
\noindent 
then the couplings in \sdecomp include the
standard charged minimal couplings of ${A_+}$, ${A_-}$, ${\psi_+}$, and
${\psi_-}$ to the U(1) field ${\tilde A}_{\mu 3}$.  The amplitude to create
unit electric flux is therefore given by summing the loop contributions to
\fetint in which one of these charged particles circulates once around the
${\sigma}$-direction; higher encirclings yield more flux.  Therefore we
need the contribution of the charged state windings to the one-loop amplitude.
This immediately follows by reading off the spectrum
from the second through fourth terms of \sdecomp in the backgrounds 
\initstate and \clfield\!\!.  We begin by defining 
\eqn\phatdef{\phat^2 = g^2 p^2 + 4 g^4 e^2\ .}
In the bosonic sector we find the massless, neutral fields
\eqnar\neutf{
{\tilde X}_i^3,&\ i=1,\cdots,8;\quad m^2=0 \nonumber \\
 \\[-\the\baselinetruc]\cr
A_\mu^3,&~~~~~~~~~~~~~~~~~~~~~m^2=0\cr\nonumber 
}
\noindent and the charged fields 
\eqnar\bosmass{
{\tilde X}^\pm_i\, ,\,i=2,\cdots,8\quad;&\quad m^2=r^2\equiv \phat^2 
\tau^2 + a^2 + g^2b^2\cr
{{1}\over
%{\sqrt{\ehat^2+p^2}}
\phat} (gp {\tilde A}^{9\pm} {-} 2eg^2 {\tilde
X}^{1\pm})\quad;&\quad m^2=r^2 
\nonumber\\ \\[-\the\baselinetruc] \cr  
{\tilde A}^{0\pm} + {{i}\over %{\sqrt{\ehat^2+p^2}}
\phat} (2eg^2 {\tilde A}^{9\pm} + gp
{\tilde X}^{1\pm})\quad;&\quad m^2=r^2+2 %{\sqrt{p^2+\ehat^2}}
\phat\cr
{\tilde A}^{0\pm} - {{i}\over %{\sqrt{\ehat^2+p^2}}
\phat} (2eg^2 {\tilde A}^{9\pm} + gp
{\tilde X}^{1\pm})\quad;&\quad m^2=r^2-2 \phat %{\sqrt{p^2+\ehat^2}}
.\cr \nonumber
}
For the fermions, we have 16 massless uncharged states, 8 charged states
with masses $m^2 = r^2 + %\sqrt{p^2+\ehat^2}
\phat$, and 8 charged states with masses $m^2 =
r^2- %\sqrt{p^2+\ehat^2}
\phat$, as in \BeBe. Finally, including the ghosts gives one
complex, uncharged field ${C^3}$ with $m^2=0$ and one complex charged
field ${C^\pm}$ with $m^2=r^2$.

All of the charged fields are minimally coupled to the background field
${\bar A}_9 \equiv {\bar A}_\sigma$. 
At one loop level, we have
\eqn\oneloopamp{
{\cal A}_1 \rm {(a,e)} =  \int \prod_I {\cal
D}{\Phi_I}e^{iS^{(2)}[{\Phi_I}]}, 
}
where I labels the charged fields enumerated above (the
uncharged contributions cancel), arbitrary winding is allowed, and where
$S^{(2)}$ is the quadratic part of the action \sdecomp including the
coupling to ${\bar A}^3_\sigma$ through ${\bar D}$. Working with phase
shifts, we then have
\eqnar\lnamp{
i\delta_1= {\rm ln} {{\cal A}_1} = \sum_I {\rm ln} \sum_n \int_n
{\cal D}{\Phi_I} e^{iS^{(2)}[\Phi_I]} =
\qquad\qquad\qquad\qquad\qquad
\cr
- {\sum_I} (-{1})^{F_I}{\sum_n}{\int_0^\infty}{{dS}\over{S}}
\int_n{\cal D}{\tau}{\cal D}{\sigma}\,
\exp{i\int_0^S ds
\bigl\lbrack{\dot\sigma}^{\mu2}/2 - {\bar A}^3_\mu {\dot\sigma}^\mu
({s}) - {m^2_I(\tau)/2}\bigr\rbrack}.\cr
}
Here we have used the functional integral representation in
terms of the first-quantized trajectory ${\sigma^\mu}({s}) = ({\tau}
({s}),{\sigma}({s}))$, $n$ is the winding number about the cylinder, 
and ${F_I}$ denotes fermion number of the
field. 

For general winding ${n}$ the functional integrals in \lnamp are readily
rewritten in terms of functional determinants. For example, with
${m^2}={r^2}$ we have
\eqn\Pintrg{
{\int_n}{\cal D}{\tau}{\cal D}{\sigma}\,{e^{i\int_0^S ds\bigl\lbrack
{\dot\sigma}^{\mu2}/2 - g^2(p^2\tau^2+b^2)/2 -
{\bar A}_\sigma^3{\dot\sigma}({s})\bigr\rbrack}}
=e^{- ina+i n^2/2S-ig^2b^2S/2}\Delta(p,e,S), 
}
where
\eqn\Detexp{
\Delta(p,e,S) = {\rm
det}^{-1/2}\left(
\begin{matrix}
\partial^2_s-g^2p^2&-2g^2e\partial_s\cr
2g^2e\partial_s&-\partial_s^2
\end{matrix}
\right)\ .}
\noindent 
Combining such expressions and defining $S=2T$ then gives
\eqn\pephase{
i\delta_1(a,p,e)=\sum_n\,{e^{- ina}}\int_0^\infty {{dT}\over{T}}
{e^{in^2/4T-ig^2b^2T}}\Delta(p,e,T)
\biggl\lbrack-6-2{\cos}(2gp T)
+8{\cos}
({gp T})\biggr\rbrack.}
Recalling the quantization rule \equant\!\!, we see that as long
as $gp\gg 1$ the electric
background only contributes at higher order in $g$; neglecting this, the
determinant is readily evaluated using 
\eqnar\deteval{
{\rm det}^{{1}\over{2}}i(-\partial^2_s+g^2p^2)&=&{gp}\,{\prod^\infty_{n=1}}
\Biggl\lbrack\Biggl({{2n\pi}\over{S}}\Biggr)^2 +g^2p^2\Biggr\rbrack =
{2}{\sinh}\left({gpT}\right), 
\nonumber \\ \\[-\the\baselinetruc] \cr 
{\rm det}^{{1}\over{2}}i(-\partial^2_s)&=&{\sqrt{2\pi T\over i}}\ .\cr
\nonumber}
Here we used $\zeta$-function regularization. 

The phase shift then becomes
\eqnar\onephase{
i\delta_1(a,p,e)={1\over 2} \sqrt{i\over 2\pi}\sum_n\,e^{- 
ina}\int_0^\infty {{dT}\over{T}}
{e^{in^2/4T-ig^2b^2T}}
{1\over \sqrt{T} \sinh(gpT)}
\nonumber \\ \\[-\the\baselinetruc] \cr
\times \biggl\lbrack-6-2{\cos}(2gpT)
+8{\cos}
({gpT})\biggr\rbrack. \cr \nonumber  
}
In the full amplitude ${\cal A}_1=\exp\{i\delta_1\}$, 
the coefficient of the term
$e^{- ina}$ is the amplitude to make a transition from a state with electric
flux $e$ to $e+n$:
\eqn\ampdecomp{{\cal A}_1(a,e)= \sum_n e^{- ina} \langle e+n|e\rangle\ .}
We have found that this is independent of $e$ to order $g^2$. The
amplitude for a change by one unit of charge (e.g.\ two gravitons to
$D{\bar D}$ pair), as well as the effective interaction Hamiltonian, can be
derived from these expressions in the range
$p\ll b^2$.  There the integrand in \onephase can be expanded in $pT$ 
to find
\eqn\phasap{i\delta_1(a,p,e)\approx - {g^3p^3\over 2} \sqrt{i\over \pi} 
\sum_n e^{- ina} \int {dT} T^{3/2} e^{in^2/4T -ig^2b^2T}\ ;}
the leading order $D{\bar D}$ production amplitude is just the coefficient
of $e^{- ia}$ in the series.  From
\phasap and \phham we can also work backwards to extract the effective
Hamiltonian.  We find
\eqn\phasham{ H_{int}= -{15\over 8} g^{-3}p^4\sum_k {1\over 
\left[r^2 + (a+2\pi k)^2/g^2\right]^{7/2}}\ ,}
in agreement with \Hint\!\!.

\newsubsection\arbitraryn{Generalization to arbitrary ${N}$}

The one-loop calculation of the preceding subsection is readily
generalized to the case where the incoming and outgoing particles have
arbitrary (though discretized) ${p_{11}}$, 
or equivalently, ${N}$. In this case there are a
variety of different boundary conditions that may be placed on the
${N}{\times}{N}$ blocks. Two that we will consider are the trivial
boundary condition,
\eqn\trivbc{
X (\sigma + 1) = X (\sigma)\ ,}
and the single long string boundary condition,
\eqn\stringbc{X (\sigma + 1) = V^{-1} X(\sigma) V,
}
where ${V}$ is given in \cycle\!\!.

In the case of two incoming states with momenta $N_1,N_2,$ we write
\eqn\Xexpan{
X^i={\bar X}^i + {\tilde X}^i,
}
\noindent where $X^i$ is an $(N_1+N_2){\times}(N_1+N_2)$ matrix. In
particular, the background is taken to be
\eqn\Xbackg{
{\bar X}^1 ={{p\tau}\over{2}}\left\lgroup
\begin{matrix}
I/N_1&0\cr
0&-I/N_2\cr
\end{matrix}
\right\rgroup
\equiv {{p\tau}\over{2}}\,T_D \qquad
{\bar{X^2}} ={{b}\over{2}}\,T_D , 
}
where we have split the matrix into $N_1 {\times} N_1$ and
$N_2\times N_2$ blocks corresponding to two ``clusters,'' and $I$
represents the corresponding identity matrices.

A useful decomposition of the fluctuations ${\tilde X}^i$ is in terms of
the matrices
\eqn\genmatr{
T^{a_1}=\left\lgroup
\begin{matrix}t^{a_1}&0\cr
0&0\cr
\end{matrix}
\right\rgroup,\ 
T^{a_2}=\left\lgroup
\begin{matrix}0&0\cr
0&t^{a_2}\cr
\end{matrix}\right\rgroup,
}
where ${t^{a_i}}$ are hermitian generators of $SU(N_i);
T^{\alpha_1\alpha_2}_+, T^{\alpha_1\alpha_2}_-$, which have matrix
elements
\eqnar\Matrdefs{
(T^{\alpha_1\alpha_2}_1)_{\beta_1\beta_2}&={\sqrt{2}}\delta_{\alpha_1\beta_1}
\delta_{N_1+\alpha_2\beta_2},\nonumber\\ \\[-\the\baselinetruc] \cr
(T_2^{\alpha_1\alpha_2})_{\beta_1\beta_2}&={\sqrt{2}}
\delta_{N_1+\alpha_1\beta_1}
\delta_{\alpha_2\beta_2};\cr\nonumber
} 
and $T_D$:
\eqn\XTexpan{
{\tilde X}^i= {{\tilde X}_D\over{2}}T_D + {{\tilde X}_{a_1}\over{2}}T^{a_1} +
{{\tilde X}_{a_2}\over{2}}T^{a_2} + 
{{\tilde X}_{+\alpha_1\alpha_2}\over{2}}T_+^{\alpha_1\alpha_2} +
{{\tilde X}_{-\alpha_1\alpha_2}\over{2}}T_-^{\alpha_1\alpha_2}. }
Following the preceding subsection (and working with $e=0$ for
simplicity) we find that the charged states now have an extra $N_1N_2$
in their multiplicities, and have masses
\eqnar\Nmasses{
{\tilde X}^i_{\pm\alpha_1\alpha_2}:\quad m^2&={r^2\over 4\nu^2}\cr\cr
{\tilde A}^9_{\pm\alpha_1\alpha_2}:\quad m^2&={r^2\over 4\nu^2}
\nonumber\\  \\[-\the\baselinetruc] \cr
{\tilde A}^0_{\pm\alpha_1\alpha_2}+{\tilde X}
^1_{\pm\alpha_1\alpha_2}:\quad m^2&={r^2\over 4\nu^2} + {gp\over \nu}\cr\cr
{\tilde A}^0_{\pm\alpha_1\alpha_2}-{\tilde X}
^1_{\pm\alpha_1\alpha_2}:\quad m^2&={r^2\over 4\nu^2} - {gp\over \nu}\cr
\nonumber
}
\noindent where
\eqn\nudef{ {1\over \nu }= {1\over N_1} +{1\over N_2}\ .}
Likewise, the charged fermions and ghosts have masses as in
the $N=2$ case with the trivial rescalings to 
\eqn\pbreplace{
{\bar p} = {p\over 2\nu}\ ,\ 
 {\bar b} ={b \over 2\nu} . 
}
Therefore, in the case of trivial boundary conditions the amplitude (and
Hamiltonian) is
exactly as computed in \onephase with the only 
difference being multiplication by $N_1N_2$ and replacement of $p$ and
$b$ as in \pbreplace\!\!. Note that $2{\bar p} = {{p}\over{N_1}} +
{{p}\over{N_2}}$ is simply relative velocity of the two
clusters, and ${\bar b} = {{1}\over{2}} ({{b}\over{N_1}} + 
{{b}\over{N_2}})$ is precisely the impact parameter between the
clusters.

In the case of long-string boundary conditions, this result is modified.
Now
\eqn\lsbc{
X(\sigma+2\pi)=\left\lgroup
\begin{matrix}
V^{-1}_1&0\cr
0&V^{-1}_2\cr
\end{matrix}
\right\rgroup
X(\sigma)\left\lgroup\begin{matrix}V_1&0\cr
0&V_2\cr\end{matrix}\right\rgroup 
}
and in particular the charged off-diagonal blocks satisfy
twisted boundary conditions
\eqnar\TwBC{
X_+ (\sigma + 1)&=V_1^{-1}X_+ (\sigma) V_2, \nonumber\\ 
\\[-\the\baselinetruc] \cr
X_- (\sigma + 1)&=V_2^{-1}X_- (\sigma) V_1\ .\cr\nonumber
} 
\noindent 
The matrices $V$ can be diagonalized by working on basis
vectors
\eqn\Newbasis{
w_k={{1}\over{\sqrt{N}}}\left\lgroup
\begin{matrix}
1\cr
\lambda^k\cr
\lambda^{2k}\cr
\vdots\cr
{\lambda}^{(N-1)k}\cr\end{matrix}
\right\rgroup\ ,\ \lambda=e^{2\pi i/N},\ k\in\Zvet
}
and in this basis simply give phases ${\lambda^k}$. Thus the
amplitude \onephase  is modified to 
\eqnar\Nonephase{
i\delta_1= \hf \sqrt{i\over 2\pi}
\sum^{N_i}_{\alpha_i=1}\sum_n\,\, 
e^{-in\left[2\pi(\alpha_1/N_1-\alpha_2/N_2)+a\right]}
\int^\infty_0 \,\,{{dT}\over{T}}
{e^{in^2/4T-ig^2{\bar b}^2T}}{1\over \sqrt{T} \sin(g{\bar p} T)}
\nonumber \\ \\[-\the\baselinetruc] \cr 
\times 
\biggl\lbrack -6-2\,{\cos} (2g{\bar p}T) + 8\,\cos(g{\bar
p}T)\biggr\rbrack\ , \cr\nonumber 
}
and the interaction Hamiltonian takes the form
\eqn\intham{H_{int}\approx {g^4p^4\over 2\pi i}
\sum_{n,\alpha_1,\alpha_2} 
e^{-in\left[2\pi(\alpha_1/N_1-\alpha_2/N_2)+a\right]}
\int {dT} T^2
e^{in^2/4T - ir^2T}\ .
}
For non-zero $n$, the supergravity 
correspondence no longer holds when $ N>2 $: the matrix string 
then yields a different result.
%due to the fact that we have to sum over 
%the fractional Fourier frequencies $n_1$ and $n_2$, which has the
%effect of restricting the minimal exchanged D-particle number
%to $N_1N_2$ (in case $N_1$ and $N_2$ are relatively prime). 
In \intham the expression in the summation only gives a non-zero 
contribution  when the integer $n$ is a multiple of both $N_1$ and $N_2$. 
Hence we see that the minimal exchanged D-particle number 
between two long strings of length $N_1$ and $N_2$ must be proportional to
$N_1N_2$ (if the lengths are relatively prime), else the amplitude will 
be simply zero. 

This leads us to the conclusion that the long strings do not give an effective
means of creating D-particles. For two strings to create a minimally charge
D-pair, the strings apparently first need to each emit a minimal length string
%fractionate into a collection 
%of $N_1$ respectively $N_2$ short strings, 
such that two short strings of 
both collections can exchange a single D-particle. In the SYM language,
this last process is effectively an $SU(2)$ process, where correspondence
with 11-dimensional supergravity is found. It is important to note, however,
that the electric flux thus created must then subsequently spread out
over the complete $U(N_i)$ gauge group, since otherwise it would not carry
the SYM energy appropriate for the massive D-particle with $p^+ = N_i/R$.

Furthermore, in the sector with a fixed $p^+$ momentum, we now have an
improved
idea of what state contributes most to D-charge production: it is the state
with trivial boundary conditions, \trivbc\!\!, corresponding to a collection of
minimal length strings.  Since this is the state that yields amplitudes
agreeing with low-energy supergravity in the limit $g\rightarrow 0$, it is
apparently this state (or a bound version of it when finite $g$ effects are
taken into account) that dominates the wave-function of the graviton in the
small $g$ region, rather than the state with the long
string boundary conditions \stringbc\!\!.

\newsubsection\nonamerange{Ranges of validity}

In this section we will give a preliminary discussion of the relevant
scales and ranges of validity of the calculations of the preceding
sections.  This analysis is preliminary in that the systematics of the
perturbation theory for the Yang-Mills Lagrangian \mstaction has not
been performed at the level of that for pure matrix theory \BBPT\ and
additional subtleties are possible.  We leave such analysis for future work.
For simplicity we will consider the case where the $p^+$ momentum of the
two incoming states are comparable, $N_1\sim N_2\sim N$.  Our arguments
readily generalize to the probe situation $N_1\gg N_2$. 

We begin by considering the expansion of the action about a classical
background as in \sdecomp\!\!; 
such a treatment is relevant both for corrections
to the saddle-point solutions of section four as well as for the systematic
treatment of D-pair production.  

This expansion is governed by the Yang-Mills coupling $\gym$, and naively one
expects the condition $\gym\ll1$ for corrections to be small.  However, as
mentioned in the introduction, the Yang-Mills coupling is scale dependent
and one expects the relevant scale to be set by the physics one is
considering.  

For example, in the scattering with background \initstate\!\!, 
loops of the charged, 
massive states of the YM theory play a central role.  One either has a
loop localized on the cylinder, whose calculation leads to the ${\cal
O}(v^4/r^6)$ supergravity potential, or the loop can encircle the cylinder
leading to the D-pair production that we have computed.  These massive states
receive masses of minimum size $b/g_s$ through the Higgs effect, setting the
length scale $\ell_b\simeq g_s/b$.  At this scale, we expect the relevant
dimensionless parameter to be 
\eqn\gymb{ \gym \ell_b \simeq {1\over b }\ .}
Smallness of this parameter thus requires 
\eqn\bcondit{b\gg1\ .}
For the case of pair creation, there is another requirement arising from
the condition that the back reaction due to the created electric field be
small.  One way of stating this is to require that the YM energy be large
as compared to the energy stored in the electric field,
\eqn\YMenerg{ {p^2}\gg \gym^2\ .}

Finally, in the case of the string interactions of section four, we see from
\rlimitp that the relevant scale is set by the parameter $|B|$, and is given
by 
\eqn\Bscale{ \ell_{inst} \simeq \left({g_s\over |B|}\right)^{2/3} \simeq
\left({g_s^2 N\over p^2 \sin \theta}\right)^{1/3} 
 \ .}
At this scale the dimensionless coupling is given by
\eqn\Bymc{ \gym \ell_{inst} \simeq \left({N\over g_s p^2 \sin
\theta}\right)^{1/3}\ .}
Another condition to apply the methods of section four is that the size of
the instanton be small as compared to the size of the cylinder,
$\ell_{inst}\ll 1$, or 
\eqn\smallinst{ p^2 \gg {g_s^2 N\over \sin\theta}\
.}

It is certainly possible to simultaneously satisfy the conditions 
\bcondit\!\!, \YMenerg\!\! and \smallinst\!\!, 
as well as the more stringent condition $\gym\ll1$, 
for finite $N$ and large $s\sim p^2$.  If all important corrections are
governed by expansions in the parameters of \gymb and \Bymc\!\!, 
then it appears possible to even push the calculations into the range
$g_s\roughly<1$.  

A more complete analysis can be performed in the large $g_s$ (large $R_9$)
case 
in the restricted energy range
\eqn\Matrange{ {1\over g_s} \ll  E \ll g_s \ .}
The
lower bound corresponds to the energy threshold to create D charge, and the
upper bound is the energy to create winding states wrapping $x^9$.  In
between these bounds the theory can be effectively described by matrix
theory DLCQ quantized in 10 dimensions.  

As explained in \BBPT, the matrix expansion is an expansion in 
terms of the form
\eqn\matexpp{\left({N \over \Mpl^3r^3}\right)^L 
\left(v^2 \over R^2\Mpl^6 r^4 \right)^n, 
}
where $L$ counts loops and $\Mpl$ is the eleven dimensional 
Planck mass.  
The terms with $L=n$ are readily identified with
terms in the corresponding supergravity expansion, and the small parameter
justifying this expansion is \BBPT\BGL 
\eqn\sugrapar{ {N v^2\over R^2 \Mpl^9 r^7}\ll 1  .}
This has a simple physical interpretation, which is easily seen by
estimating the net transverse momentum transfer due to the potential
\eqn\approxpot{{N^2 v^4\over R^3 r^7}\ ;}
this gives 
\eqn\transvp{ \Delta p_\perp = {N^2 v^3\over R^3 b^7}\ .}
The condition \sugrapar is then easily seen to be $\Delta p_\perp \ll p$,
or equivalently $\theta\ll 1$ where $\theta$ is the scattering angle.
%Combining this with the threshold condition \YMenerg then yields
%
%\eqn\gpsll{g_s p \gg 1\ , }
%
%in correspondence with \Matrange.

Expansion terms with $n>L$ are then suppressed for 
\eqn\nexpp{ p^2 \ll g_s^{-2}N^2 r^4  \ ,
} 
and terms with $n<L$ for 
\eqn\Lexpp{r \gg (Ng_s)^{1/3} \ .}
It is unclear whether the latter condition is strictly necessary; the first
term in this expansion vanishes \BeBe\BBPT, and the other terms
have been conjectured to vanish in \BGL.

To better understand these conditions, we convert them into statements
relating the Mandelstam parameter $s\sim p^2$ and the angle $\theta$.  It
is easily seen that condition 
\nexpp\!\!, using \transvp\!\!, becomes 
\eqn\ntheta{s \ll N^2 \left({N\over \theta}\right)^{4/3} g_{s}^{-14/3}
}
and the condition \Lexpp\ becomes 
\eqn\Ltheta{s\gg g_s^{7/3} N^{10/3} \theta\ .}
Comparing \smallinst with \ntheta and \Ltheta, we see that within the
energy range given by \Matrange the instanton and D-particle production
calculations are not obviously simultaneously valid. However by
relaxing the restrictions by giving up \Lexpp (and \Ltheta\!\!), 
which maybe justified due to a non-renormalization theorem, it seems 
that it is possible that the calculations are simultaneously valid. 
Outside the energy range \Matrange 
we appeal to the preceding (less rigorous) analysis which
suggests that these calculations are indeed simultaneously valid at large
$s$, and may even be extendable to $g_s\roughly < 1$.  It is partly on this
basis that we will, in the next section, consider the consequences of
combining these two calculations.

\newsection\nonamedis{Discussion and conclusions}

We begin this section by recalling several observations from our preceding
discussion. The first is that, as pointed out in section \arbitraryn\!\!, 
string
scattering only efficiently produces D charge if the strings break off at
least one minimal length string.  Furthermore, sections four and five
discussed saddle-point configurations that are expected to make important
contributions to high energy, fixed angle scattering.  Combining these
yields a picture of how the important non-perturbative process of D-charge
production can arise in high-energy string scattering.

The analysis of Gross and Mende \grossmende\ found saddle-points believed to
dominate scattering at high energy.  These saddle-points have a common
structure at arbitrary genus, and the contributions of these
saddle-points grows with the genus suggesting the relevance of
non-perturbative effects.  We have found a new version of their analysis in
which a mechanism appears that can cut off this growth.  The cutoff
originates from the minimal string length, which is in our language the
minimal $p^+$.  String fragmentation is 
stopped when the string breaks into the maximal
number of minimal-length strings.  
%\footnote{In this sense, the matrix string formalism 
%behaves very similar to the discretized models of string theory,
%advocated in particular by C. Thorn. We
%thank S. Shenker  for drawing our
%attention to this similarity.}

It is precisely in the context of minimal length string scattering that we
have found that D-charge pair production can become an important effect.
We therefore have a very nice picture in which the instantons of section
four and five lead to maximal fragmentation of the strings, and this is 
followed by
the production of D-charge via the process of section six.  
Here we expect that the size \rlimitp of the instanton,
as well as the corresponding minimal distance \dlimit\!\!,
may be an important ingredient in determining the size of
both these effects.

From the
stringy viewpoint this is an intrinsically non-perturbative process.  This
is suggestive that there is in fact a basic connection between these two
processes, and in particular that the non-perturbative production of
D-charge is an important correction to the high-energy scattering analysis
of \grossmende.  While we believe that, by combining the various ingredients 
presented in this paper, it may be possible to obtain definite quantitative
estimates of these corrections, we leave further analysis of this connection 
for future work.

Next we turn to several other observations and connections.

First, recall that Banks and Susskind \BaSu\ previously considered
the $D{\bar D}$ system in the context of perturbative string theory.  There
they found an instability with unknown outcome.
In the present framework we have been able to treat the same system
analytically, at least in the large $g_s$ limit, without signs of
pathology.  In principle, the matrix string calculations appear to extend
to arbitrary $g_s$.  
One might hope that some extrapolation of our approach could shed
further light on the discussion of \BaSu.

It is an interesting conceptual question under which circumstances
one needs to include the virtual effects of D-particles propagating in loops.  
Although in the literal sense of an expansion about $g_s=0$ they do not
contribute, since they have infinite mass there, there is clearly a 
strong sense in which D-particles can be found in intermediate states 
when $g_s$ is finite.  
Indeed, intermediate states with D-charge are distinguished
from other intermediate states only by the presence of electric flux, and
there is no apparent reason why these should be suppressed at finite $g_s$.
In fact, looking at the results of section \hbox{\dpairprod\!\!,}
leads one to suspect that 
it may be possible to extend the matrix string interactions as
discussed in chapter two,  
to include the possibility of electric flux ``pair'' creation. 
The eleven dimensional symmetry of M-theory, in particular,  
suggests as a possible generalization of the DVV string-interaction vertex,
an expression of the form $V_{int} = V_{twist} \delta(A_{12})$ 
(with $A_{12}$ the difference between the U(1) gauge fields on the two 
strings that are created). With this choice of vertex, the couplings
between string and all $n$-D-particle bound states are all of the same 
strength. This would suggest that there may possibly exist a systematic
semi-classical expansion in string theory -- generalizing the standard
perturbative expansion -- in which the D-particle-loop contributions play
the role of instanton-like corrections. Indeed, in other recent 
studies of non-perturbative contributions to string scattering 
amplitudes \Greo\Gret\GGV\GrVh\ it was suggested that D-particle loops 
are related via T-duality to D-instanton contributions in IIB string 
theory. It clearly would be interesting to see if the suggestive
formulas obtained in these works can possibly be reproduced via the 
matrix string methods developed in this paper.

To conclude, we have succeeded in using the matrix string approach to
begin an investigation of 
aspects of high energy string scattering, and in particular to
begin to explore the role of important non-perturbative (from the string
viewpoint) processes such as D-charge production.  Further investigation
along these lines is expected to unravel a rich structure at substringy
scales, and may shed further light on the short distance structure and
fundamental degrees of freedom and dynamics of M-theory.

\listrefs

\newpage \ifodd\thepage \relax \else \phantom{bla} \newpage \fi 
\lhead[\fancyplain{}{\sc Acknowledgements}]{\fancyplain{}{}}
\thispagestyle{empty}
\noindent 
\\[50pt]
{\phantom{tk} \hfill 
\Huge\rmfamily\bfseries\slshape Acknowledgements}
\testclino
\write\cfile{\bf Acknowledgements &&\bf \thepage\noexpand\\[1.5mm]}
\noindent
\\[2cm]
On this last page I would like to thank a number of people
for their contribution to this thesis. First of all I want to 
thank my advisor Herman Verlinde. 

Herman, I admire your knowledge of physics and in particular your 
knowledge of string theory. Also, I appreciate your creativity
and optimistic and friendly character very much. 
Besides that I got the opportunity to visit workshops, conferences
and universities abroad, with financial support of your NWO 
Pionier grant. Thanks a lot for everything.  

It was a pleasure for me to write an article together with Steve
Giddings and Herman Verlinde. I thank them for enabling me to make 
my own (small) contribution to a better understanding of the relation 
between matrix string theory and light-cone string theory. 
  
I thank Jae-Suk Park for sharing his enthusiastic ideas and knowledge
with me. I admire his erudition and I enjoy our conversations
about physics and beyond very much. 
Also, I would like to thank Boksun Han for delicious Korean dinners. 

The discussions with Christiaan Hofman and Gysbert Zwart 
were very important for me. Gysbert and Christiaan: thanks a lot. 

I benefitted from useful discussions with Bernd Schroers, Sander 
Bais, Tom Wynter and Boris Pioline. 

Bernd Schroers, Bert-Jan Nauta and Ronald van Elburg are acknowledged 
for reading parts of the manuscript of this thesis. 

I thank Alain Verberkmoes for his patience and his willingness to
solve problems I had with the amazing world of LaTeX, TeX,
Postscript and Mathematica. In particular I thank him for producing 
figure \alain. 
 
Large parts of this thesis were written during two stays in Princeton.
I thank the department of physics of Princeton University for 
hospitality. Furthermore I thank the Spinoza Institute, Utrecht
University, for hospitality. 

I thank the proprietors and crew of {\it small world coffee} 
in Princeton for kindly supplying me with coffee of reasonable quality. 

Last I thank my family and friends for their support.

\write\cfile{\noexpand\end{tabular}}
\closeout\cfile
\end{document}